\documentclass[a4paper,11pt]{article}
\usepackage{amsmath}
\usepackage{amssymb}
\usepackage{tikz}
\usepackage{pgfplots}
\pgfplotsset{compat=newest}
\usepackage[T2A]{fontenc}      
\usepackage[utf8]{inputenc}
\usepackage[russian, english]{babel}

\usepackage{graphicx}
\usepackage{gensymb}
\usepackage[top=3cm, bottom=3cm, right=2cm, left=2cm]{geometry}
\usepackage{matlab-prettifier}
\usepackage{caption}
\usepackage{subcaption}
\usepackage{longtable}
\usepackage{pdflscape}

\newcommand{\m}[1]{\mathbf{#1}}
\newcommand{\A}[4]{\mathcal{A}^{(#1,#2)}_{#3,#4}\left(\mathbf{U}^{(#3,#1)} - \mathbf{U}^{(#4,#2)}\right)}

\usepackage{placeins}

\usepackage{hyperref}

\usepackage[compress]{cite}

\title{\bf 
Global lattice models of a rotating planet - view on shake, rattle and roll}

\author{{\bf W.E. Barnfield$^1$, A. Kandiah$^1$, V. Frid$^2$, I.B. Movchan$^3$, A.B. Movchan$^{1}$
} \\ \vspace{0.02cm} \\
$^1$Department of Mathematical Sciences, University of Liverpool, Liverpool 
L69 7ZL, UK \\
$^2$Sami Shamoon College of Engineering, Jabotinsky 84, 77245 Ashdod, Israel \\ 
$^3$St Petersburg Mining University, St Petersburg 199106, Russia
}

\begin{document}

\maketitle


\begin{abstract} 

 A new approach, based on the analysis of a spectral problem for a discrete gyroscopic system, to modelling earthquakes on a rotating planet is presented in this paper. The eigenmodes of gyroscopic three-dimensional icosahedron-dodecahedron lattices are used for the discrete approximations of Earth's vibrations.  Related model of the gyroscopic centred icosahedron lattice reveals new results about the motion of Earth's core. The vibrations of the tectonic plate boundaries, inducing shear and longitudinal motions of the ground, are described through the oscillations of the nodal elements within the discrete gyroscopic models. The theoretical analysis is complemented with the illustrative numerical examples linked to natural resonant vibrations observed during large earthquakes.

\end{abstract}



\section{Background and key observations}
Analysing seismic activities on a global scale presents formidable challenges, particularly in capturing the complex interplay of vibrational modes across Earth’s tectonic framework. Existing models \cite{rundle2003statistical, moczo2014finite, lu2011large, minster1974numerical, shearer2019introduction} often do not  account for the dynamic interconnections between distant fault systems and the rotational influences of the planet itself.
The term ``geodynamic'' can be seen being used in the framework of creep models (cold flow), where the inertia terms are omitted from the governing equations (see, for example, \cite{ghosh2019role, bahadori2019geodynamic}); such models, despite being referred as ``geodynamic'',  do not describe seismic vibrations.  On the other hand, the experimental data, used in  \cite{ghosh2019role, bahadori2019geodynamic}, show significant tectonic changes occurring within the latitude band between $30^\circ$ N and $40^\circ$ N, which is directly linked to the rotational eigenmodes of vibrations of Earth, as explained in our present paper.
   
\begin{figure}[h]
    \centering
    \begin{subfigure}[h]{0.45\textwidth}
        \includegraphics[width=\textwidth]{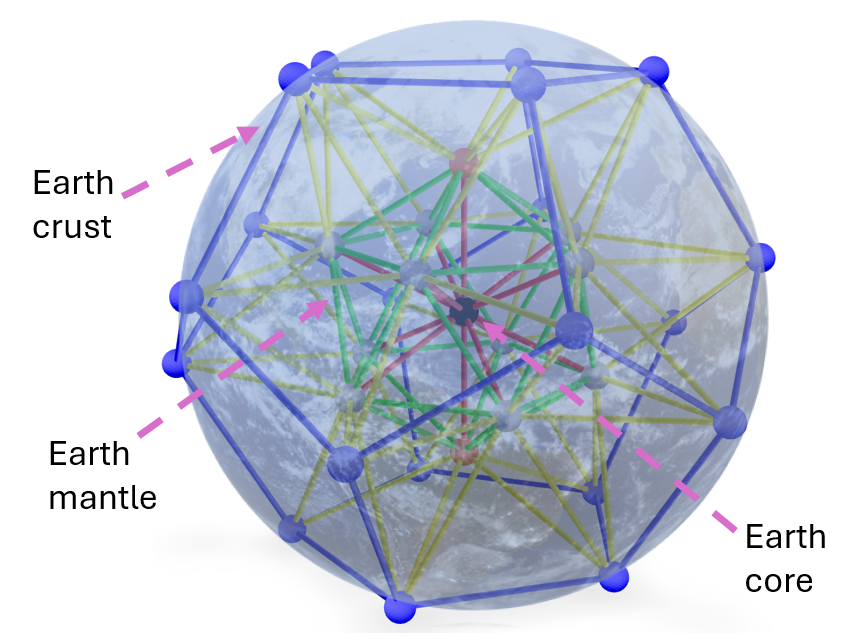}
        \caption{} \label{partaphysicalEarthicosahedron}
    \end{subfigure} \hspace{.5in}
    \begin{subfigure}[h]{0.33\textwidth}
       \includegraphics[width=\linewidth]{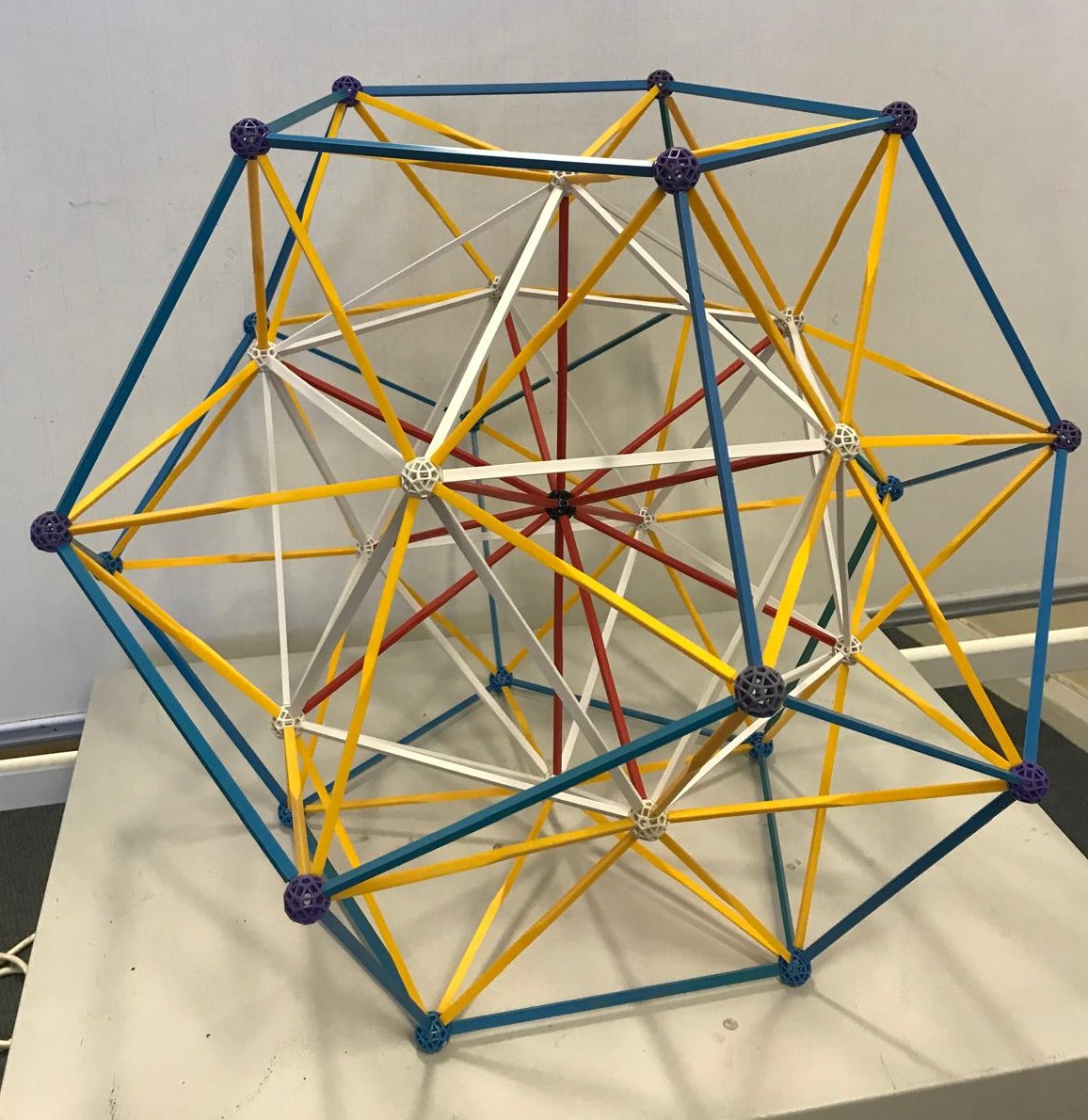}
       \caption{}\label{partaphysicalEarthicosahedron2}
    \end{subfigure}
       \caption{\small (a) The chiral elastic icosahedron-dodecahedron lattice model of Earth, incorporating the core, mantle and crust inertial elements. 
       (b) The geometrical/physical model of the icosahedron-dodecahedron lattice, built by the authors, which incorporates the dodecahedral outer structure, the core, and intermediate icosahedral frame. }
       \label{physicalEarthicosahedron}
\end{figure}

 By discretising Earth's structure into three-dimensional gyroscopic lattice networks, as shown in Fig. \ref{physicalEarthicosahedron}, 
this approach identifies eigenmode patterns that simulate the full spectrum of tectonic vibrations, including shear, longitudinal, and core-induced oscillations. These nodal vibrations not only reflect localised seismic responses but also reveal systemic links between regions, explaining how clusters of earthquakes can emerge from the redistribution of energy across interconnected tectonic elements. 

\FloatBarrier

Creep models, describing strain rates over long time intervals, appear to be 
combined with GPS data and  glacio-isostatic adjustment (GIA) framework.  
Recent efforts to model the stress distributions contributing to intraplate earthquakes, particularly in regions such as Central and Eastern North America, underscore both the ambition and complexity of tectonic research in low-strain continental settings. The paper \cite{ghosh2019role} presents an attempt, based on a creep model, to capture the slow deformation, employing GPS-derived strain rates and incorporating glacio-isostatic adjustment (GIA) processes into the model framework. Although this approach offers valuable insights into the possible sources of crustal stress, certain methodological constraints, such as the limited resolution of GPS data below $1$ mm/year and the application of continental stress models to oceanic domains, highlight areas where further refinement is required. The reliance on the stress tensor interpretations and inversion methods also points to an opportunity for expanding the mathematical and physical depth of tectonic plate modelling. Also,  \cite{ghosh2019role} mentions a significant discrepancy in the evaluation of the contractional strain rates  within the latitude band between $32^\circ$ N and $40^\circ$ N for the ICE-6G and CORS GPS models, which confirms the importance of rotational dynamics in the interpretation of data, especially in the latitude band associated with the eigenmodes of vibration (as shown in Fig.  \ref{balls} and Fig. \ref{balleig}).

To address the global challenge of modelling the dynamics of interplate seismotectonics, we present a mathematically rigorous and fundamentally distinct gyroscopic modelling framework based on the Earth's natural elastic vibrations. This spectral lattice methodology is based on the icosahedron-dodecahedron framework, which discretises the planet into interconnected nodal elements exhibiting gyroscopic behaviour, allowing for the analysis of seismic vibration modes and dynamic interactions, which take into account the rotation of the planet. 
We investigate the connections between chiral waves on discrete three-dimensional gyroscopic multi-structures and continuum geophysical models. 
Spectral analysis of multi-scale gyroscopic systems is not a conventional tool used in geophysics. However, this effective tool has proved to be very powerful in the analytical description of chiral waves linked to the rotational dynamics of Earth and other planets, such as Saturn and Jupiter (see, for example, \cite{kandiah2024controlling, kandiah2025dispersion, kandiah2025dispersion2}).

The natural elastic vibrations of the Earth are noted to occur  in the millihertz frequency range
\cite{deen2017first}, while the frequencies of seismic waves span a wider range \cite{aki1980attenuation}; 
there is a notion of the ``hum of the Earth'' linked to low-frequency vibrations due to the Earth's expansion and contraction, as discussed in
\cite{webb2007earth, suda1998earth, kobayashi1998continuous, tanimoto2005oceanic}. 
We develop a new spectral model 
focused on three-dimensional lattices, which discretise the Earth's structure into interconnected nodal elements with gyroscopic properties that approximate the 
effects linked to the rotation of the planet, 
and the vibrations during large-scale earthquakes. This novel framework is analytical and provides an approximate method for modelling global tectonic interactions with computational efficiency and improved scalability.
The new results provide groundbreaking insights on the formation of earthquakes, via the analysis of spectral problems for three-dimensional gyroscopic elastic lattice systems.
 In two-dimensional approximations, the study 
 presented in \cite{kandiah2025dispersion, kandiah2025dispersion2},  incorporated gyroscopic elements in discrete lattice strip models 
 linked to 
flows in atmospheric and oceanic wave phenomena subjected to the Coriolis force. 

\begin{figure}[h!]
    \centering
    \begin{subfigure}[h]{0.71\textwidth}
        \centering
        \includegraphics[width=\textwidth]{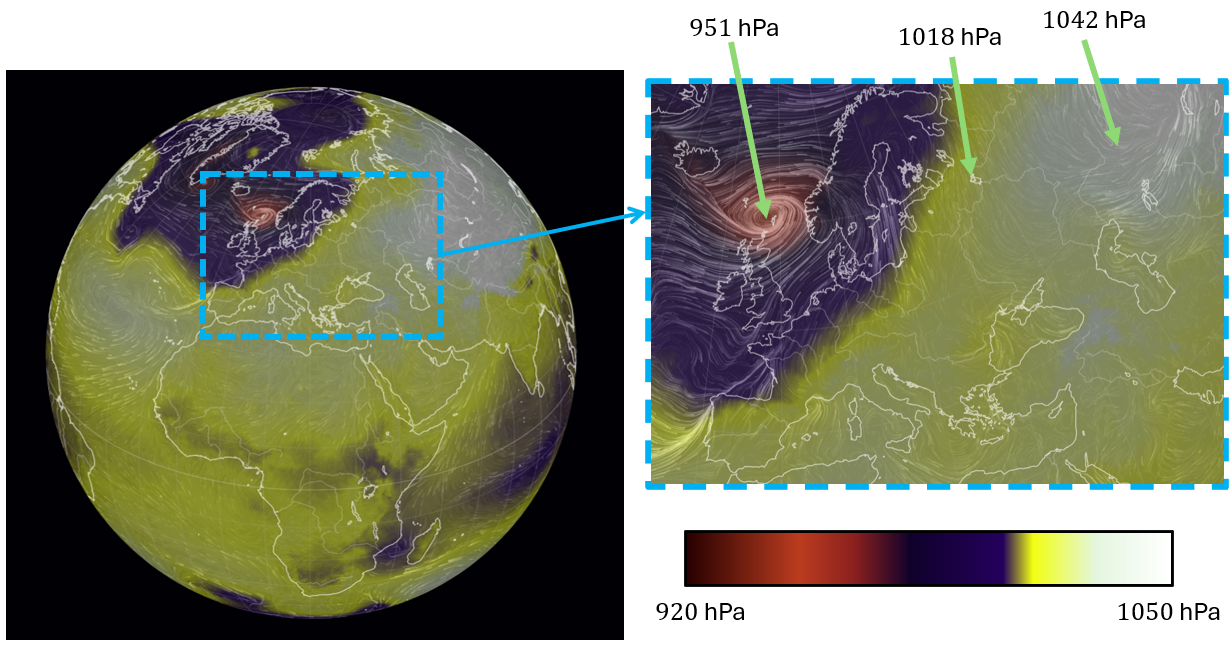}
        \caption{} \label{oceanographic1a}
    \end{subfigure} \\
    \begin{subfigure}[h]{0.4\textwidth}
        \centering
       \includegraphics[width=\linewidth]{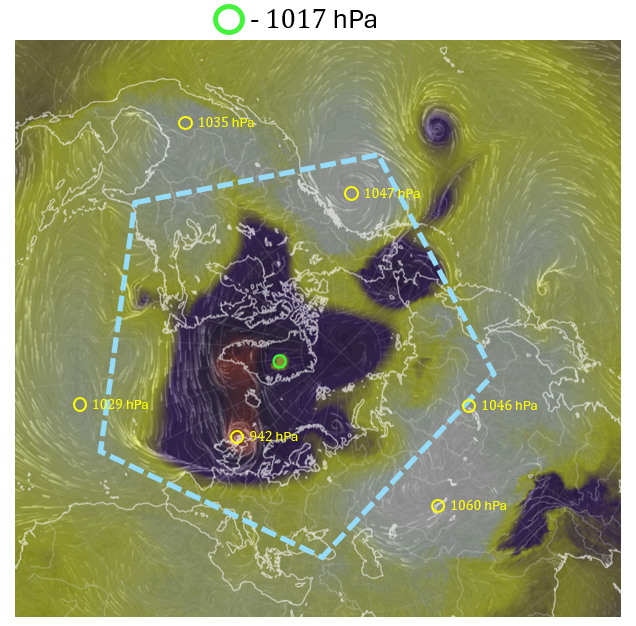}
       \caption{} \label{atmosphericoas1}
    \end{subfigure}
       \caption{\small (a) Mean sea level pressure on the surface of the Earth on 24/01/25 at 21:00 UTC, analysis uses the data available at  \url{https://earth.nullschool.net/\#2025/01/24/2100Z/wind/surface/level/overlay=mean\_sea\_level\_pressure/orthographic=20.28,30.85,336/loc=-158.496,58.055}. (b) Stereographic projection illustrating the surface wind and the mean sea level pressure of the Earth on 24/01/25 at 09:00 UTC. The localised high pressure region, corresponding to $1017$ hPa, is located at $74.35^{\circ}$ N and $37.84^{\circ}$ W. The analysis uses the data available at \url{https://earth.nullschool.net/\#2025/01/24/0900Z/wind/surface/level/overlay=mean\_sea\_level\_pressure/stereographic=43.38,90.47,517/loc=-37.839,74.346}. }
       \label{stormsocean}
\end{figure}
\FloatBarrier

\subsection{Record of extreme pressure changes connected to earthquakes}
In the recent months, several major seismic phenomena occurred across the globe over a relatively short period of time (October 2024 to March 2025). These seismic events have led to unusual wave patterns recorded by seismologists, as well as to observation of standing waves in the atmosphere. The paper \cite{carbone2021mathematical} makes a connection between major seismic clusters, being active at the same time, and formation of wave patterns in the atmosphere.

One of such events was recorded on the 24th January 2025 at 21:00 UTC, where the mean sea level pressure values were varying abruptly from 951 hPa to 1042 hPa, as shown in Fig. \ref{oceanographic1a}. Fig. \ref{atmosphericoas1} shows the stereographic projection 
and differences in pressure regions twelve hours earlier, at 09:00 UTC on the 24th January $2025$, showing even more extreme variations from 942 hPa to 1060 hPa, and demonstrating the gradual changes in the position of the polar axis of the Earth. This figure also includes  the approximate pentagonal pattern of the atmospheric vortex waves, which are influenced by the jet stream, rotation of the planet, and variations in pressure gradients.
Despite extensive studies of earthquake phenomena, the wide range of examined models rely on numerical simulations and empirical observations of geophysical phenomena \cite{de2016statistical, arrowsmith2010seismoacoustic, rundle2003statistical, anagnos1988review, kagan1994observational, moczo2014finite, lu2011large, minster1974numerical}, which often struggle to capture the complexity of Earth's properties in modelling of seismic events,  
including the rotational dynamics of the Earth.

Special attention shall be given to Fig. \ref{atmosphericoas1}, where extremely low pressure of 942 hPa is shown in the North of Ireland. The storm \'{E}owyn, which affected Ireland, the Isle of Man and the North of Scotland appears to be the result of the large pressure gradient variations, induced by earthquakes. In this context, the atmospheric conditions observed in January 2025 resembled those of January 1884, with both periods preceded by very strong seismic activities around the globe  that induced standing waves and regions of extreme high and low pressure in the atmosphere. In both cases, extremely low values of mean sea level pressure were recorded; 926.5 hPa was recorded in the North of Scotland in January 1884 \cite{the1884, historic1930}. We emphasise the pentagonal pattern, shown in Fig. \ref{atmosphericoas1}; this is an important  feature of the Earth vibrations, as discussed below.

\subsection{Icosahedron-dodecahedron shapes in vibration patterns}
\begin{figure}[h!tb]
\centering
\begin{subfigure}[h!]{0.42\textwidth}
		\centering
		\includegraphics[width=\linewidth]{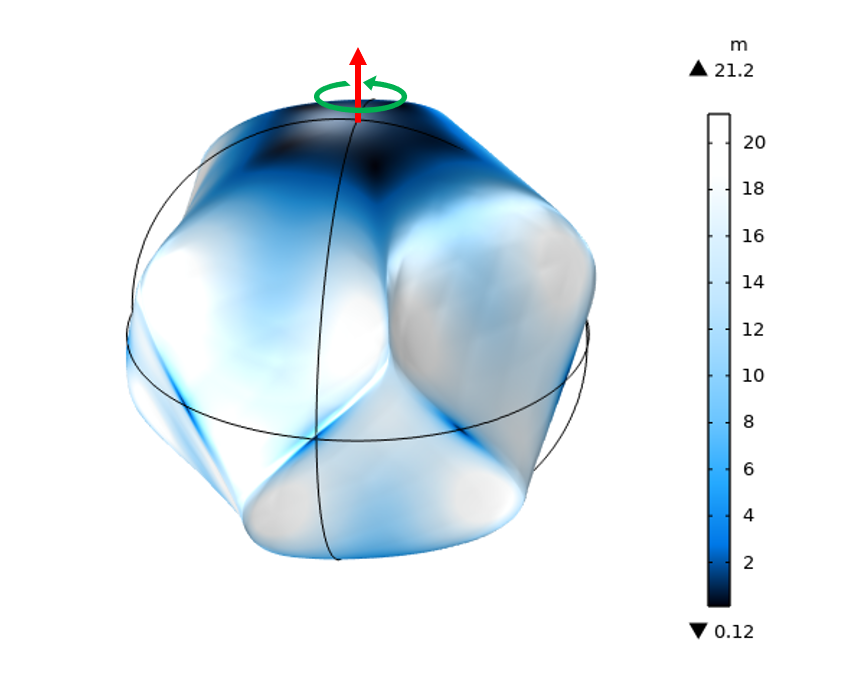}
		\caption{$f = 7.7873\times 10^{-4}$ Hz}  \label{balleig4}
	\end{subfigure} \hspace{.5in}
	\begin{subfigure}[h!]{0.41\textwidth}
		\centering
		\includegraphics[width=\linewidth]{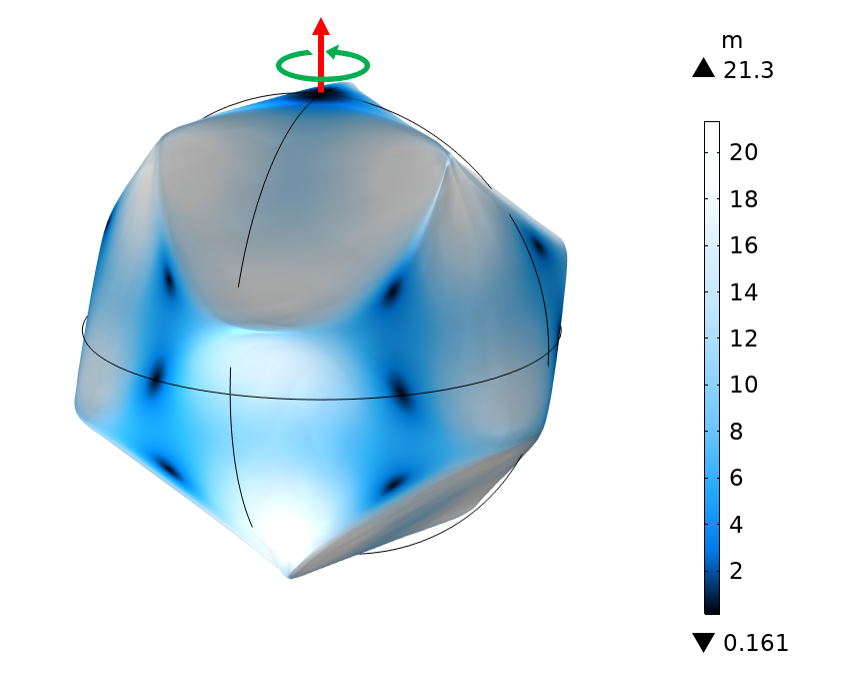}
        \caption{$f = 7.8063\times 10^{-4}$ Hz} \label{balleig3}
	\end{subfigure} 
	\caption{\small Vibration modes of the isotropic elastic rotating ball with rotational symmetry about its rotational axis, the $z$-axis. The material and geometric properties of the ball are the same as in the examples shown in Fig. \ref{balleig}. (a) Eigenmode for the rotating elastic ball with five-fold rotational symmetry with the oscillation frequency $f=7.7873 \times 10^{-4}$ Hz, and (b) eigenmode for elastic ball with three-fold rotational symmetry with the vibration frequency $f=7.8063\times 10^{-4}$ Hz.  These vibration modes resemble the  icosahedron and dodecahedron shapes.} 
    \label{balls}
\end{figure}

The study of wave phenomena on spherical geometries has been an important area of research in the scientific community, with applications spanning geophysics and materials science. Among these, physically chiral waves \cite{kandiah2025dispersion, kandiah2024controlling, carta2019wave, carta2018elastic, allison2024mechanical, garau2018interfacial, carta2014dispersion, brun2012vortex, carta2020one, kandiah2023effect, carta2017deflecting, tallarico2017tilted, kandiah2023gravity}, characterised by their handedness and asymmetry properties, present a rich area of research, particularly when analysed in the context of continuum domains and discrete geometric structures. Fig. \ref{balls} shows two vibration modes of a rotating elastic ball, of the size and inertial properties corresponding to the Earth. Such eigenmodes do not occur for an elastic ball without rotation - the rotational effects, which result in the {\em physical chirality}, are the essential part of the computational model, which is described in detail in Section \ref{rotelas1a}. The computations shown in Fig. \ref{balls}, Fig. \ref{earthicosahedr} and Fig. \ref{balleig} have been produced using COMSOL Multiphysics $6.2.$

Fig. \ref{balleig4} shows the vibration mode with the five-fold rotational symmetry, which approximates an icosahedron, whereas Fig. \ref{balleig3} shows the vibration mode with the three-fold rotational symmetry, which resembles a dodecahedron with twelve pentagonal faces.

Normal mode theory has been fundamental in seismology \cite{carbone2021mathematical, artru2004acoustic, alterman1959oscillations, sato2012seismic, benioff1961excitation, jeans1923propagation}, providing a method to examine the oscillations of the Earth as well as the long wavelength seismic waves from major earthquakes. Although the advancement of numerical methods is decreasing the necessity of normal mode calculations, we show that the modes of the three-dimensional lattice models, which take into account vibrations of the Earth structure subjected to rotation, approximate the earthquake-induced dominant seismic wave frequencies, which is reaffirmed by the spectral analysis of several large scale earthquakes. Our spectral analysis produces low (infrasound level) eigenfrequency and global vibration modes of the planet. The results of the model are compared with the observational data for major earthquakes, during the last 50 years.
The comparison is remarkably impressive, and it provides a fully justified insight for unusual wave patterns observed recently in connection with a series of earthquakes in January-March 2025. 
We would also like to cite the book \cite{lachugin2005earth}, which describes interesting  philosophical ideas and geometrical observations on the Earth, including a lattice grid, based on the icosahedron-dodecahedron framework.

\FloatBarrier

\subsection{Earth's rotation and discrete gyroscopic lattice systems}
In the beginning of 2025, 
the BBC published a report suggesting that the spin of the Earth core has changed its orientation, and the core itself has changed its shape, as discussed in \cite{hughes2025}.
The effects of Earth's rotation on its global-scale vibrations are connected to the three-dimensional Coriolis force field.  In particular,  the paper \cite{gubbins1981rotation} includes a model of the rotation and oscillation of the Earth core, in the context of the continuum approach involving two solid spheres. The paper \cite{montagner2008normal} discusses the Earth's free oscillations in the absence of rotation, which are categorised into spheroidal and toroidal modes. 
The articles \cite{montagner2008normal, roult2010observation} provide important practical connections between the vibration modes of Earth and the seismic data linked to the Sumatra-Andaman earthquake of 2004. It was shown that Earth's rotation leads to gyroscopic frequency splitting, which was also studied in a different context for  mechanical gyroscopic systems in \cite{kandiah2024controlling}. 
The recent articles \cite{yang2023multidecadal, vidale2025annual} presented 
measurements and statistical analysis of global seismic anomalies, and made a conjecture of changes in the rotation patterns of Earth's core. 
Here, we show that this conjecture should be interpreted as a change in the eigenmode of vibration rather than a change in the orientation of spin of Earth's core (see Section \ref{sec:icosahedron}).

We introduce a centred icosahedron elastic lattice with gyroscopic properties at each nodal element to model the vibration of the Earth's core. It is shown that although the orientation of the gyroscopic spinners is chosen to be uniform across the three-dimensional lattice, the motion of the central node can exhibit various patterns. The concept, which is used as the base for the model, is linked to the dynamics of chiral gyroscopic multi-structures \cite{chiralElasticChain, ChiralMultiStructures}. 
In addition, for a more detailed analysis, which includes Earth's core, mantle and crust, the gyroscopic elastic multi-structure, which approximates the earthquake-induced vibrations of Earth, is combined of two dual Plato's polyhedra: icosahedron and dodecahedron, as shown in Fig. \ref{physicalEarthicosahedron}. For visualisation, the three-dimensional centred lattice has been constructed, as shown in Fig. \ref{physicalEarthicosahedron}(b), which also includes the tetrahedral connections between the interior (icosahedron) and exterior (dodecahedron) sections. It will be shown that such discrete gyroscopic models can approximate the typical 
shear motions of the tectonic plates, 
including the convergent-divergent and strike-split type earthquakes.
This classification can be achieved through the spectral analysis of the discrete gyroscopic lattice approximation, which is elegant and has never been done in the past. 



The structure of the paper is as follows. 
Section \ref{resultso} presents the discussion of the problem formulation for the icosahedron and icosahedron-dodecahedron lattices with gyroscopic elements and the comparison between the vibrations of the nodal elements within the structures and observed geophysical phenomena. The analogies between the discrete lattice models and the transient geophysical events are discussed in Section \ref{discusso}. The fundamental principles of the Coriolis effect due to the rotation of the planet are shown to be equivalent to the gyroscopic force in a lattice framework. 
Section \ref{methodo} deals with 
the methods used in the current study for the earthquake data analysis. This section also includes the study of the characteristics of the eigenmodes for a rotating elastic ball. Finally, the concluding remarks are provided in Section \ref{summaro}.

\section{Results}\label{resultso}
This section presents novel results on the modes of three-dimensional lattices with gyroscopic elements that induced a coupling between the displacement components of the nodal points. 
 The analysis of such lattices, which include the icosahedron and icosahedron-dodecahedron lattices, is linked to the dynamics of the oscillations of the Earth. We show that the discrete models provide a very good approximation to the frequencies of the ground vibrations and tectonic plate motions generated during major earthquakes.

\subsection{Icosahedron lattice and the discrete Earth approximation} \label{sec:icosahedron}

\begin{figure}[h]
    \centering
       \begin{subfigure}[h]{0.3\textwidth}
        \centering
       \includegraphics[width=\linewidth]{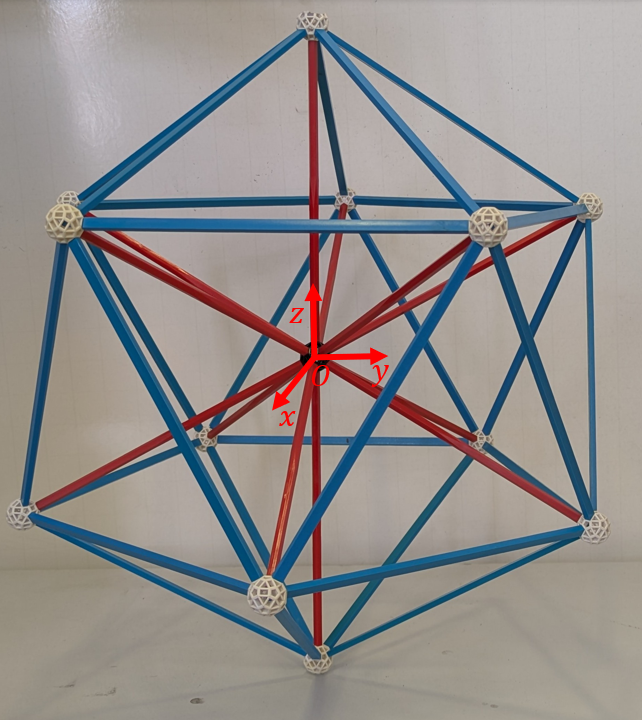}
       \caption{} \label{icosaslttice}
    \end{subfigure} ~~~~~~~~~~~
    \begin{subfigure}[h]{0.4\textwidth}
        \centering
        \includegraphics[width=\textwidth]{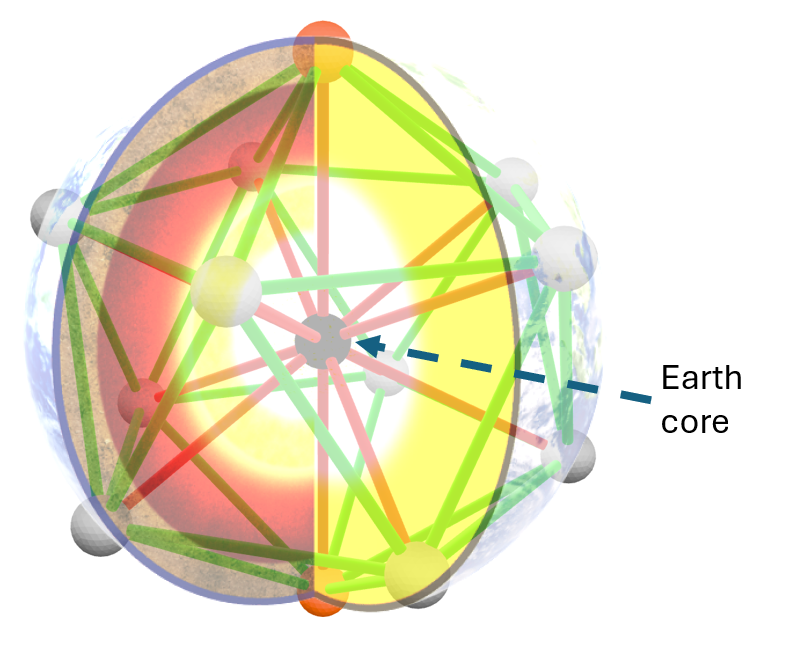}
       \caption{~~~~~~~~~~~} \label{earthisocas}
    \end{subfigure} 
       \caption{\small (a) The geometrical/physical model of the icosahedron lattice, built by the authors, with 13 nodal masses and 42 elastic links. (b) 
       Icosahedron chiral lattice model of Earth, with the Earth core represented by the central inertial junction.
       }
       \label{Earthicosahedron}
\end{figure}

The process of triangularising a sphere is an important step in geometric modelling, computational mathematics, and various applications such as mesh generation and vibration analysis \cite{subich2018higher, williamson2007evolution, williamson1968integration, satoh2014non}. Polyhedra are a good approximation to a sphere, and provide an effective theoretical framework for studying various physical models due to its highly symmetric structure and the simplicity of its faces. 
The vertices of a regular icosahedron can be positioned on a sphere, making it a natural candidate for approximating a spherical surface \cite{fisher1943world}. The discrete centred icosahedron lattice model is shown in Fig. \ref{icosaslttice}. In the context of geophysical phenomena, the icosahedron lattice provides an approximate representation of the Earth as illustrated in Fig. \ref{earthisocas}, where the central nodal point of the icosahedron is similar to the core of the Earth and the icosahedron frame itself effectively models the crust and mantle of the Earth. The general form of the equations of motion for the icosahedron lattice as well as the vibration analysis of the structure are discussed in the subsequent sections.

Vibrations of a centred icosahedron lattice subjected to gyroscopic forces, are considered in connection with the motion of the rotating  Earth core. The derivation of the governing equations for the chiral lattice  is included in the Appendix.

We use the following notations: $m_{i}$ is the mass of the $i$-th node in the lattice, $\omega$ is the radian frequency ($f= (2 \pi)^{-1} \omega$ is the frequency in Hz), 
${\bf U}_{i}$ is the Fourier transform of the displacement vector of the respective masses, 
$k_{ij}$ is the stiffness of the elastic link between node $i$ and its neighbouring node $j,$ $k_{i c}$ is the elastic stiffness of the connection between node $i$ and the central node $c,$ ${\bf e}_{ij}$ is the unit vector along the direction from $i$ to $j$ in the equilibrium position, ${\bf e}_{ic}$ is the unit vector in the direction from $i$ to the central node $c$ at equilibrium and ${\bf k}$ is the unit vector along the $z$-axis. 
We note that ${\bf e}_{ i j } = - {\bf e}_{j i}$ and ${\bf e}_{i c} = - {\bf e}_{c i}.$ The angular speed, corresponding to rotation of Earth is 
$\Theta = 7.2921159 \times 10^{-5}$ s$^{-1},$ as discussed in Section \ref{icos:lattzxsa} of the Appendix. 

\FloatBarrier

\subsubsection{Eigenmodes and motion of the Earth Core}

The dynamics of a centred gyroscopic icosahedron lattice structure 
is studied in connection 
with oscillations of the Earth. We assume that masses assigned to the junctions, corresponding to vertices of the icosahedron are equal, 
that is $m_i = m_0, ~ i=1,\ldots, 12$, whereas the central nodal point is assigned the mass $m_c$. 
After applying the Fourier transform with respect to time to the equations of motion (as derived in the Appendix), 
we have (summation index notations are not used here)
\begin{equation}
- m_0 \omega^2 {{\bf U}_i} = \sum_{j=1}^{5} k_{ij} \big( {\bf e}_{ij} \cdot ( { \bf U}_j - {\bf U}_{i}) \big)  {\bf e}_{ij}  + k_{i c} \big( {\bf e}_{ic}\cdot ({\bf U}_{c} - {\bf U}_{i}) \big) {\bf e}_{i c} - 2 i \omega m_0 \Theta 
{\bf k} \times {\bf U}_{i}, ~~~i=1,\ldots, 12, 
\label{centraleq1a}
\end{equation}
and
\begin{equation}
- m_c \omega^2 {\bf U}_c 
=\sum_{n=1}^{12} k_{cn } \big( {\bf e}_{cn} \cdot ( { \bf U}_n - {\bf U}_{c})  \big) {\bf e}_{c n}  - 2 i \omega m_c \Theta 
{\bf k} \times {\bf U}_{c}. \label{centraleq2a}
\end{equation}
It is noted that the icosahedron lattice with gyroscopic forces at each nodal point is a passive gyroscopic system, where there is no dependence between the gyricity and the radian frequency of the vibrations.  Following \cite{kandiah2025dispersion, kandiah2024controlling, carta2019wave}, we use the terms ``active gyroscopic systems'' and ``passive gyroscopic systems''.
Here, the gyroscopic passive lattice exhibits motions determined purely by its physical chiral properties. The analysis of the active gyroscopic lattice system is provided in Section \ref{sec:math_formulation}.

The centred icosahedron lattice models the interactions between neighbouring nodes as well gyroscopic effects due to the presence of gyros at each vertex. In the present analysis, we show that 
the individual motions of the nodal elements can differ, thereby demonstrating that a rotational action can result in varied movements of the components within a body. The icosahedron lattice provides a simplified model of the Earth, where the central nodal point of the lattice approximately resembles the Earth's inner core. The motion of Earth's inner core is a fascinating phenomenon that has been studied through seismic data and experimental analysis \cite{vidale2025annual, wang2024inner, chao2000coseismic}. The inner core is a solid ball of iron and nickel that can rotate at different rates relative to the surface of the Earth \cite{yang2023multidecadal, cambiotti2016residual, mccarthy1996path}. Although experimental studies have suggested that the rotation and movement of the inner core varies over time \cite{yang2023multidecadal, song1996seismological, wang2022seismological, yao2019temporal, greiner2000influence, zhang2005inner, mathews1991forced, song1997anisotropy}, which consider seismic wave observations and spectral analysis, there is no universally accepted model characterising the evolution of Earth's inner core.  In the context of Earth's internal dynamics, this shows that the Earth's inner core can follow a wide range of motions that is not always aligned with its direction of rotation.

\subsubsection{
Visualisation of vibration modes for the 
rotating Earth's core}
Fig. \ref{earthicosahedr} illustrates the typical vibrations of the physically chiral centred icosahedron lattice, consisting of $13$ nodal points connected by elastic links, where each nodal point is subjected to a gyroscopic force resulting from the presence  of spinners attached to every junction point and is aligned with the $z$-axis.
In particular, the gyroscopic force
is orthogonal to the particle's velocity and the vertical $z$-axis,
as detailed in Section \ref{icos:lattzxsa} of the Appendix. 
The physical and geometrical parameters of the centred icosahedron lattice in the illustrative examples are chosen as follows. 

We assume the masses of the nodal points located on the vertices of the icosahedron are $m_0=3.36015 \times 10^{23}$ kg, which considers the combined mass of the mantle and crust within the Earth distributed over the $12$ nodes of the icosahedron, while the central mass of the structure is $m_c=1.94142 \times 10^{24}$ kg, taking into account the mass of the Earth's core \cite{planetHandbook}. Considering the discussion of the gyricity vector as presented in Section \ref{icos:lattzxsa} of the Appendix, the gyricity parameter chosen in the examples for the nodes on the vertices of the icosahedron is such that 
$2 m_0 \Theta = 4.900520648277 \times 10^{19}$ kg/s, while the gyricity parameter for the central node is such that  
$2 m_c \Theta = 2.8314119301156 \times 10^{20}$ kg/s. 
Noting that the Earth's rigidity is higher in the solid layers, such as the lithosphere and lower mantle, 
we set the stiffness of the elastic links coinciding with the edges of the icosahedron as $k_{ij}=10^{21}$ N/m, while the stiffnesses of the inner elastic links representing the outer core are chosen as $k_{ic}=10^{20}$ N/m. In the icosahedron lattice, the elastic links are assumed to be massless, and the lengths of the inner links are chosen as $6.371\times 10^{6}$ m, in connection with the mean radius of the Earth \cite{planetHandbook}, while the lengths of the edges of the icosahedron are $ 6.698865832 \times 10^{6}$ m. 

We note that the gyricity associated with the central nodal element, representing Earth's core, is the same for all vibration modes. However, the actual motion of the core may be different for different modes, as shown in Fig. \ref{earthicosahedr}.

 \subsubsection*{Changing direction of motion with a fixed direction of rotation for the Earth's core}
\begin{figure}[h]
	\centering
	\begin{subfigure}[h!]{0.4\textwidth}
		\centering
		\includegraphics[width=\linewidth]{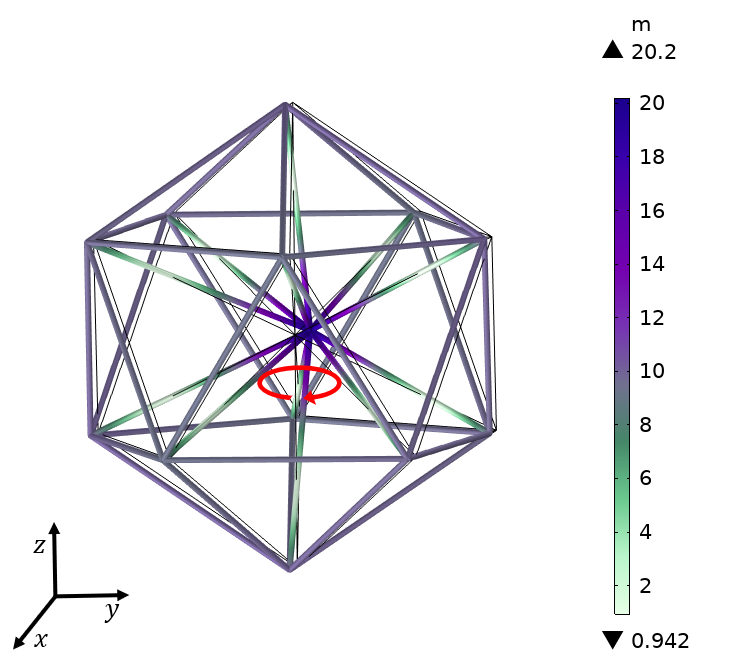}
		\caption{ $f= 0.002737609$ Hz}\label{icos3a}
	\end{subfigure} 
	\begin{subfigure}[h!]{0.4\textwidth}
		\centering
		\includegraphics[width=\linewidth]{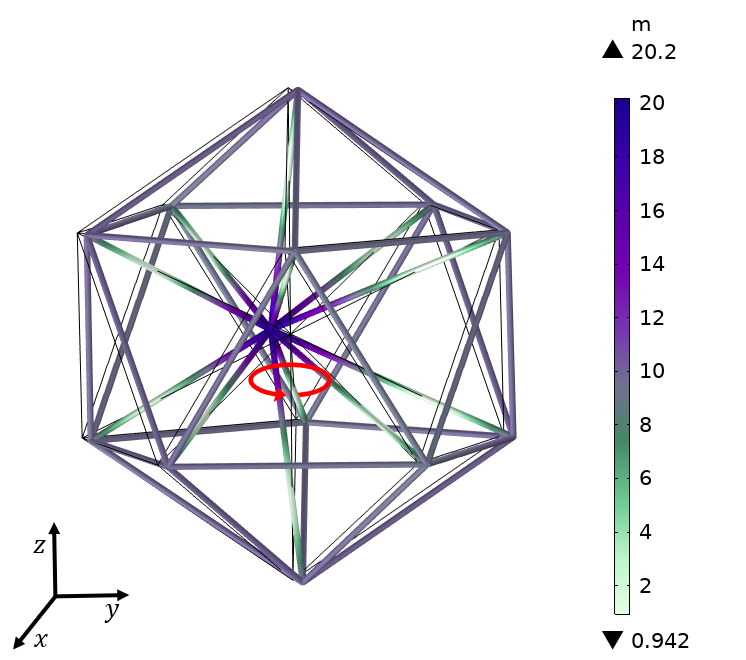}
		\caption{$f = 0.002760794$ Hz} \label{icos3b}
	\end{subfigure}
	\\
	\begin{subfigure}[h!]{0.4\textwidth}
		\centering
		\includegraphics[width=\linewidth]{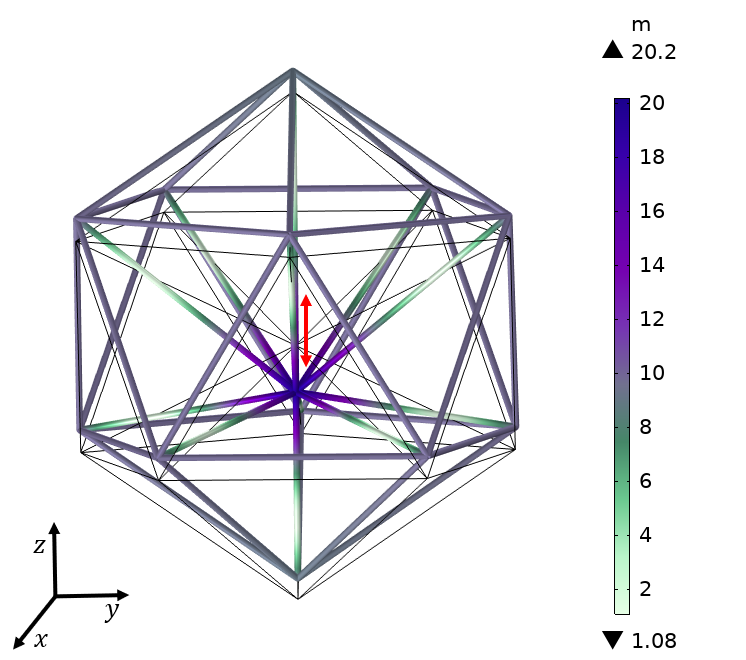}
		\caption{$ f = 0.002749177$ Hz}\label{icos3c}
	\end{subfigure} 
	\begin{subfigure}[h!]{0.4\textwidth}
		\centering
		\includegraphics[width=\textwidth]{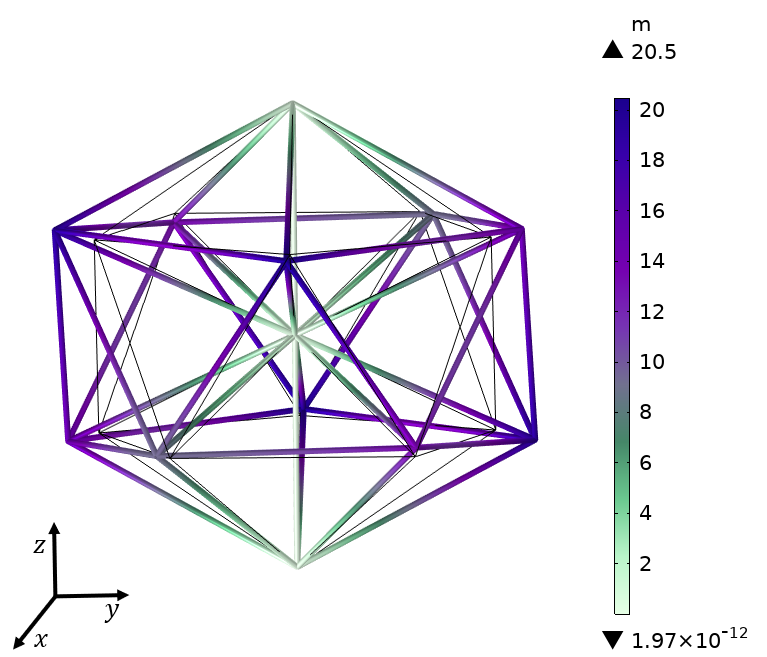}
		\caption{$ f = 0.007081973$ Hz} \label{icos3d}
	\end{subfigure}
	\caption{\small Typical eigenmodes of the passive gyroscopic icosahedron lattice. 
    The mode shapes are plotted for the following frequencies: (a) $f= 0.002737609$ Hz, (b) $f = 0.002760794$ Hz, (c) $f = 0.002749177$ Hz and (d) $f = 0.007081973$ Hz. The arrows show the direction of motion of the central nodal point.} 
	\label{earthicosahedr}
\end{figure}

Fig. \ref{icos3a} shows the mode of the centred gyroscopic icosahedron lattice
where all the nodal elements move in a clockwise direction when viewed from the positive direction of the $z$-axis, each following an elliptical trajectory with the frequency $f = 0.002737609$ Hz.
The elliptical paths traced by the nodal points positioned on the vertices of the icosahedron differ in magnitude and eccentricity compared to the elliptical path of the central nodal point. This occurs due to the prescribed gyricity and physical parameters for the nodal points of the lattice. 
Fig. \ref{icos3b} shows that when $f= 0.002760794$ Hz, the central nodal point as well as the nodes at the poles of the icosahedron move with a \emph{phase shift} in the counterclockwise direction. 
In this case the orientation of the spinners is opposite to the direction of motion of the nodes located along the central axis of the icosahedron. In particular, this motion differs from the clockwise trajectories of the nodes positioned along the vertical axis of the icosahedron associated with the frequency mode shown in Fig. \ref{icos3a}. This demonstrates that although the gyroscopic force in both illustrative examples is the same for each nodal point, the masses can follow different paths compared to the direction of spin of the gyros. A similar effect where the direction of motion of a structure differs from the direction of spin of its attached spinner was also observed for an elementary gyropendulum \cite{kandiah2023effect, kandiah2024controlling}, which incorporates a gyroscopic spinner attached to a rod, where the introduction of the gyroscopic force through a gyro resulted in two frequencies of the vibrations; the low-frequency mode led to the gyropendulum motion in the opposite direction compared to the orientation of spin of the spinner, while for the high-frequency mode, the direction of motion of the structure was the same as the orientation of spin of the spinner. 


\FloatBarrier

\subsubsection*{Vertical motion of the Earth's core and centrally symmetric modes}

Fig. \ref{icos3c} displays the eigenmode of the icosahedron lattice with $f = 0.002749177$ Hz, where the dominant direction of motion for the central mass and the nodes at the poles is in the longitudinal direction (along the $z$-axis). 
Fig. \ref{icos3d} illustrates the mode where the velocities of the masses along the central vertical axis of the icosahedron lattice are negligibly small, and the masses on the upper and lower layers move along elliptical paths, 
with the frequency $f = 0.007081973$ Hz. 
Regions of localisation in equatorial bands were also observed for vibrations of a rotating elastic ball in Section \ref{rotelas1a}. 


 The examples presented in this section demonstrate that for different vibrational frequencies of the centred gyroscopic icosahedron lattice, the direction of motion of the nodal points can change, whereas the orientation of spin for the gyroscopic spinners on each node remains the same, i.e. in the clockwise direction when viewed from the positive $z$-axis. Additionally, the central node, representing the Earth core, can move in a different direction and path compared to the masses at the vertices of the icosahedron. In the geophysical context, 
this implies that the spin of the Earth core may remain unchanged, but the change of the eigenmode of vibration may lead to a different path of the centre of mass of the Earth core.  

 \subsection{Earth as the icosahedron-dodecahedron multi-structure: vibration modes versus earthquakes} \label{sec:math_formulation}
As discussed in Section \ref{sec:icosahedron}, the icosahedron provides an effective lattice model to triangulate and analyse the vibrations of the Earth. However, the Earth is composed of multiple internal layers with differences in composition, physical and chemical properties as well as dynamic interactions between them. Given these challenges, we construct a discrete multi-layered lattice by extending the icosahedron lattice introduced in Section \ref{sec:icosahedron}. 
The new lattice multi-structure is composed of an icosahedron lattice and a dodecahedral structure, where the two layers are connected by tetrahedral links; we refer to such model as the icosahedron-dodecahedron lattice. In this section, we show that the icosahedron-dodecahedron lattice provides a valuable approximation to the tectonic plate motions induced by large earthquakes as well as the frequencies of the ground motions, while taking into account the stratified regions of the Earth with simplified physical properties.   

The illustrative figure of the icosahedron-dodecahedron lattice is presented in Fig. \ref{partaphysicalEarthicosahedron}, as well as the approximation of the multi-structure with the various Earth layers. Fig. \ref{partaphysicalEarthicosahedron2} shows the physical model built by the authors of the paper. The icosahedron-dodecahedron lattice consists of gyroscopic forces acting at each nodal point associated with the inertial force of the rotating planet in the non-inertial reference frame, and varying stiffness properties across the massless elastic links connecting the nodes which are linked to the stiffness of the Earth's materials. Additionally, compared to the passive gyroscopic icosahedron lattice studied in Section \ref{sec:icosahedron},
here we consider an active gyroscopic system so that the rate of spin of individual gyroscopes are tuned with the frequency of the icosahedron-dodecahedron lattice vibrations.  
The centred gyroscopic icosahedron-dodecahedron lattice provides a representation of the rotating Earth, where the central nodal point is the inner core, the links between the central node and the icosahedron resemble the outer core, the icosahedron approximates the outer core, the links between the icosahedron and dodecahedron make up the mantle, which is a mixture of semi-liquid and solid material \cite{o2005mantle, mcdonough1995composition}, and the dodecahedron models the crust of the Earth. Through this approximate comparison, we demonstrate that the icosahedron-dodecahedron lattice provides a simplified yet efficient model to approximate the elastic vibrations of the Earth.

\subsubsection{Spectral problem for the active gyroscopic lattice system}\label{Sparsity1a}

In this section, we formulate the matrix problem for the nodal elements within the centred gyroscopic icosahedron-dodecahedron lattice and derive the characteristic equation for the lattice through which the vibration frequencies of the structure are determined.  We consider an active gyroscopic lattice system, where the gyricity parameter of the gyroscopic spinners in the lattice model is proportional to the radian frequency of the vibrations, 
which results in controlled frequencies that can be adjusted to influence the oscillations of the structure as well as its characteristic equation. 
The full problem formulation and governing equations of the lattice are provided in Section \ref{sec:time-harmonic_equationszxasx} of the Appendix. The icosahedron-dodecahedron lattice is shown to posses five-fold rotational symmetries in connection with the characteristic properties of geophysical atmospheric models (see Fig. \ref{stormsocean}).

\begin{figure}[h!]
    \centering
    \begin{subfigure}[h]{0.4\textwidth}
        \centering
        \includegraphics[width=\textwidth]{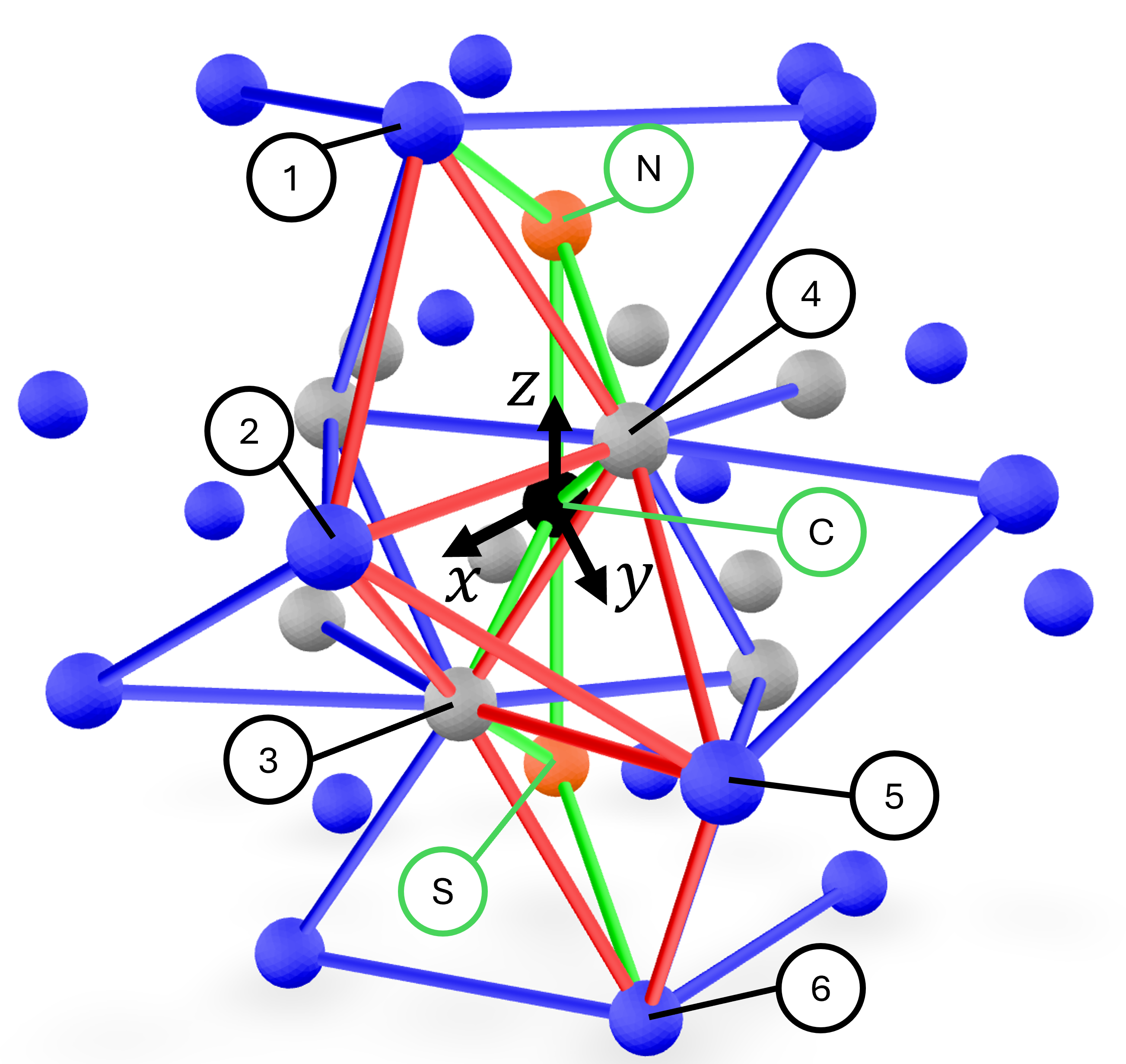}
        \caption{}
        \label{fig:unitcell}
    \end{subfigure} 
    \hspace{.5in}
    \begin{subfigure}[h]{0.4\textwidth}
        \centering
       \includegraphics[width=\linewidth]{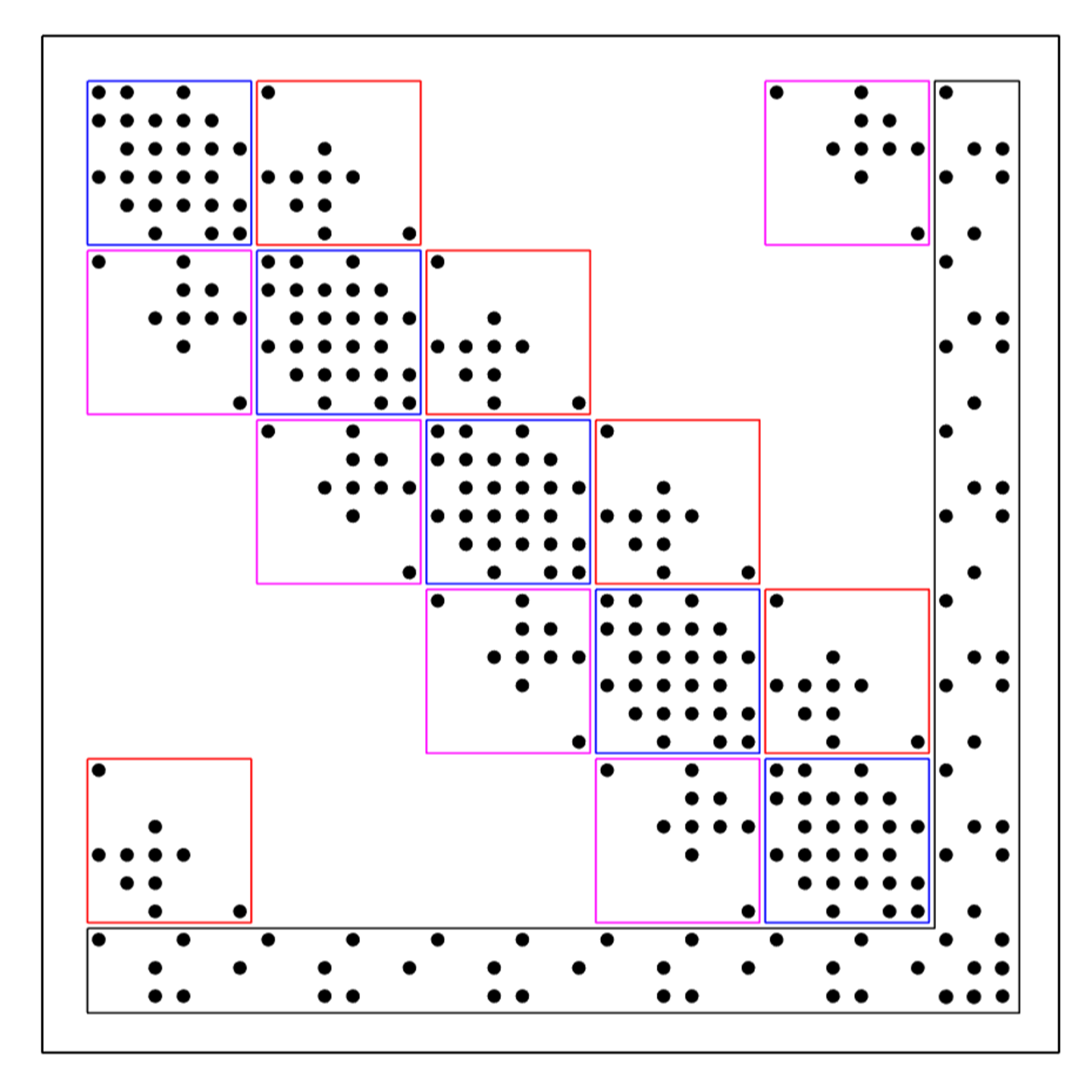}
       \caption{}
       \label{fig:stiffness_sparsity}
    \end{subfigure}
       \caption{\small (a) Diagram of the icosahedron-dodecahedron lattice multi-structure with the reference cell highlighted (red), where the nodal points of the cell are labelled. The masses aligned with the vertical axis of the structure, corresponding the nodes along the centre, north and south of the lattice, are also shown. (b) Sparsity plot of the matrix $\mathbf{C}-\omega^2(\mathbf{M}+\mathbf{A}),$ where the matrices $\mathbf{C}, \mathbf{M}$ and $\mathbf{A}$ are defined in the text below. The $6\times 6$ matrix blocks correspond to the six numbered nodes in the reference cell shown in Figure \ref{fig:unitcell}. The blue blocks denote the $n$-th reference cell, the red blocks are associated with the $(n+1)$-th cell, the pink blocks correspond to the $(n-1)$-th reference cell and the black blocks represent the additional cell containing the north (N), central (C) and south (S) nodes of the icosahedron-dodecahedron lattice.} 
       \label{fig:models}
\end{figure}
\FloatBarrier

We note that the icosahedron-dodecahedron lattice possesses a five-fold rotational symmetry about its vertical axis, and thus it is convenient to introduce the reference cell of the structure as shown in Fig. \ref{fig:unitcell}. Then the icosahedron-dodecahedron lattice can be generated by finite rotations of the reference cell about the vertical axis. Additional details on the construction of the reference cell are presented in Section \ref{sec:time-harmonic_equationszxasx} of the Appendix. The system of equations governing the motions of the nodal elements in the icosahedron-dodecahedron lattice, which can be described by the nine equations associated with the labelled nodes shown in Fig. \ref{fig:unitcell} (see Section \ref{sec:time-harmonic_equationszxasx} of the Appendix), can be written in the matrix form as follows
\begin{equation}
[\mathbf{C}-\omega^2(\mathbf{M}+\mathbf{A})]\mathbf{U} = \mathbf{0}, \label{eq:matrixeigvalseqn}
\end{equation}
where $\mathbf{C}$ is the stiffness matrix characterising the elastic interactions between neighbouring nodes in the icosahedron-dodecahedron lattice, $\omega$ is the radian frequency, $\mathbf{M}$ is the diagonal matrix with entries corresponding to the mass of each node, that takes the form
\begin{equation}
\mathbf{M} = \text{diag}[\mathbf{M}^{(1)},\mathbf{M}^{(2)},\dots,\mathbf{M}^{(N)}], \quad \mathbf{M}^{(i)} = 
\begin{pmatrix}
    m_i & 0 & 0 \\
    0 & m_i & 0 \\
    0 & 0 & m_i
\end{pmatrix},
\end{equation}
and the hermitian matrix $\mathbf{A}$ is the gyricity matrix, which is given by 
\begin{equation}
\mathbf{A} = \text{diag}[\mathbf{A}^{(1)},\mathbf{A}^{(2)},\dots,\mathbf{A}^{(N)}], \quad \mathbf{A}^{(i)} = 
\begin{pmatrix}
    0 & i \alpha & 0 \\
    -i \alpha & 0 & 0 \\
    0 & 0 & 0
\end{pmatrix}, \label{gyricitymat}
\end{equation}
where \(\alpha\) is the chirality parameter, characterising the gyroscopic action induced by the rotating spinner. The gyricity matrix takes the above form since the axis of rotation is the axis passing through the nodes labelled by S, C and N in Fig. \ref{fig:unitcell}.  Additionally in (\ref{eq:matrixeigvalseqn}), the displacement vector ${\bf U}$ takes the form
\begin{equation}
{\bf U} = \Big(
    \mathbf{U}^{(0)}, 
    \mathbf{U}^{(1)},
    \ldots,
    \mathbf{U}^{(4)},
    \mathbf{u}^{(N)}, 
    \mathbf{u}^{(S)},
    \mathbf{u}^{(C)} \Big)^T, 
\quad \text{with}  \quad \mathbf{U}^{(n)} = \Big(
    \mathbf{u}^{(n)}_1, 
    \mathbf{u}^{(n)}_2,
    \ldots,
    \mathbf{u}^{(n)}_6
\Big)^T, 
\end{equation}
where ${\bf u}^{(k)}$ for $k=0, 1,2,3,4, N, S, C,$ are $3 \times 1$ column vectors associated with the displacement components of the nodes. In particular, the vector ${\bf U}^{(i)}$ for $i=0,1,2,3,4,$ corresponds to the displacements of the nodal elements within the reference cell (see Fig. \ref{fig:unitcell}), where four rotations of this cell generates the centred icosahedron-dodecahedron lattice. We note that the vectors $\mathbf{u}^{(N)}, \mathbf{u}^{(S)},$ and $\mathbf{u}^{(C)} $ represent the displacements for the masses in the north, south and central regions of the icosahedron-dodecahedron lattice, respectively, that are highlighted in Fig. \ref{fig:unitcell}.

The icosahedron-dodecahedron lattice is composed of $33$ nodal points as shown in Fig. \ref{partaphysicalEarthicosahedron}, where each node can move along three principal directions. Thus, the sizes of the matrices in (\ref{eq:matrixeigvalseqn}) are $99\times 99.$ A sparsity plot of the $99\times 99$ matrix $\mathbf{C}-\omega^2(\mathbf{M}+\mathbf{A})$ is provided in Fig. \ref{fig:stiffness_sparsity}, which is constructed by taking into account the rotations of the nodal elements in the reference cell (see Fig. \ref{fig:unitcell}). Due to the five-fold rotational symmetry of the icosahedron-dodecahedron lattice about its vertical $z$-axis by rotations of $2\pi/5$, the stiffness matrix has a repeated block band structure as illustrated in Fig. \ref{fig:stiffness_sparsity}. 
Similarly, the structures of the gyricity and mass matrices demonstrate a five-block repetition pattern as shown in the sparsity plot. 
This demonstrates that the motions of the $33$ nodes in the centred icosahedron lattice are fully determined by the system of equations (14)-(22) of Section \ref{sec:time-harmonic_equationszxasx} in the Appendix. 
Additionally, we note that the construction of the stiffness matrix ${\bf C}$ 
for the centred icosahedron-dodecahedron elastic truss system, shown in Fig.  \ref{partaphysicalEarthicosahedron2} is standard; the full description of the corresponding governing equations for this system in included in the Appendix.

The radian frequencies $\omega,$ describing the oscillatory motion of the nodal points in the gyroscopic icosahedron-dodecahedron lattice, are derived by numerically solving the following equation: 
\begin{equation}
    \mbox{det}[\mathbf{C}-\omega^2(\mathbf{M}+\mathbf{A})] = 0.
\end{equation}
The results of these computations are 
used in conjunction with the data of the recorded earthquakes, as discussed below.








\subsubsection{Geophysical interpretation of the three-dimensional gyroscopic lattice and comparison with the earthquake data}
\label{earthquakecomparison}
In this section we present the comparison of the re-aligned eigenmodes of the centred gyroscopic icosahedron-dodecahedron lattice and the geophysical models of recorded earthquakes. Namely, we consider the Mexico City earthquake (1985), the Kobe earthquake (1995) and the D\"{u}zce earthquake (1999). We derive the dominant frequency of these earthquakes through the application of the Fast Fourier Transform (FFT) of the displacement seismograph data and discuss the corresponding type of tectonic plate motions causing such earthquakes. It is shown that by setting the geometrical and physical parameters of the icosahedron-dodecahedron lattice to approximately resemble the Earth parameters \cite{planetHandbook}, with the optimised positions of the nodal point locations in connection with the localised earthquake regions of magnitude $7$ and above, the motions of the lattice masses and their corresponding frequencies provide a very good approximation to the fundamental vibrations during the observed earthquakes.

If the reference half plane $Oxz,~ x>0$ (see Section \ref{sec:time-harmonic_equationszxasx} in the Appendix) includes the Greenwich meridian, together with the vertices numbered 1, 2, 3 in Fig. \ref{fig:unitcell}, and the $Oz$-axis contains the Geographical  North, then the alignment, which minimises the distance between the nodal points of the icosahedron-dodecahedron lattice and the three-dimensional dataset of earthquakes, {corresponds to a counterclockwise  rotation of $9.64^\circ$ about the $Oz$-axis}, and the repositioning of the vertex N of the icosahedron to the point with the geographic coordinates $79.60^\circ$ N and $20.72^\circ$ E. 
We refer to this point as the ``{\em Tectonic North}'', which is different from the Geographical North, and may move as the tectonic plate structure evolves with time. The optimisation procedure to obtain the matching of the centred icosahedron-dodecahedron lattice with the dataset of epicentres of major earthquakes is detailed in Section \ref{opt:algozasxasx} of the Appendix; we note that this dataset includes the three-dimensional positions (latitude, longitude and the depth of the epicentres).   

\begin{figure}[h!]
\centering
    \begin{subfigure}[h!]{0.47\textwidth}
    \centering
    \includegraphics[height=6.5cm]{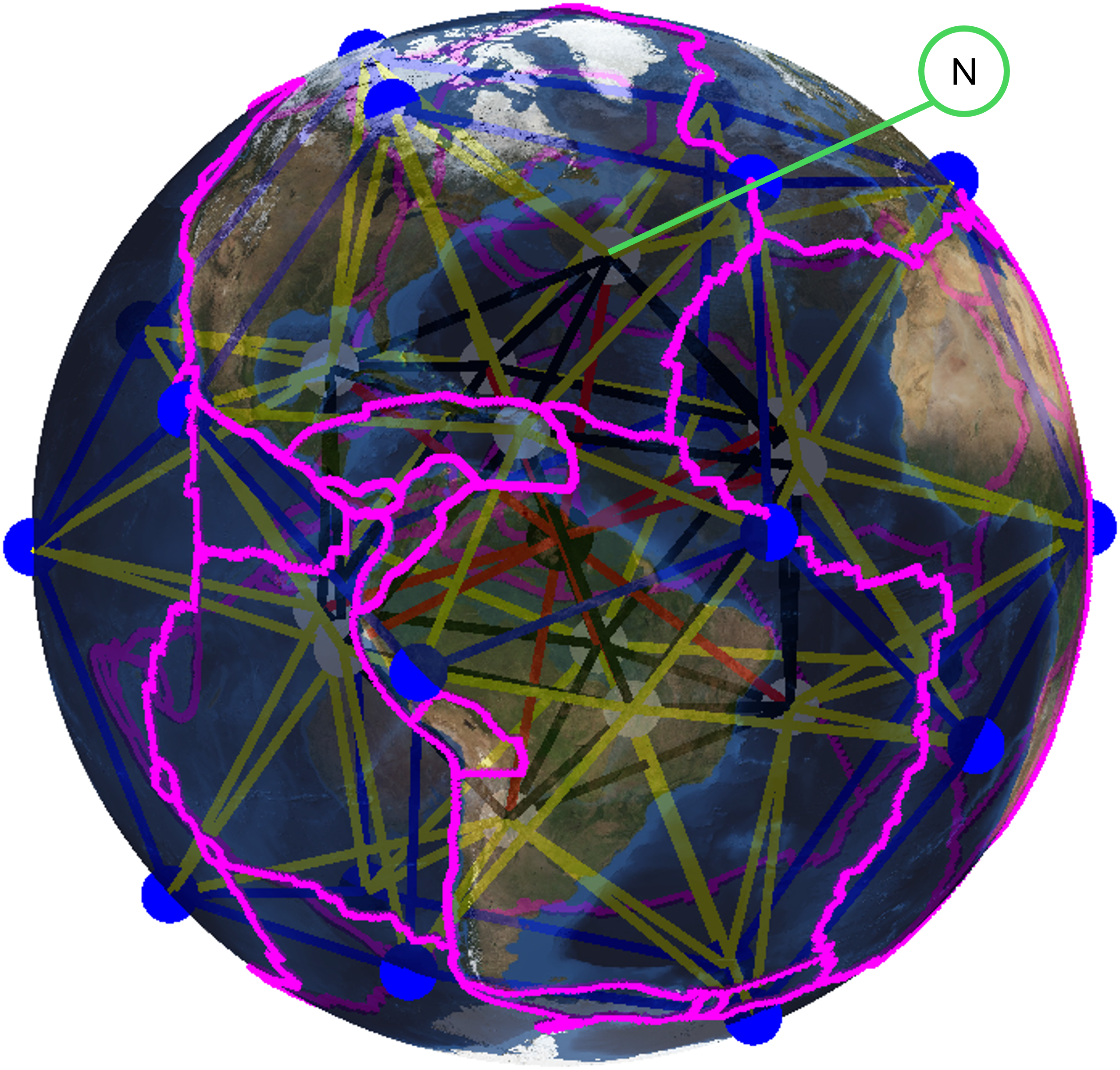}
    \caption{} 
    \label{fig:icosdodec_framework}
	\end{subfigure}
    \hfill
        \begin{subfigure}[h!]{0.47\textwidth}
		\centering
    \centerline{\includegraphics[height=6.5cm]{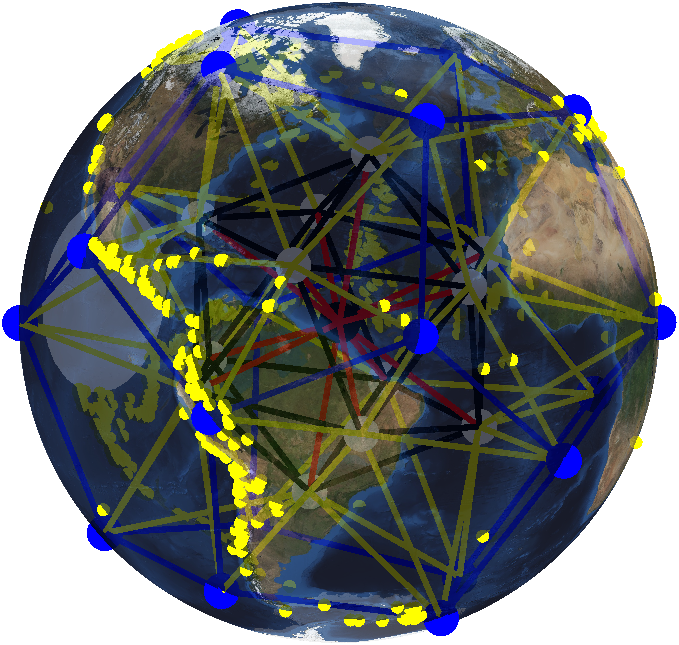}}
    \caption{}
    \label{fig:earthquakestructure}
	\end{subfigure}
	\caption{\small (a) The centered icosahedron-dodecahedron lattice inscribed by the Earth sphere, where the vertices of the dodecahedron are positioned along the surface of the Earth, with the boundaries of the tectonic plates shown by the solid lines \cite{BirdTectonicPlates}. The lattice structure inside the sphere is shown in Fig. \ref{physicalEarthicosahedron}. The ``{\em Tectonic North}" of the model is denoted by N and lies at $79.60^\circ$ N, $20.72^\circ$ E. We note the difference between the ``{\em Tectonic North}'' and the geographical North Pole, which demonstrates that the rotational polar axis of the Earth changes over time. (b) The icosahedron-dodecahedron Earth model, where the earthquakes with magnitude $7$ Mw and above are shown by the solid markers (yellow). 
    The nodal positions of the centred icosahedron-dodecahedron lattice provide a very good approximation with the locations of earthquakes. }
	\label{fig:structuremap}
\end{figure}

Fig. \ref{fig:structuremap} shows the results of the optimisation procedure, where the positions of the nodal points of the icosahedron-dodecahedron lattice are optimised to match the locations of earthquakes. Fig. \ref{fig:icosdodec_framework} illustrates the nodal points of the centred icosahedron-dodecahedron relative to the Earth and the tectonic plate boundaries \cite{BirdTectonicPlates}, which are illustrated by the solid lines (purple), while Fig. \ref{fig:earthquakestructure} shows the structure with the highlighted locations of earthquake epicentres (solid dots in yellow), where the high seismic activity regions are shown to occur at the tectonic plate boundaries \cite{frohlich1992earthquake, mccann1979seismic, berryman2012major, khattri1987great, sykes1981repeat}. The database, that we use, includes the coordinates and the depth of the earthquake epicentres, as well as their magnitudes \cite{kanamori1978quantification}.
We re-align the icosahedron-dodecahedron frame with the three-dimensional dataset of epicentres of major earthquakes (the table is included in Section \ref{data:earthquaket} of the Appendix). The earthquake dataset is obtained from the U.S. Geological Survey (USGS) earthquake catalogue \url{https://earthquake.usgs.gov/earthquakes/search/}, 
and is 
visualised in Fig. \ref{fig:earthquakestructure}. In this study, only earthquakes of magnitude $7$ and above, that have occurred in the past $50$ years, are considered. 


\FloatBarrier

\subsubsection{Earth's vibration modes in the gyroscopic centred icosahedron-dodecahedron lattice approximation}\label{newsearthasa12a} 

\begin{figure}[h!]
\centering
    \begin{subfigure}[h!]{0.47\textwidth}
    \centering
    \centerline{\includegraphics[height=6.5cm]{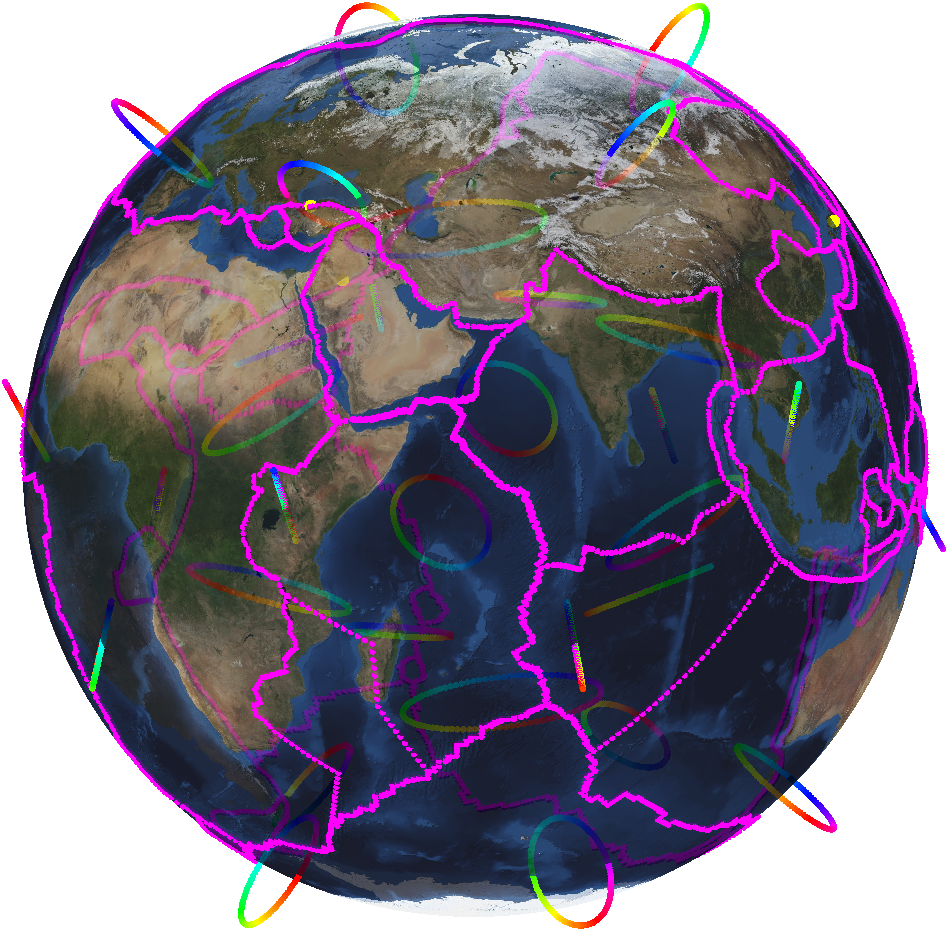}}
    \caption{}
    \label{fig:fullmap1}
	\end{subfigure}
    \hfill
        \begin{subfigure}[h!]{0.46\textwidth}
		\centering
    \centerline{\includegraphics[height=6.5cm]{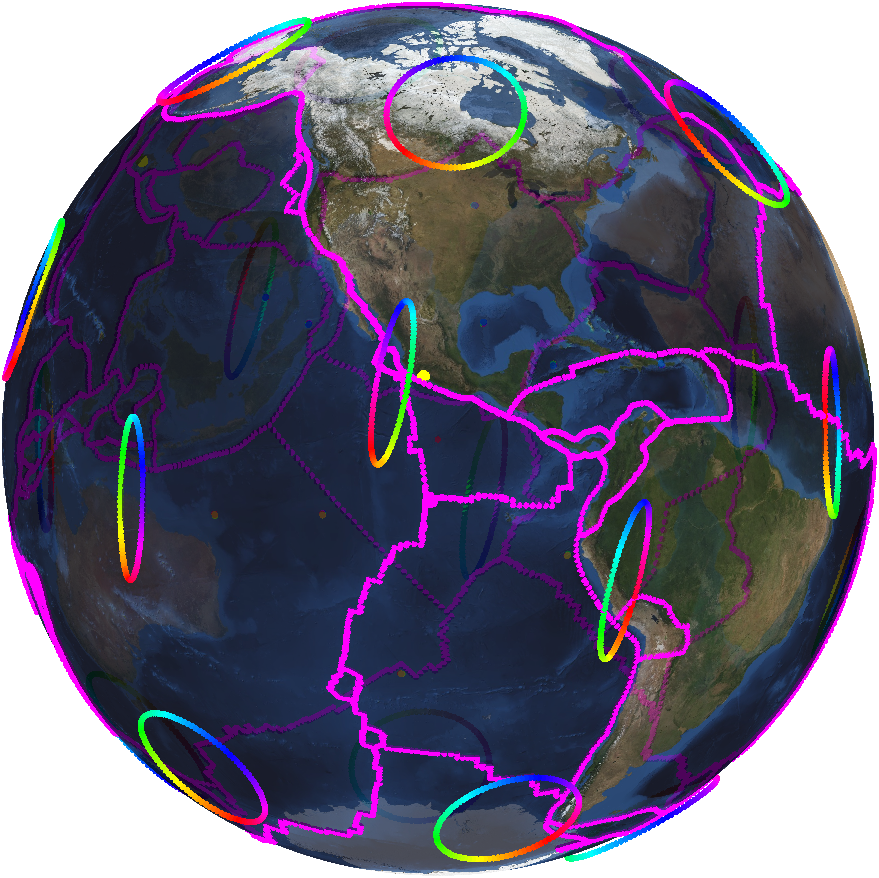}}
    \caption{}
    \label{fig:fullmap2}
	\end{subfigure}
	\caption{\small Typical modes of the icosahedron-dodecahedron lattice, where the trajectories of the nodal points are shown relative to the Earth model. The solid lines represent the tectonic plate boundaries. The frequency of oscillations for the nodal elements of the lattice are as follows: (a) $f=0.0197565217194899$ Hz and (b) $f=0.113211349862919$ Hz. Here and in the following figures, the colour code (red$\to$yellow$\to$green$\to$blue) of nodal trajectories is used to show the direction of motion along an elliptical path. }
	\label{fig:fullmap_main}
\end{figure}

Two examples illustrating the eigenmodes of the icosahedron-dodecahedron lattice are presented in Fig. \ref{fig:fullmap_main}, where the elliptical motions of the nodal points can be observed. 
The parameter values of the icosahedron-dodecahedron lattice used in the illustrative examples are given in Table \ref{tab:earth_params}, where the notations describing the different components of the structure are defined in Section \ref{sec:time-harmonic_equationszxasx} of the Appendix. 
In particular, 
the mass of the Earth's core, mantle and the crust has been redistributed across the lattice, including the central node and the vertices of the icosahedron and dodecahedron, respectively. 
%
The four values  $\kappa_{C}, \kappa_{I}, \kappa_{B}$ and $\kappa_{D}$ (see Table \ref{tab:earth_params})  have been used to represent the stiffness coefficients of the elastic links between  the central node and icosahedron vertices, the links along the edges of the icosahedron, the tetrahedral connections between the icosahedron and the dodecahedron, and the elastic links along the edges of the dodecahedron, respectively (also, see Fig. \ref{partaphysicalEarthicosahedron2}).    
The quantities $\rho_{I}$ and $\rho_{D}$ in Table \ref{tab:earth_params} represent the radii of the circumscribed spheres of the icosahedron and dodecahedron in the icosahedron-dodecahedron lattice, respectively. The gyroscopic action introduced at each nodal element of the active discrete lattice are characterised by the chirality parameter $\alpha.$ 

In Fig. \ref{fig:fullmap1}, the trajectories of the nodal masses are aligned in the radial direction of the Earth with the oscillation frequency $f=0.0197565217194899$ Hz, while the tangentially-dominated motions of the nodal points with the vibration frequency  $f=0.113211349862919$ Hz are shown in Fig. \ref{fig:fullmap2}. 
The elliptical paths of the nodal points and their orientation relative to the tectonic plate boundaries are used to describe the representative ground motions during an earthquake. Although seismic waves propagate in different ways, they can be classified according to their magnitude and depth, as well as their distinct motions in the vicinity of the tectonic plate boundaries. In the next section, we provide a new framework for characterising the ground vibrations induced by earthquakes and their vibration frequency using the discrete icosahedron-dodecahedron lattice, which captures the fundamental vertical and horizontal tectonic plate movements.

\begin{table}[htb]
    \centering
    \begin{tabular}{l|l}
        Parameter & Value \\
        \hline
        The mass of Earth (kg): & $5.9736 \times10^{24}$ \\
        \hline
        \quad Mass of the layers as percentages of the Earth: \\
        \quad Core & 32.5\% \\
        \quad Mantle & 67\% \\
        \quad Crust & 0.5\% \\       
        \hline
        The stiffness values (N/m): \\
        \quad $\kappa_D$ & $4\cdot 10^{21}$\\
        \quad $\kappa_B$ & $2\cdot 10^{21}$\\
        \quad $\kappa_I$ & $1\cdot 10^{21}$\\
        \quad $\kappa_C$ & $1\cdot 10^{20}$\\
        \hline
        The radii (km): \\
        \quad $\rho_i$ & 3471 \\
        \quad $\rho_d$ & 6371 \\
        \hline
        The gyricity (kg): \\
        \quad $\alpha$ &  $1\cdot10^{20}$
    \end{tabular}
    \caption{\small The parameter values for the centred gyroscopic icosahedron-dodecahedron lattice model of the Earth. The parameter definitions and notations are provided in Section \ref{sec:time-harmonic_equationszxasx} of the Appendix. 
    }
    \label{tab:earth_params}
\end{table}

 An interesting observation of the discrete icosahedron-dodecahedron model is that its vertical axis, passing through the central, southern and northern nodes of the structure, is not aligned with the polar axis going through the geographic poles of Earth. In particular, the rotational axis of the Earth can drift due to axial precession, geophysical events and changes in climate patterns \cite{rochester1973earth, eubanks1993variations, chao1996seismic}. This phenomenon also includes the Chandler wobble resulting in deviations of the Earth's axis of rotation \cite{dahlen1975influence, gross2000excitation, brzezinski2002oceanic}. The position of the discrete three-dimensional icosahedron-dodecahedron lattice relative to the Earth, which was obtained through the optimisation, approximately captures this change in the rotational axis of the Earth. 
 
 Fig. \ref{atmosphericoas1} illustrates the significant variations in the pressure region in January $2025$ during the storm \'{E}owyn, accompanied by a large number of recorded seismic events. Additionally, we observe the high pressure zone located in Greenland, which is not aligned with the geographic North Pole, 
 corresponding to the approximate centre of rotation in connection with the atmospheric vortex waves moving around the pentagonal pattern.  
 In particular, the localised high pressure region shown in Fig. \ref{atmosphericoas1} is located at $74.35^{\circ}$ N and $37.84^{\circ}$ W, which approximates the location of the northern nodal point of the icosahedron-dodecahedron lattice. Through the illustrative examples presented here, we also demonstrate that the motion of the geographic North Pole can be viewed as the oscillations of the polar node in the discrete icosahedron-dodecahedron lattice, labelled by N in Fig. \ref{fig:icosdodec_framework}, where the vibrations of the lattice structure approximately capture the variations in Earth's rotational behaviour.

\FloatBarrier

\subsubsection{Case study I: Mexico City Earthquake, 1985}\label{Mexico:earthquake}
The Mexico City $8.1$-magnitude earthquake of 1985 was a major seismic event \cite{castro2016review, flores2007seismic, beck1986factors} that caused severe damage to the Greater Mexico City area and many casualties. This catastrophic geological event occurred due to a convergent-type earthquake, where the Cocos and North American tectonic plates moved toward each other and collided. The Cocos plate subduction process at these boundaries lead to an accumulation of stress, which was released as seismic waves linked to shear modes in the vertical plane. In this section, we demonstrate that the modes of the centred gyroscopic icosahedron-dodecahedron lattice, distinguished by the trajectories of the nodal elements, approximate the shear-type motions of the tectonic plates during the earthquake.

\begin{figure}[h!]
\centering
    \begin{subfigure}[h!]{0.4\textwidth}
    \centering
    \includegraphics[height=6cm]{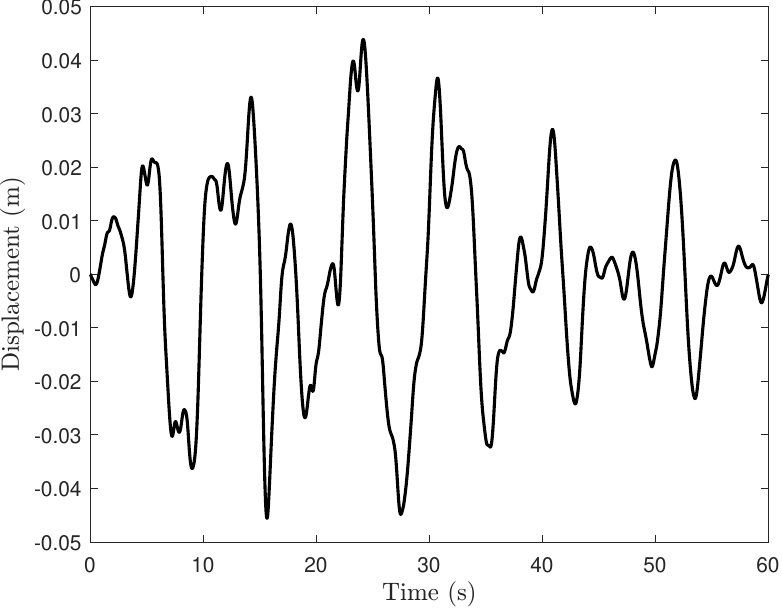}
    \caption{} 
    \label{fig:Mexico_seism}
	\end{subfigure}
    ~~~~~~~~~~~~~~~~
        \begin{subfigure}[h!]{0.4\textwidth}
		\centering
    \centerline{\includegraphics[height=6cm]{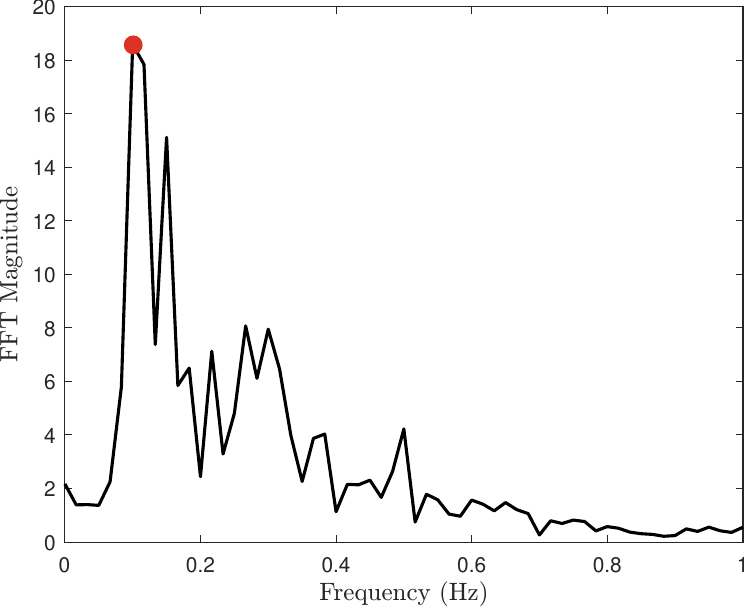}}
    \caption{}
    \label{fig:Mexico_seism_FFT}
	\end{subfigure}
	\caption{\small Seismograph data for the Mexico City earthquake in $1985$. (a) The seismograph recordings of the displacement variations during the earthquake for $60$ seconds and (b) the Fast Fourier Transform of the displacement seismograph. The (red) dot corresponds to a frequency of $0.1$ Hz, associated with the highest frequency value in the domain.   
    }
   \label{fig:Mexico_sesim_main}
\end{figure}
\FloatBarrier

\begin{figure}[h!]
\centering
    \begin{subfigure}[h!]{0.35\textwidth}
    \centering
    \includegraphics[height=5cm]{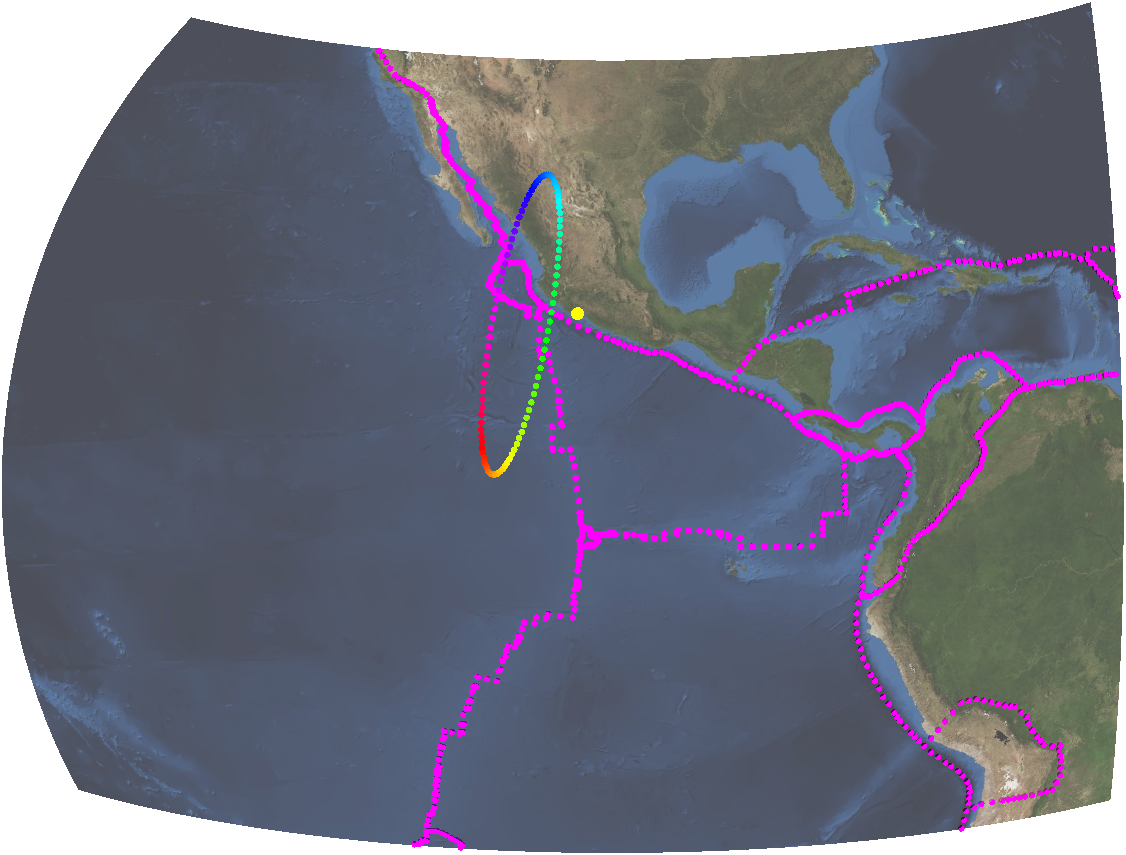}
    \caption{} 
     \label{fig:mexico_mode40}
	\end{subfigure}
    ~~~~~~~~~~~~~~~~
        \begin{subfigure}[h!]{0.35\textwidth}
		\centering
    \centerline{\includegraphics[height=5cm]{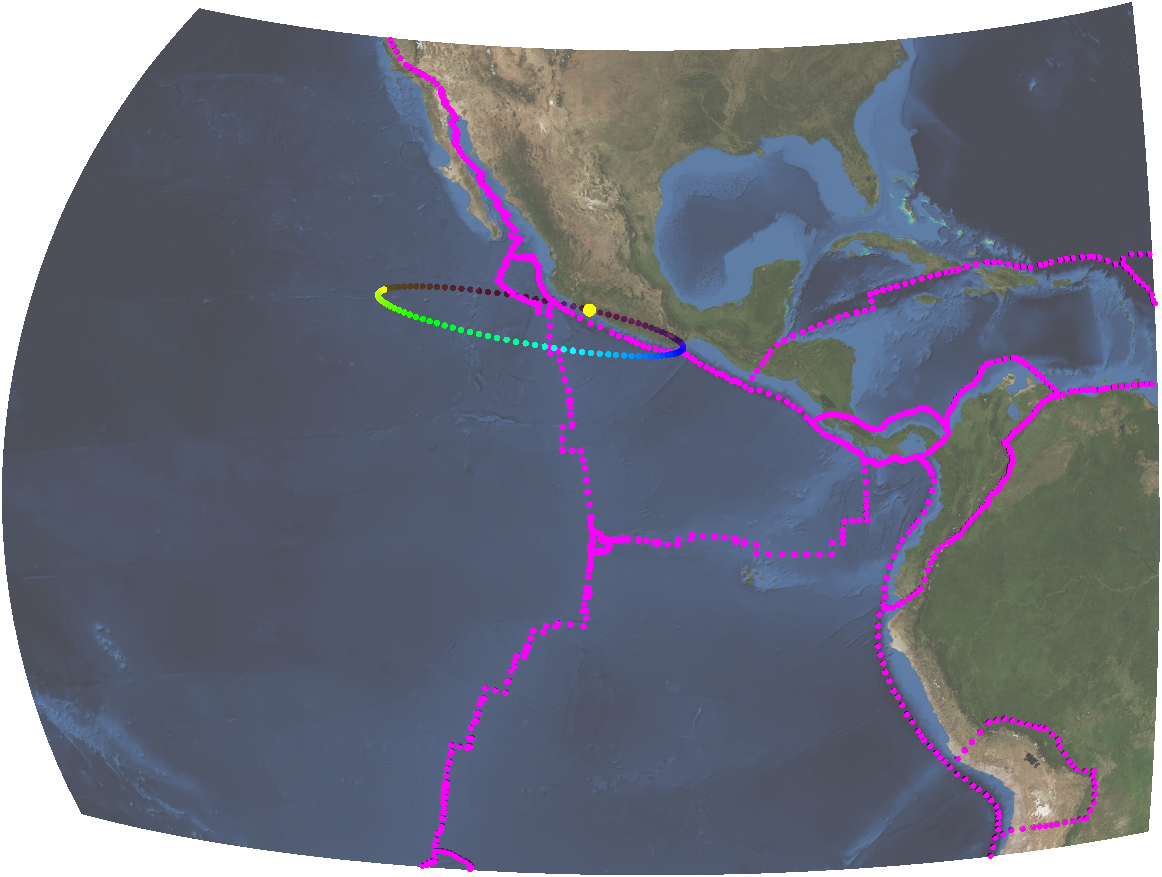}}
    \caption{}
     \label{fig:mexico_mode52}
	\end{subfigure}
	\caption{\small Nodal trajectory of the icosahedron-dodecahedron lattice in the vicinity of Mexico City. The geometrical and physical parameters of the lattice are presented in Table \ref{tab:earth_params}. The elliptical paths and orientation of motion for each node approximates the vertical shear motions induced by the convergent plate boundaries during the Mexico City earthquake. The oscillations frequencies of the icosahedron-dodecahedron lattice are: (a) $f=0.113211349862919$ Hz and (b) $f=0.154226571944373$ Hz.}
  \label{fig:mexico_modes}
\end{figure}

A sample of the displacement seismograph recordings from the Mexico city earthquake is presented in Fig. \ref{fig:Mexico_seism} for the duration of $60$ seconds.
The corresponding results of the Fast Fourier Transform (FFT) amplitude applied to the recorded seismogram dataset is plotted in Fig. \ref{fig:Mexico_seism_FFT}, which shows the dominant frequency range of the ground motions at approximately $0.1$ Hz, and a secondary peak at $0.15$ Hz. This demonstrates that the seismic waves had periods lasting several seconds, which resonated with the city's buildings leading to widespread damage \cite{flores2007seismic}.
We use the discrete icosahedron-dodecahedron lattice, detailed in Section \ref{sec:math_formulation}, to model the corresponding eigenmodes of the structure in connection with the shear tectonic plate motions during the Mexico city earthquake. The eigenmodes of the icosahedron-dodecahedron lattice, visualised around Mexico City, are shown in Fig. \ref{fig:mexico_modes}, with the highlighted nodal trajectories for two frequency values; $f=0.113211349862919$ Hz in Fig. \ref{fig:mexico_mode40} and $f=0.154226571944373$ Hz in Fig. \ref{fig:mexico_mode52}. 
It is noted that there is an eigenmode of the icosahedron-dodecahedron lattice equivalent to the example shown in Fig. \ref{fig:mexico_mode52} with $f=0.163409518561657$ Hz, which exhibits a clockwise motion, instead of a counterclockwise motion, due to the presence of gyroscopic effects. The nodal point, subjected to a gyroscopic action, follows elliptical paths perpendicular to the interface boundary between the Cocos and North American tectonic plates, representing the convergent-type earthquake mode during the Mexico city earthquake. The nodal trajectories of the masses for the discrete lattice model resemble the vertical shear modes associated with seismic waves induced by the Cocos plate subduction. Furthermore, we note that the choices of the geometrical and physical parameters for the icosahedron-dodecahedron lattice approximating the Earth values, provided in Table \ref{tab:earth_params}, result in a very good agreement with the frequency ranges of the recorded seismic data. 

\FloatBarrier

\subsubsection{Case study II: Kobe, Japan Earthquake, 1995}\label{Kobe1995}
In this section, we present the seismological data recordings of the Kobe, Japan earthquake in 1995 of magnitude $7.2$ \cite{katayama2006earthquake, morozov2021mechanism} (also known as the Great Hanshin earthquake), and the discuss the corresponding tectonic plate motions and ground displacement frequencies in connection with the vibration modes of the icosahedron-dodecahedron lattice. In comparison with the Mexico city 1985 earthquake caused by the convergent plate boundary, as described in Section \ref{Mexico:earthquake}, the Kobe earthquake was caused by a strike-slip fault movement along the Nojima fault \cite{wald1996slip, li1998delineation}. The earthquake involved the release of the accumulated stress due to the interactions between the Eurasian plate, Philippine Sea plate and Pacific plate, leading to widespread destruction \cite{katayama2006earthquake}. The strike-slip tectonic motion occurs when two plates slide horizontally past each other, rather than vertically. Through analysing the motions of the nodes in the discrete lattice model, we show that the vibration modes of the masses follow movements aligned with the tectonic plates which resemble the dominant horizontal shear seismic vibrations produced during the Kobe earthquake.       
\begin{figure}[h!]
\centering
    \begin{subfigure}[h!]{0.4\textwidth}
    \centering
    \includegraphics[height=6cm]{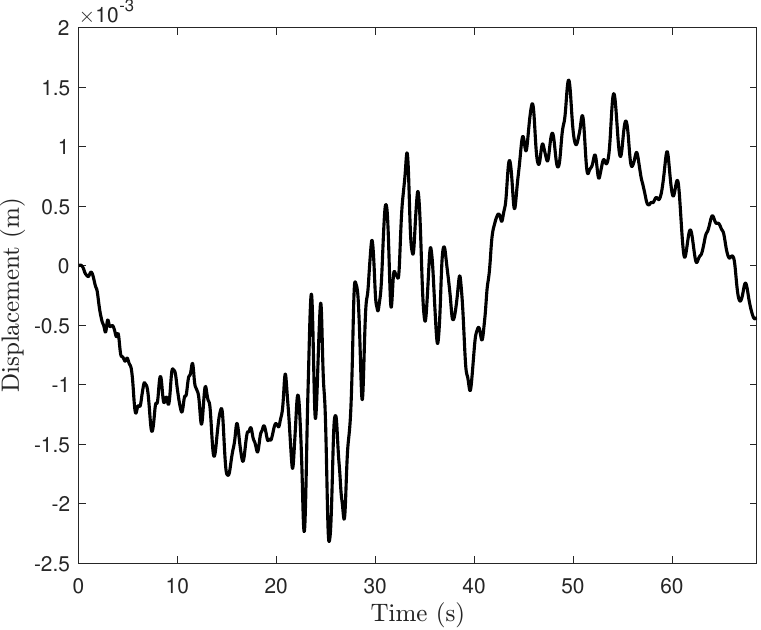}
    \caption{} 
    \label{fig:japan_seismo}
	\end{subfigure}
       ~~~~~~~~~~~~~~~~
        \begin{subfigure}[h!]{0.4\textwidth}
		\centering
    \centerline{\includegraphics[height=6cm]{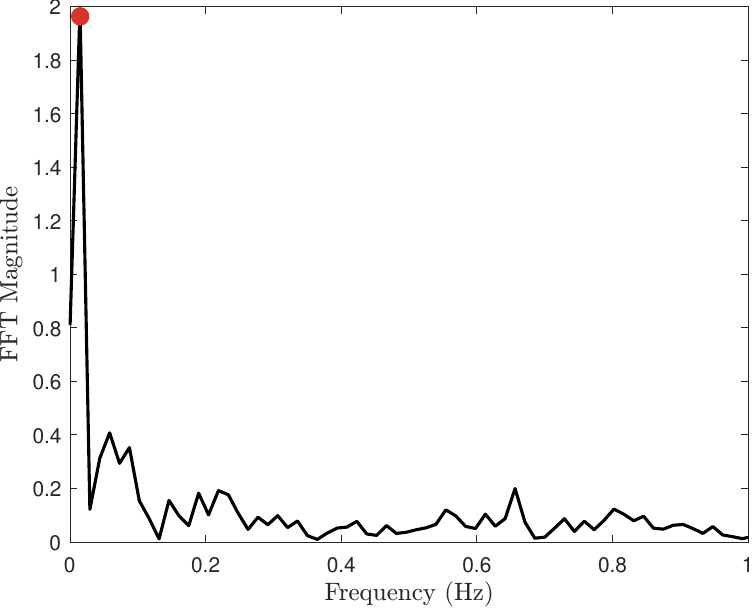}}
    \caption{}
      \label{fig:japan_FFT}
	\end{subfigure}
	\caption{\small Seismograph data for the Kobe, Japan $1995$ earthquake. (a) Seismograph dataset of the ground displacement for a time interval of $68.56$ seconds and (b) the Fast Fourier Transform of the displacement seismograph. The (red) dot highlights the highest frequency value of $0.015$ Hz of the ground motions induced by the earthquake.}
	\label{fig:japan_main}
\end{figure}
\FloatBarrier

Prior to the discussion of the modes for the icosahedron-dodecahedron lattice in connection with the tectonic plate motions, we derive the frequency range of the propagating seismic waves during the Kobe earthquake. Fig. \ref{fig:japan_main} presents the seismograph recording of the ground displacements generated by the earthquake, for a duration of $68.56$ seconds, as well as the output of the Fast Fourier Transform performed on the displacement data, which shows that the dominant frequency of the seismic waves is $0.015$ Hz. To approximately model the fundamental ground motions influenced by the Kobe earthquake with the frequency $0.015$ Hz, we analyse the eigenmodes of the centred gyroscopic icosahedron-dodecahedron lattice.

\begin{figure}[h!]
\centering
    \begin{subfigure}[h!]{0.4\textwidth}
    \centering
    \includegraphics[height=6cm]{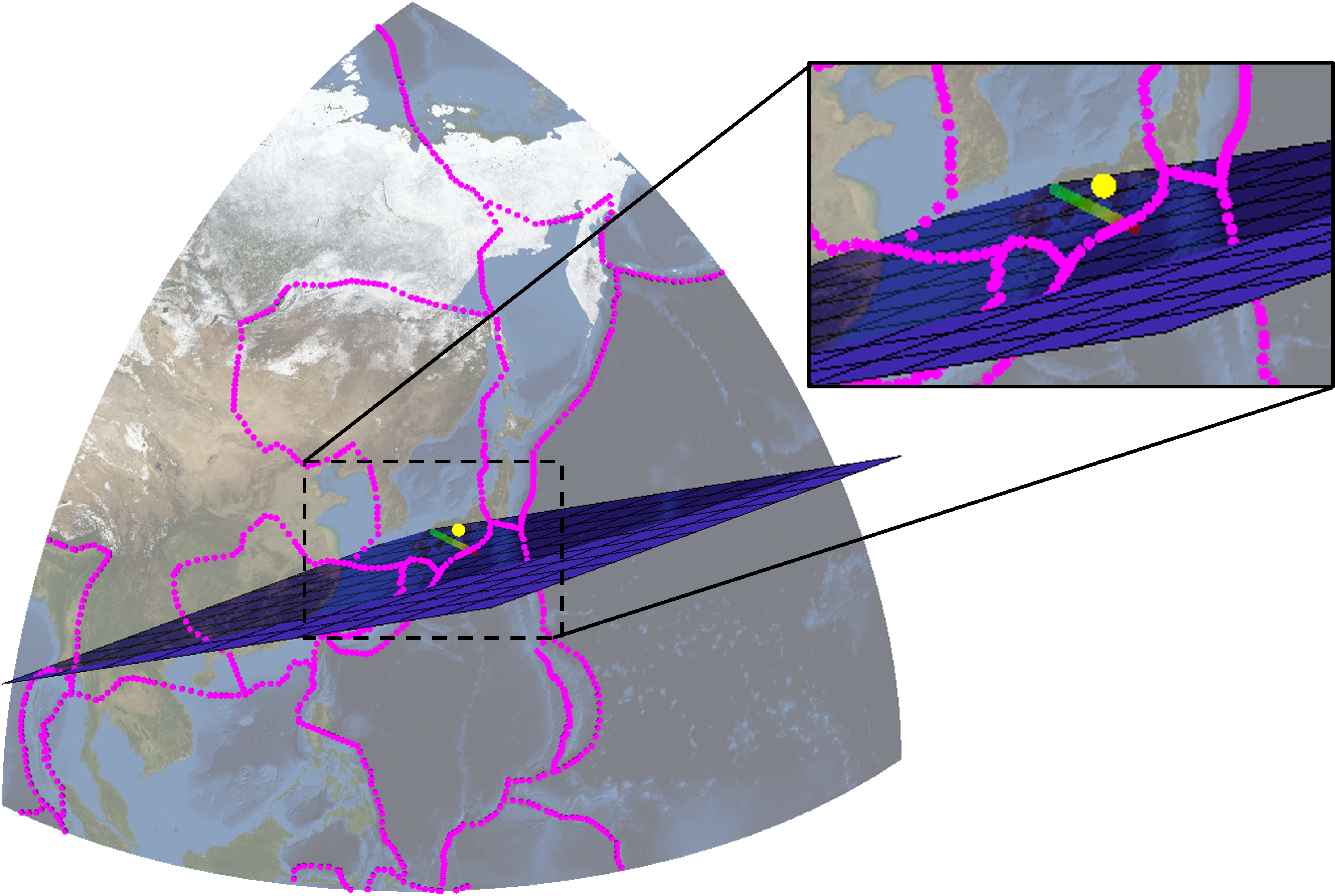}
    \caption{} 
   \label{fig:japan_mode18}
	\end{subfigure}
        ~~~~~~~~~~~~~~~~
        \begin{subfigure}[h!]{0.4\textwidth}
		\centering
    \centerline{\includegraphics[height=6cm]{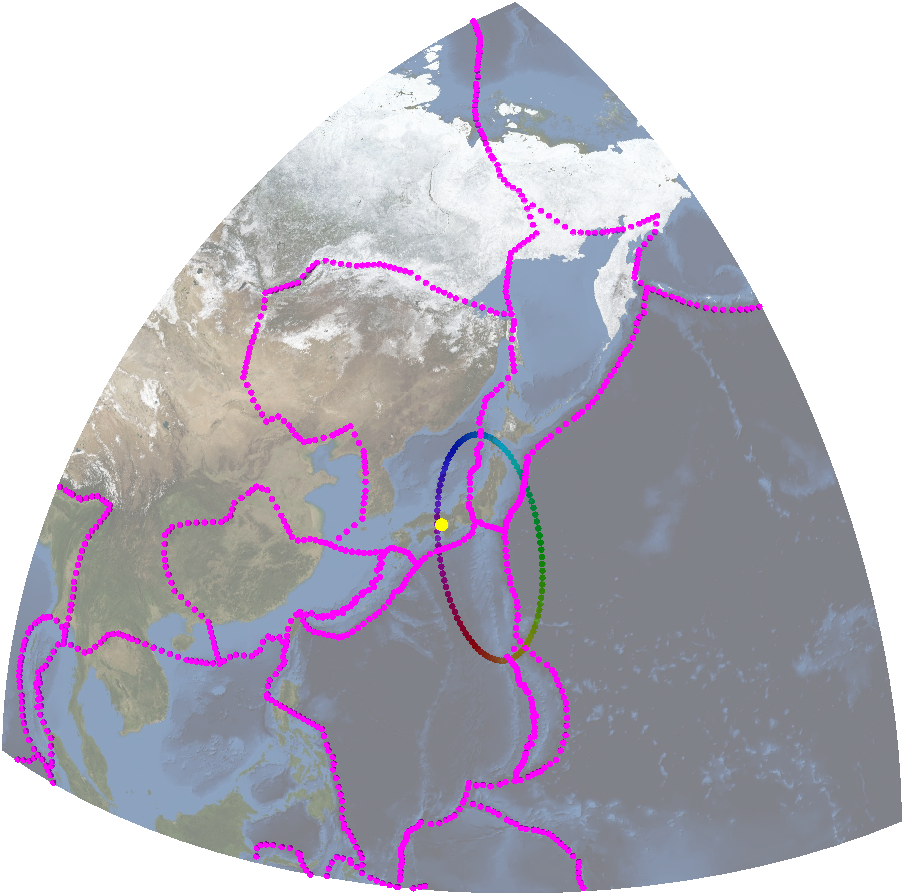}}
    \caption{}
    \label{fig:japan_mode19}
	\end{subfigure}
	\caption{\small Nodal trajectory of a mass in the icosahedron-dodecahedron lattice in the vicinity of the tectonic plates surrounding Japan. The elliptical motions of the nodes illustrate the horizontal shear motions of the Philippine Sea Plate and Eurasian plate, which resulted in the east-west strike-slip fault during the Kobe earthquake in 1995. The epicentre of the earthquake is highlighted by the marker (in yellow), while the tectonic plate boundaries are represented by the solid lines (in purple). The eigenmodes are plotted for two frequency values (a) $f=0.0161863480766603$ Hz and (b) $f=0.0171120475587316$ Hz.}
	\label{fig:japan_modes}
\end{figure}

Fig. \ref{fig:japan_modes} shows the normal modes for a nodal point of the icosahedron-dodecahedron lattice for two different frequencies of oscillation. The mass is situated on the vertex of the icosahedron and is in the vicinity of the tectonic plates around Japan, which are shown by the solid lines. The epicentre of the $1995$ Kobe earthquake is illustrated by the solid dot. The parameter values of the discrete lattice structure are chosen to be the same in the examples shown in Fig. \ref{fig:mexico_modes}. In Fig. \ref{fig:japan_mode18}, the trajectory of the nodal point follows a narrow elliptical path, with the vibration frequency $f=0.0161863480766603$ Hz, aligned with the radial direction of the Earth. To show the orientation of the ellipse we plot the corresponding plane, which contains this ellipse, in Fig. \ref{fig:japan_mode18}. In particular, this mode resembles the horizontal and vertical shear-type tectonic plate movements during the Kobe earthquake. The nodal trajectory with the frequency $f=0.0171120475587316$ Hz is shown in Fig. \ref{fig:japan_mode19}, where although the motion of the node is elliptical in a comparable manner to the example shown in  Fig. \ref{fig:japan_mode18}, the major axis of the ellipse is aligned in the tangential direction to the surface of the Earth. The mode illustrated in Fig. \ref{fig:japan_mode19} provides the approximate analogy with the shear motions of two tectonic plates sliding past each other, which contributed to the development of the Kobe earthquake. It is also observed that, at the frequencies of the normal modes shown in Fig. \ref{fig:japan_modes}, the motions of the nodal masses in the centred icosahedron-dodecahedron lattice exhibit a vertical shear-type oscillation. This movement is intrinsically connected to the horizontal displacements and shear stress characteristics that define a strike-slip type earthquake. Additionally, the frequencies of the normal modes in both illustrative examples approximate the dominant frequency of the seismic waves generated by the strike-slip fault during the Kobe earthquake (see Fig. \ref{fig:japan_FFT}). Thus, the modes of the discrete icosahedron-dodecahedron lattice, with the approximate Earth parameters, provide a very good agreement with the experimentally observed earthquake phenomena. 

\FloatBarrier

\subsubsection{Case study III: D\"{u}zce, Turkey Earthquake, 1999}\label{turk1}

The D\"{u}zce, Turkey earthquake of 1999 was a $7.1$ magnitude earthquake \cite{motosaka2002ground, utkucu2003slip} that resulted in extensive damage of buildings and collapse of walls. Analogously to the 1995 Kobe earthquake, induced by a strike-slip fault movement as described in Section \ref{Kobe1995}, the D\"{u}zce earthquake was the result of a strike-slip event along the North Anatolian fault \cite{duman2005step, akyuz2002surface}, causing the Anatolian microplate and the Eurasian plate to slide against each other. This tectonic setting makes Turkey highly seismically active, with frequent earthquakes occurring along the North Anatolian Fault. In this section, we derive the dominant frequency ranges for the sequence of earthquakes during the D\"{u}zce seismic event, and show that such ground motions can be approximated by the modes of the icosahedron-dodecahedron lattice.    

\begin{figure}[h!]
\centering
    \begin{subfigure}[h!]{0.4\textwidth}
    \centering
    \includegraphics[height=6cm]{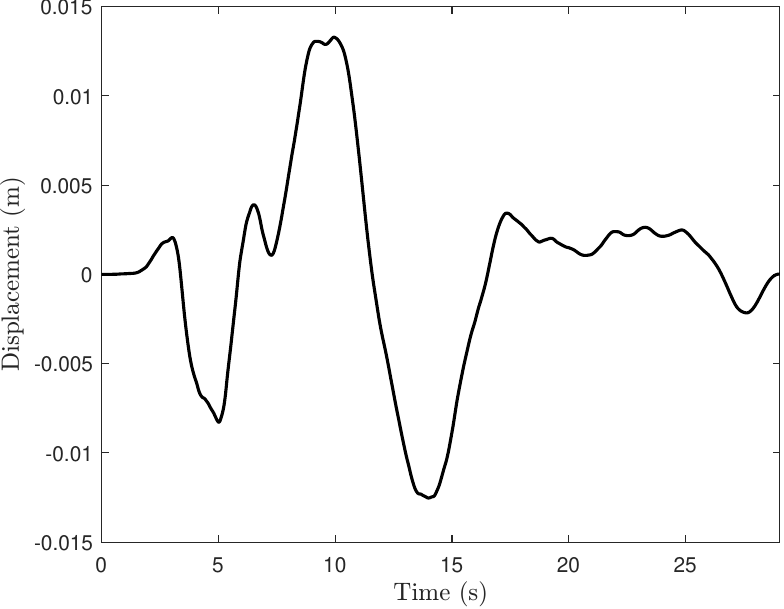}
    \caption{} 
    \label{fig:Turkey_seism}
	\end{subfigure}
       ~~~~~~~~~~~~~~~~
        \begin{subfigure}[h!]{0.4\textwidth}
		\centering
    \centerline{\includegraphics[height=6cm]{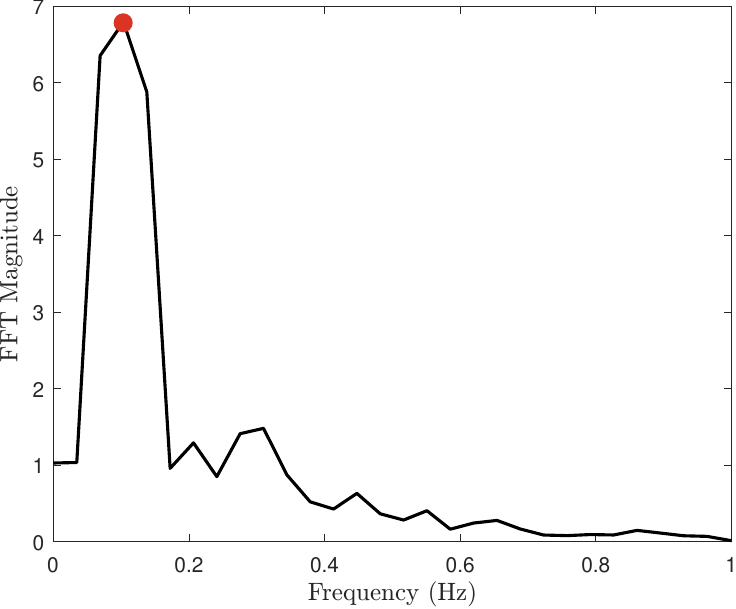}}
    \caption{}
    \label{fig:Turkey_seism_FFT}
	\end{subfigure}
	\caption{\small Seismograph data for the D\"{u}zce earthquake of 1999. (a) The seismograph recording of ground displacement during the D\"{u}zce earthquake for a time interval of $29.04$ seconds, and (b) the frequency-domain representation of the displacement seismograph, obtained using the Fast Fourier Transform. The (red) dot represents the highest frequency value of the ground vibrations during the earthquake, which occur in the range $0.02$ Hz to $0.07$ Hz.}
	\label{fig:Turkey_seism_main}
\end{figure}

The seismic recordings of the ground displacements during the D\"{u}zce earthquake for the duration of $29.04$ seconds is provided in Fig. \ref{fig:Turkey_seism}. 
By applying the FFT algorithm on the seismograph recording, we obtain the frequency domain of the seismic dataset as shown in Fig. \ref{fig:Turkey_seism_FFT}, where the frequency ranges of the seismic waves are along the $x$-axis and the magnitude spectrum of the FFT amplitude is along the $y$-axis. This analysis shows that the dominant frequencies of the ground vibrations during the D\"{u}zce earthquake are in the range $0.02$ Hz to $0.07$ Hz. 

\FloatBarrier

\begin{figure}[h!]
\centering
    \begin{subfigure}[h!]{0.35\textwidth}
    \centering
    \includegraphics[height=5cm]{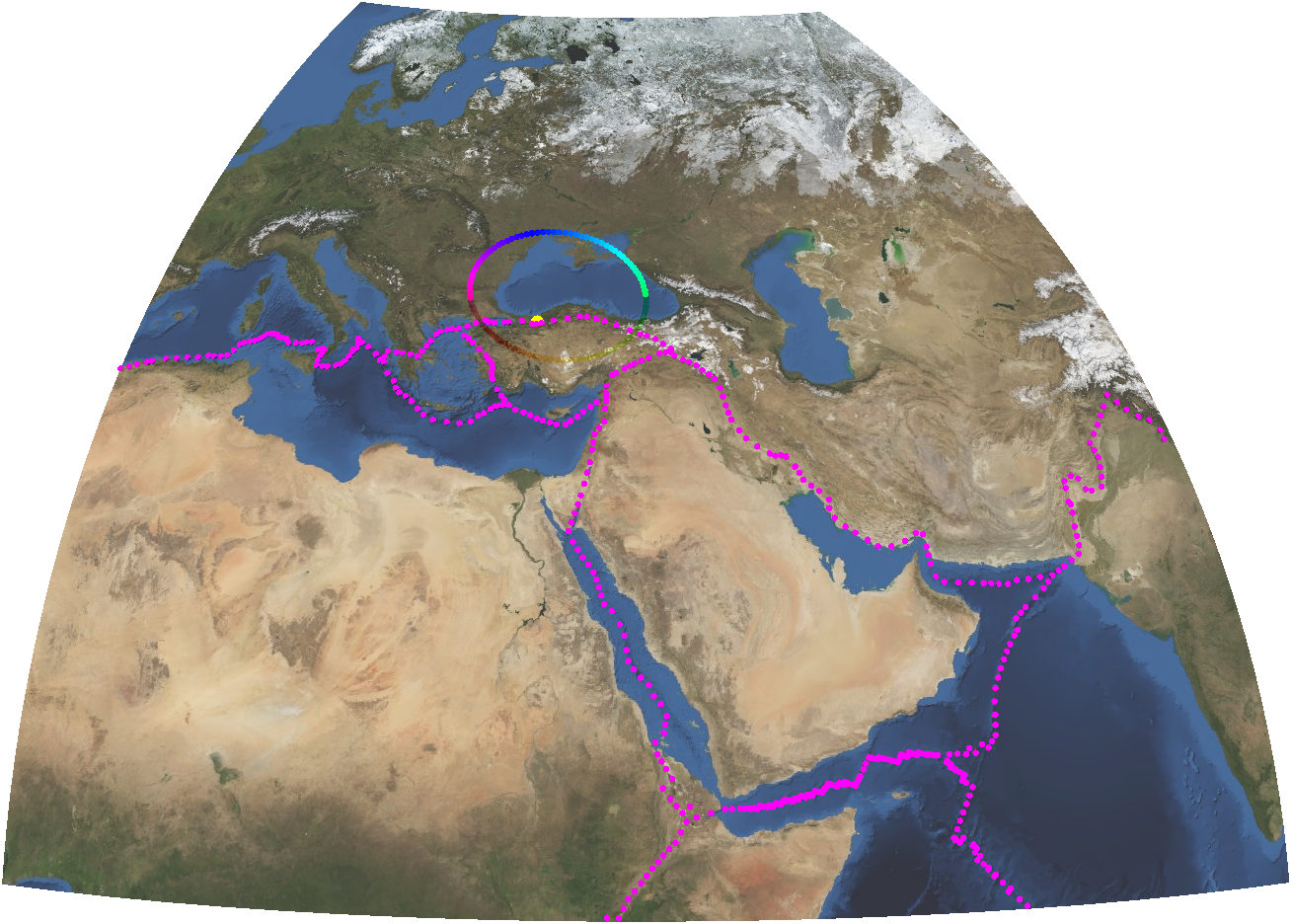}
    \caption{} 
   \label{fig:turkey_mode29}
	\end{subfigure} 
       ~~~~~~~~~~~~~~~~
        \begin{subfigure}[h!]{0.35\textwidth}
		\centering
    \centerline{\includegraphics[height=5cm]{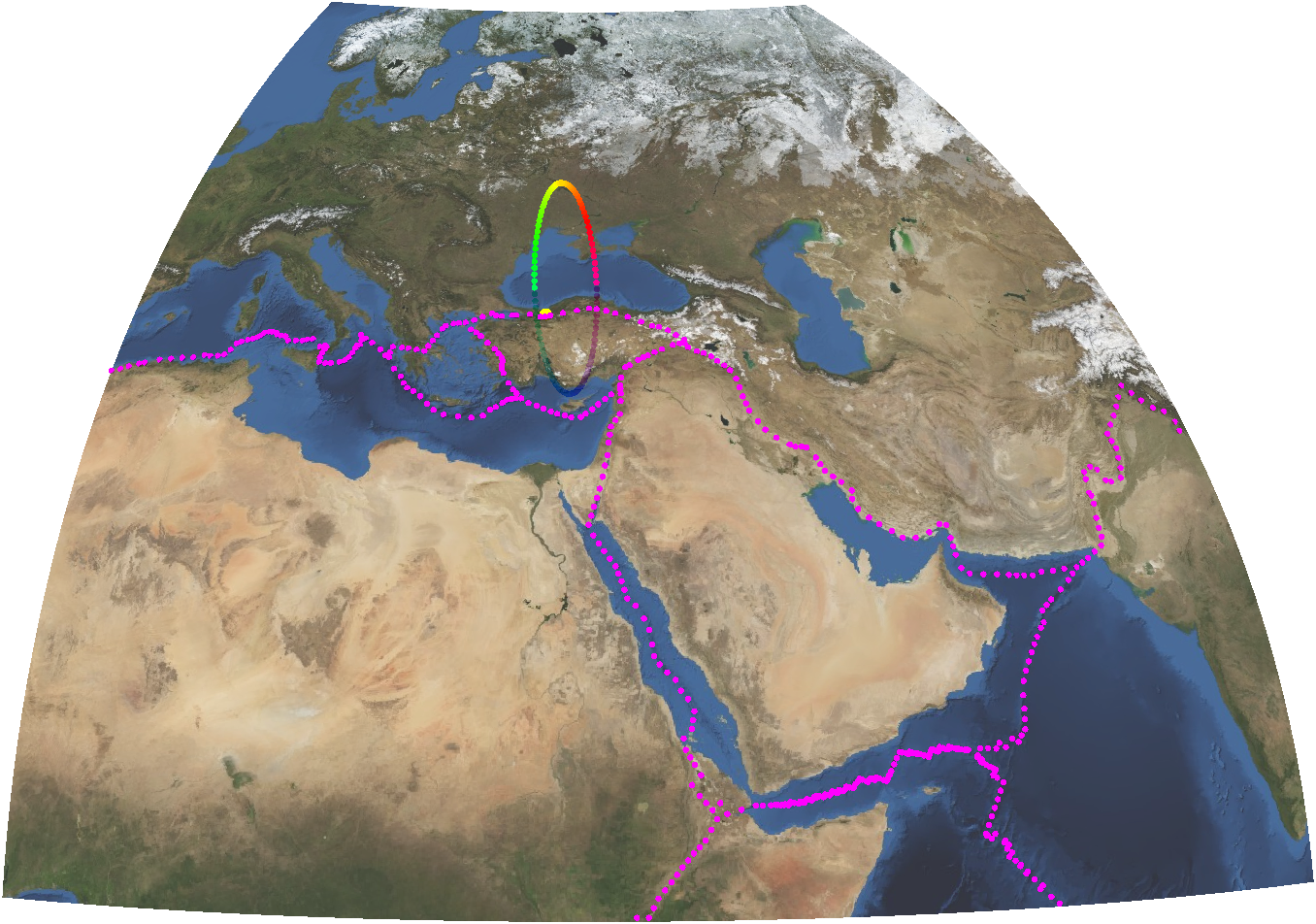}}
    \caption{}
    \label{fig:turkey_mode33}
	\end{subfigure}
	\caption{\small Trajectory of a node on the icosahedron-dodecahedron lattice associated with the $1999$ D\"{u}zce earthquake in the Anatolian fault zone. The frequencies of the icosahedron-dodecahedron lattice modes are as follows: (a) $f=0.0197565217194899$ Hz and (b) $f=0.0199599149067435$ Hz.}
	\label{fig:turkey_modes}
\end{figure}

We follow a similar analysis to the studies presented in Sections \ref{Mexico:earthquake} and \ref{Kobe1995}, and show that the eigenmodes of the icosahedron-dodecahedron lattice yield a simplified model approximating the frequencies and type of tectonic plate motions during the D\"{u}zce earthquake. In particular, the geometrical and physical parameters used for the discrete lattice are presented in Table \ref{tab:earth_params}, which are chosen in connection with the Earth parameters. The icosahedron-dodecahedon lattice vibration modes corresponding to the frequencies $f=0.0197565217194899$ Hz and $f=0.0199599149067435$ Hz are shown in Fig. \ref{fig:turkey_mode29} and Fig. \ref{fig:turkey_mode33}, respectively. The illustrative examples show the elliptical trajectory of the nodal point, aligned 
perpendicular to the surface of the Earth. The motion shown in Fig. \ref{fig:turkey_mode29} is dominated by the horizontal movement of the node, while Fig. \ref{fig:turkey_mode33} demonstrates the localised vertical displacements of the node. The combined characteristics of the two modes display the combined effects of the horizontal and vertical shear motions between the tectonic plates in the vicinity of Turkey. As noted previously, the D\"{u}zce earthquake of 1999 occurred due to the motion of the Anatolian and the Eurasia plates, resulting in a strike-slip fault motion. The particular shape of the vibration mode for the centred discrete lattice model suggests that the tectonic plates exhibited shear motions as they slid against each other during the D\"{u}zce earthquake, which was precisely the observed result of the geophysical phenomena. 

\FloatBarrier









\section{Discussion}\label{discusso}

\subsection{Rotation of the Earth's core}
The discrete lattice structures studied in this paper provide a novel framework for modelling the rotational changes in Earth's inner core. Through this study, it was shown that although the gyroscopic forces can be uniformly distributed across each nodal point of the lattice, the trajectories of the masses, which includes the central node, can follow different paths. In the context of Earth's interior, this analysis demonstrates that the Earth's inner core can rotate with the same orientation relative to the crust and mantle, but its motion can be in a different direction. By incorporating gyroscopic spinners at each node of the discrete lattice, these lattice models simulate the inertial effects and Coriolis force within the core. This approach provided an approximate analogy with the continuum phenomena describing variations in the inner core's rotation and its influence on seismic wave propagation. 
The study presented in this paper provides an analytical method for modelling the changing Earth inner core's motion influenced by long-term earthquake cycles. 


\subsection{Global view of earthquakes in the novel gyroscopic lattice framework }
There have been extensive studies and spectral analysis of earthquake events with an emphasis on experimental models and simulations of tectonic plate movements (see, for example \cite{de2016statistical, arrowsmith2010seismoacoustic, rundle2003statistical, anagnos1988review, kagan1994observational, moczo2014finite, lu2011large, minster1974numerical}). In particular, computational earthquake models can depend on probabilistic methods, which involve simplifying assumptions and lack deterministic physical mechanisms characterising geodynamic phenomena, while finite element methods can be computationally costly. In the present paper, we have presented a fully analytical mathematical framework to approximate the fundamental dynamics of tectonic plates and the frequencies of the ground vibrations induced by seismic activity through discrete three-dimensional gyroscopic lattice models. 
We observed that the vibration modes of the lattice capture the movements of tectonic plates, with a very good comparison with observations of major earthquakes. Earthquakes also exhibit variations in the frequency distributions of the seismic waves and distinct tectonic plate motions, which are the characteristics inherent to lattice models. This shows the applicability of studying lattice models in modelling different earthquake types and the resulting plate movements. 
By integrating findings on the core's rotation and global earthquake distributions, the lattice-based approaches provide a tuned representation in modelling the mechanics of large earthquakes. The recent seismic events of 2025 reinforce the relevance of these models, demonstrating their potential in refining the current mathematical models of earthquake phenomena.

\section{Methods}\label{methodo}

The governing equations of motion of the icosahedron-dodecahedron lattice are included in Section \ref{sec:time-harmonic_equationszxasx} of the Appendix. 
 In the present study, we consider the lattice model with five-fold rotational symmetry for convenience in constructing the reference cell. The five-fold rotational symmetry patterns were also observed for the atmospheric vortex waves on Earth as shown in Fig. \ref{atmosphericoas1}, as well as in the shape of the jet streams \cite{kandiah2025dispersion2}. 
This is consistent with the work \cite{ lognonne1998computation}, which highlighted the connection between the large seismic events and the formation of standing waves in the atmosphere. 
An important feature of the discrete models is the presence of gyroscopic spinners, associated with the inertial force of the rotating planet, and the radially multi-layered structure, which approximates the different properties of the Earth's layers. The lattice models studied in this paper present a theoretical approximation to the fundamental motions of the tectonic plates during seismic processes, with the characterisation of the ground oscillation frequencies and different earthquake-types. Additional details including seismographs and Fast Fourier Transforms 
are also provided in this section.

\subsection{Seismograph measurements and fast Fourier transforms} 
The raw earthquake data analysed in Sections \ref{Mexico:earthquake} - \ref{turk1} was obtained from the U.S. Geological Survey (USGS) earthquake catalogue \url{https://earthquake.usgs.gov/earthquakes/search/}, which provides real-time earthquake records and ground motion data. 
This data is important for earthquake hazard assessment, structural resilience modelling and tectonic activity research, which support scientists and engineers in analysing earthquake risks. To obtain the frequency components of the ground motions during earthquakes, the Fast Fourier Transform (FFT) is applied to the recorded seismic dataset, converting the signal from the time domain to the frequency domain. The FFT shows the frequency distribution and the dominant frequencies corresponding to the resonant modes of the seismic ground vibrations. Additionally, the frequencies of the centred gyroscopic icosahedron-dodecahedron lattice showed a very good agreement with the frequencies of the earthquake-induced ground vibrations. The frequency-domain representations are also used to assess the earthquake impact, predict potential ground motion hazards and refine computational models for seismic risk evaluation.




\subsection{Eigenmodes of vibrations for a rotating elastic ball}\label{rotelas1a}
The natural oscillations of the Earth are described by equations derived from the principles of elasticity, fluid dynamics and gravitational interactions. In the absence of rotation, the Earth's eigenmodes of free oscillation correspond to the spheroidal and toroidal modes. When the Coriolis terms are introduced, the interaction between these modes yields a coupled oscillatory system, where novel vibrational patterns are present. Eigenfrequencies and eigenmodes of the rotating Earth differ significantly from the corresponding vibration modes without rotation, as discussed in \cite{dahlen1975influence}, where the effects of the gyroscopic frequency split and Chandler wobble, as well as other features, were highlighted. Additional novel properties of the modes for rotating elastic bodies are analysed in the present work, with the physical scales and parameters approximating the Earth values. 
The approximate polyhedra modes of the elastic ball are analogous to the discrete three-dimensional lattices studied in this paper.

To simplify our analysis, we assume small linear elastic deformations and that the Earth is isotropic with uniform material properties in all directions. In this approximation, the linearised time-harmonic equations governing the elastic vibrations of the Earth are 
\begin{equation}
-\rho \omega^2 {\bf U}({\bf x}) + 2 i \rho \omega {\boldsymbol{\frak{S}}} \times {\bf U}({\bf x}) = \mu \nabla^2 {\bf U}({\bf x}) + (\lambda + \mu) \nabla (\nabla \cdot {\bf U}({\bf x})), \label{elas0139as}
\end{equation}
where $\rho$ is the density, $\omega$ is the radian frequency of the oscillation, ${\bf U}$ is the amplitude of the displacement vector of an element on the Earth, ${\bf x}$ is the position vector of an elastic element, $\boldsymbol{\frak{S}}$ is the angular velocity vector of the rotating body and $\lambda$ and $\mu$ are the Lam\'e constants, which are given by 
\begin{equation}
\mu = \frac{E}{2(1 + {\bf \nu})}, ~~ \lambda = \frac{E \nu}{(1+ \nu) (1 - 2 \nu)},
\end{equation}
with $\nu$ being the Poisson ratio and $E$ the Young modulus. The gyroscopic action is introduced by the rotation of the planet, through the Coriolis force terms. Additionally, we set $\boldsymbol{\frak{S}}=(0, 0, \Omega_{\text{E}})^T,$ where $\Omega_{\text{E}}$ is the angular speed of the Earth, where the positive direction of spin is counterclockwise as we look from the positive direction of the $z$-axis, the axis passing through the poles of the Earth. 

\begin{figure}[h!tb]
	\centering
	\begin{subfigure}[h!]{0.41\textwidth}
		\centering
		\includegraphics[width=\linewidth]{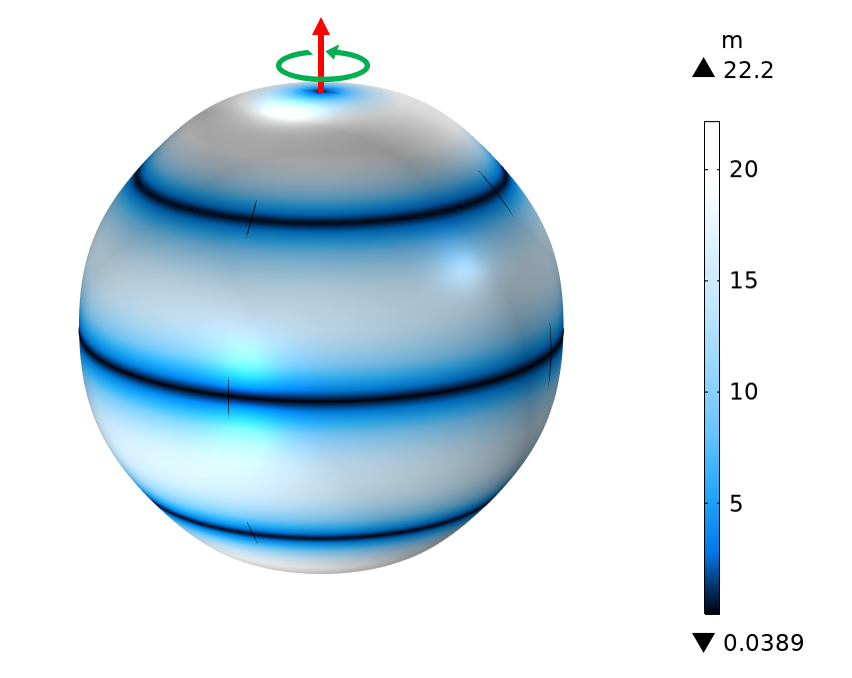}
		\caption{ $f= 3.5181 \times 10^{-4}$ Hz} \label{balleig1}
	\end{subfigure}  \hspace{.5in}
	\begin{subfigure}[h!]{0.41\textwidth}
		\centering
		\includegraphics[width=\linewidth]{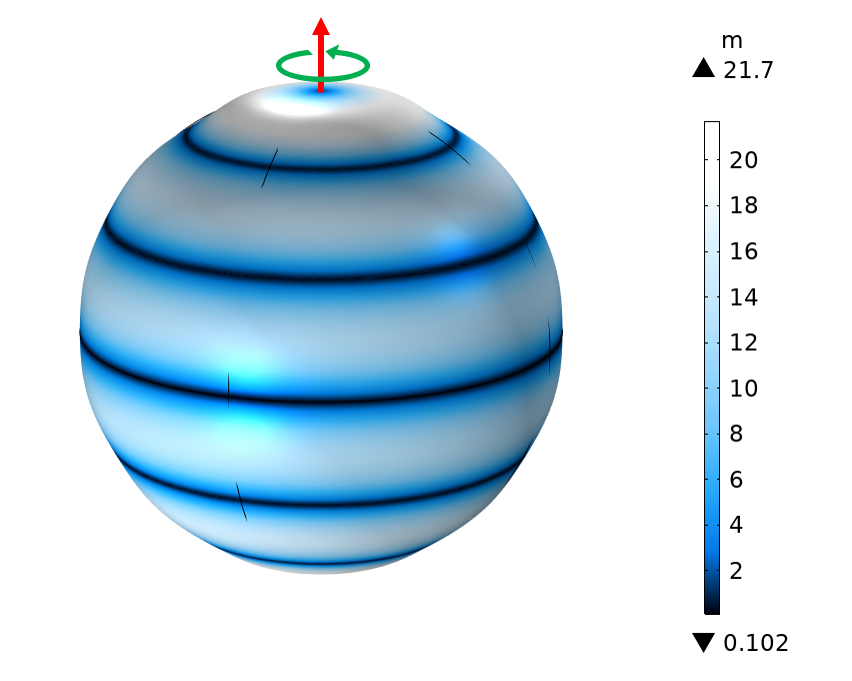}
		\caption{$f =5.1179 \times 10^{-4} $ Hz}   \label{balleig2}
	\end{subfigure}
	\caption{\small Eigenmodes of an isotropic elastic rotating ball with small displacement bands parallel to the equatorial region. The Young modulus, Poisson ratio and mass density of the ball are $110$ GPa, 0.3 and $5515$ kg/m$^{3}$, respectively. The ball is rotating about its vertical axis, which passes through its centre and the poles, with the angular speed of $7.2921\times 10^{-5}$ rad/s. The frequencies of the vibrations are: (a) $f=3.5181 \times 10^{-4}$ Hz and (b) $f = 5.1179 \times 10^{-4}$ Hz. 
    }
	\label{balleig}
\end{figure}

Figures \ref{balls} and \ref{balleig} include the illustrative simulations for elastic vibrations of the rotating Earth. In particular, Fig. \ref{balleig4} and Fig. \ref{balleig3}, show the modes, which resemble the symmetries of the icosahedron and dodecahedron, whereas the modes in Fig. \ref{balleig1} and Fig. \ref{balleig2} show the band structure, parallel to the equator. The geometrical and material parameters are chosen to provide an approximation based on the Earth's composition and the elastic properties of its materials \cite{planetHandbook}. The parameters, used in the computations,  are as follows:
 \begin{itemize}
 \item The angular frequency $\Omega_{\text{E}} = 7.2921\times 10^{-5}$ rad/s, where the rotational axis is the vertical $z$-axis. 
 
 \item The elastic ball is assumed to be isotropic, with the Young modulus $E = 110$ GPa, the Poisson ratio $\nu = 0.3$ and the mass density $\rho = 5515 $ kg/m$^{3}.$ 
 
 \item The radius of the ball is $R=6371 \times 10^{3}$ m, representing the mean radius of the Earth \cite{planetHandbook}.
\end{itemize}

\subsubsection{Axially symmetric modes of the rotating elastic ball}

For low frequency vibration modes of the rotating elastic ball, we observe bands parallel to the equator as shown in Fig. \ref{balleig1} and Fig. \ref{balleig2}, where the frequencies are $f=3.5181 \times 10^{-4}$ Hz and $f = 5.1179 \times 10^{-4}$ Hz, respectively. In particular, changing the frequencies of the oscillations can result in a different number of bands. This feature occurs due to the Coriolis term introducing anisotropy in the displacement field, and it is absent for the non-rotating ball. These bands highlight the areas where the material remains relatively undisturbed, while forming a boundary between oscillating regions, which occurs due to the combined effects of the elastic restoring forces and the Coriolis force. On the Earth, there is a similar feature connected to the atmospheric circulation patterns; they are the Hadley, Ferrel and Polar cells \cite{schneider2006general}, that are also approximately parallel to the equator. 

\subsubsection{
Icosahedron- and dodecahedron-like vibration modes}

The eigenmodes of vibration for the rotating elastic ball can also resemble polyhedra with rotational symmetries as shown in Fig. \ref{balleig3} and Fig. \ref{balleig4}. The eigenmode shown in Fig. \ref{balleig4} displays a five-fold rotational symmetry about the principal vertical axis of the system, with approximate pentagonal patterns of the poles surrounded by five approximate triangular faces as well as ten points with small displacements localised near the equator. In this case, $f = 7.7873\times 10^{-4}$ Hz and the mode of the rotating elastic ball resembles an icosahedral-like structure. Furthermore, in Fig. \ref{balleig3}, the mode shape of the rotating ball approximates a dodecahedral-like structure oscillating with the frequency $f=7.8063\times 10^{-4}$ Hz, with three-fold rotational symmetry relative to the vertical $z$-axis. 
The frequencies of these polyhedral eigenmodes depend on the rotational speed and material properties of the ball.

The properties of the isotropic elastic solid ball discussed in this section have important implications in geophysics and engineering, providing insights into the dynamics of rotating elastic bodies at different scales. It is demonstrated that the icosahedron and dodecahedron occur naturally in the analysis of the eigenmodes of the rotating elastic ball. 
This further motivates the idea of studying the Earth's vibrations through the use of three-dimensional lattice models with gyroscopic elements. The analysis presented in this section highlighted that the eigenfrequencies of the continuum elastic system are linked to resonances observed in geophysical phenomena, such as standing waves generated by earthquakes. 
\begin{figure}[h!tb]
	\centering
		\centering
		\includegraphics[width=0.35\linewidth]{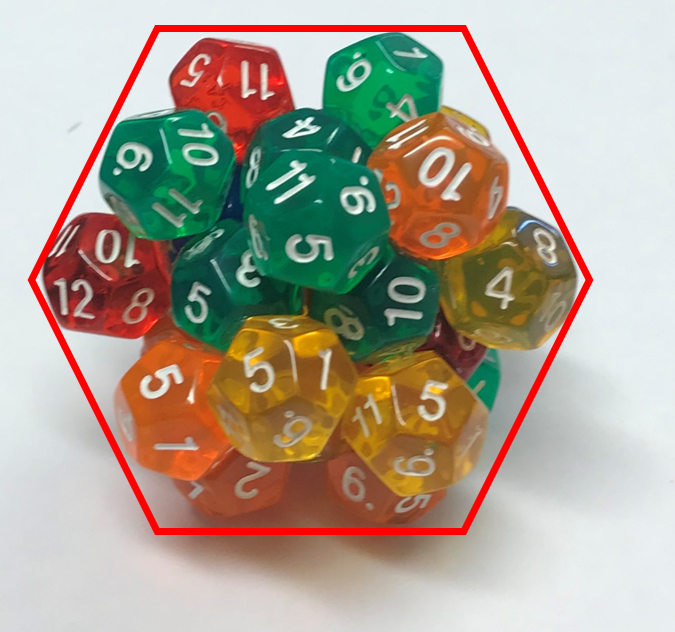}
		\caption{\small A three-dimensional cluster of dodecahedrons.}
    \label{Dod}
\end{figure}
\FloatBarrier

\section{Conclusions and outlook}\label{summaro}
The results of this paper combine geophysical phenomena and mathematical modelling to present a new understanding of the Earth's dynamics through a three-dimensional spectral lattice framework with gyroscopic elements. The analysis demonstrated a novel application of chiral centred discrete lattices in approximating the large-scale elastic vibrations of the rotating Earth. The correspondence between the dynamics of the nodal elements in the  discrete chiral lattice and tectonic plate structure provides important insights into the fundamental ground motions and seismic wave frequencies during large earthquakes. 

The discrete lattice spectral models presented in this paper overcome challenges inherent to continuum-based transient simulations, offering a multi-scale framework for modelling seismic events. Beyond the problems of geophysics, we also note that the clusters of dodecahedron-like systems are observed at a molecular level for supercooled water molecules \cite{loboda2010theoretical}.
Fig. \ref{Dod} shows a three-dimensional cluster of dodecahedrons (model built by the authors), which appears in the two-dimensional projection as an object embedded into a hexagon. Similarly, many cellular structures, observed in nature as honeycomb systems, may in fact be linked to the dodecahedron-like clusters. With added gyricity, this gives a potential new direction of further development of the spectral study, presented here.     

\subsection*{Acknowledgements}
W.E.B. is grateful for EPSRC funded 2024 Summer Internship at the Department of Mathematical Sciences, University of Liverpool. A.K. gratefully acknowledges the financial support of the EPSRC through the Mathematics DTP grant EP/V52007X/1, project reference 2599756. A.B.M. and V.F. are grateful to the British Council for supporting the academic visit of V.F. to the University of Liverpool in 2024.

\appendix
\section*{Appendix}

\numberwithin{equation}{section} 

\subsection*{This Appendix includes:}
\begin{itemize}
\item[--] Supplementary text with the governing equations for the centred gyroscopic icosahedron lattice introduced in Section \ref{sec:icosahedron}. The text also includes the derivation of the gyroscopic force acting on the nodal elements of the lattice in connection with the Coriolis force.
\item[--] Supplementary text for the problem formulation of the centred gyroscopic icosahedron-dodecahedron lattice analysed in Section \ref{sec:math_formulation}. The text includes the derivation of the governing equations for the nodal elements in the lattice and the choices of the parameter values linked with the Earth parameters. 
\item[--] Supplementary text for the optimisation used to match the centred gyroscopic icosahedron-dodecahedron lattice with the locations of earthquake epicentres in Section \ref{sec:math_formulation}.
\item[--] Supplementary tables of raw earthquake data used in Section \ref{sec:math_formulation}. 
\end{itemize}



\renewcommand{\thesection}{A\arabic{section}}

\section{Equations of motion of the icosahedron lattice}\label{icos:lattzxsa}
In this section, we discuss the vibrations of a discrete centred icosahedron lattice subjected to gyroscopic forces, and provide the connection with geophysical phenomena on Earth. We study the dynamics of the icosahedron structure through the governing equations with the linearisation assumptions. 

The equations of motion in the time-harmonic regime for the $i$-th node of the icosahedron lattice (see Fig. \ref{Earthicosahedron}) are given by (summation index notations are not used here)
\begin{equation}
m_i \frac{d^2{\bf u}_i}{d t^2} = \sum_{j=1}^{5} k_{ij} \big( {\bf e}_{ij} \cdot ( { \bf u}_j - {\bf u}_{i}) \big)  {\bf e}_{ij}  + k_{i c}  \big( {\bf e}_{ic}\cdot ({\bf u}_{c} - {\bf u}_{i}) \big) {\bf e}_{i c} +  {\bf \Omega}_{i} \times \frac{d {\bf u}_{i}}{d t}, ~~~~i=1,\ldots, 12, 
\label{centraleq1}
\end{equation}
where $m_{i}$ is the mass of the $i$-th node in the lattice, $\omega$ is the radian frequency, ${\bf u}_{i}$ is the small displacement vector of the respective masses, $k_{ij}$ is the stiffness of the elastic link between node $i$ and its neighbouring node $j,$ $k_{i c}$ is the elastic stiffness of the connection between node $i$ and the central node $c$ of the lattice, ${\bf e}_{ij}$ is the unit vector along the direction from $i$ to $j$ in the equilibrium position, ${\bf e}_{ic}$ is the unit vector in the direction from $i$ to the central node $c$ at equilibrium and ${\bf \Omega}_{i}$ is the vector characterising the gyroscopic action on the $i$-th mass. We note that ${\bf e}_{ i j } = - {\bf e}_{j i}$ and ${\bf e}_{i c} = - {\bf e}_{c i}.$ The $i$-th node of the three-dimensional lattice corresponds to a vertex of the icosahedron. 

The equation of motion in the time-harmonic regime for the central node of the icosahedron lattice is given by
\begin{equation}
m_c \frac{d^2 {\bf u}_c}{d t^2} =\sum_{n=1}^{12} k_{cn } \big( {\bf e}_{cn} \cdot  ( { \bf u}_n - {\bf u}_{c}) \big)   {\bf e}_{c n}  + {\bf \Omega}_{c} \times \frac{d {\bf u}_{c}}{d t}, \label{centraleq2}
\end{equation}
where $m_{c}$ and ${\bf \Omega}_{c}$ are the mass and the gyricity vector of the central node, respectively. Here we assume linearised motions of the central node with the displacement vector ${\bf u}_{c}$. 

How do we define the gyricity vector ${\bf \Omega}$ in  (\ref{centraleq1})-(\ref{centraleq2})? 
Consider the rotating lattice structure, shown in Fig. \ref{Earthicosahedron}, with the constant vector ${\boldsymbol{\Theta}}$ being the rotation vector. Assuming that $({\bf i}, {\bf j}, {\bf k})$ is the orthonormal Cartesian basis, corresponding to the fixed frame of reference, we also introduce the rotating Cartesian frame of reference with the basis vectors $({\bf i}'(t), {\bf j}'(t), {\bf k}'(t))$, which depend on time $t$ in such a way that 
\begin{equation} \label{rotbasis}
\frac{d {\bf i}'}{d t} = {\boldsymbol{\Theta}} \times {\bf i}',~~\frac{d {\bf j}'}{d t} = {\boldsymbol{\Theta}} \times {\bf j}',~~\frac{d {\bf k}'}{d t} = {\boldsymbol{\Theta}} \times {\bf k}'.
\end{equation}
In particular, if the axis of rotation coincides with the vertical $z$-axis, we have ${\bf k}'={\bf k}$, and ${\boldsymbol{\Theta}}= \Theta {\bf k}$, where $\Theta$ is the constant angular speed of rotation, and equations (\ref{rotbasis}) become
\begin{equation} \label{rotbasis1}
\frac{d {\bf i}'}{d t} = {{\Theta}} {\bf j}',~~\frac{d {\bf j}'}{d t} = -{{\Theta}} {\bf i}',~~\frac{d {\bf k}'}{d t} = {\bf 0}.
\end{equation}
Assuming that the elastic links in the structure, shown in Fig. \ref{Earthicosahedron}a, are massless, the equations of motion for the $p$-th nodal mass 
in the discrete lattice structure rotating with the constant angular velocity $\boldsymbol{\Theta}$ about the fixed axis takes the form 
\begin{equation}
    m_{p} {\bf a}'_p 
    = {\bf F}_{p} - 2 m_{p} \boldsymbol{\Theta} \times 
    {\bf v}'_p- m_{p} \boldsymbol{\Theta} \times (\boldsymbol{\Theta} \times {\boldsymbol{r}_{p}}') , \label{coriolis1}
\end{equation}
where $m_{p}$ is the mass of the $p$-th node and ${\bf F}_{p}$ represents the internal elastic forces acting on the $p$-th node. The quantities ${\bf a}'_{p}, {\bf v}'_{p}$ and ${\bf r}'_{p}$ are the acceleration, velocity and position vectors of the $p$-th nodal point in the non-inertial reference frame, which rotates around the $z$-axis. 
The second and third terms on the right-hand side of (\ref{coriolis1}) are referred to as the Coriolis force and the centrifugal force, respectively. For the Earth, 
the vector $\boldsymbol{\Theta}$ is directed towards the North Pole with $\Theta = 7.2921159 \times 10^{-5}$ s$^{-1}$. In the linearised regime, we neglect the centrifugal force terms in (\ref{coriolis1}), which results in a system of governing equations (\ref{centraleq1})-(\ref{centraleq2}) similar to that of the icosahedron lattice with gyroscopic nodes.  In this case, the simplified equation (\ref{coriolis1}) takes the form 
\begin{equation}
    m_{p} {\bf a}'_p 
    = {\bf F}_{p} - 2 m_{p} \boldsymbol{\Theta} \times 
   {\bf v}'_p ,
\label{coriolis1a}
\end{equation}
where the force vector ${\bf F}_{p}$ in (\ref{coriolis1a}) represents the elastic spring forces in (\ref{centraleq1})-(\ref{centraleq2}), while the Coriolis force 
 is represented by the gyroscopic force for the discrete icosahedron lattice. In the present model we consider the three-dimensional elastic vibrations of the Earth with a constant angular velocity, 
with an emphasis on the motion of a general node in a rotating frame of reference. 

In equations (\ref{centraleq1}), (\ref{centraleq2}), the gyricity vectors ${\bf \Omega}_{i}$, for  $i=1,\ldots,12,$ and ${\bf \Omega}_c$ depend on the inertial properties of the lattice junctions. 
Taking into account equation (\ref{coriolis1a}), the gyroscopic terms in (\ref{centraleq1}) and (\ref{centraleq2}) can be re-written as 
\begin{equation}
{\bf \Omega} \times {\bf v}' = - 2 m {\boldsymbol{\Theta}} \times {\bf v}',
\end{equation}
where ${\bf v}'$ is the velocity vector in the non-inertial (rotating) frame of reference, and $m$ is the  mass of the corresponding inertial junction.
Noting that $\boldsymbol{\Theta} = \Theta {\bf k},$ we define the gyricity vectors $\boldsymbol{\Omega}_i,$ for $ ~ i=1,\ldots,12 $, and  $\boldsymbol{\Omega}_c $ in connection with the Coriolis force due to the rotation of the Earth as follows
\begin{equation}
    \boldsymbol{\Omega}_i = - 2 m_i \Theta {\bf k}, ~~~ i=1,\ldots,12; ~~~~~~~~ \boldsymbol{\Omega}_c = - 2 m_c \Theta {\bf k}. \label{gyricity0}
\end{equation}
Thus, the general form of the gyricity vector for the nodal elements in the icosahedron lattice can be written as ${\bf \Omega} = (0, 0, \Omega)^T,$ where its third (non-zero) component $\Omega$ is referred to as the {\em gyricity parameter}. 

 \section{Problem formulation and governing equations of the icosahedron-dodecahedron lattice model} \label{sec:time-harmonic_equationszxasx}
In this section, we formulate the governing equations for the nodes of the icosahedron-dodecahedron lattice by defining the reference cell of the structure as shown in Fig. \ref{fig:models}a. The reference cell generates the full icosahedron-dodecahedron lattice through finite rotations by multiples of $2\pi/5$ about the vertical $z$-axis. The position vectors of the six numbered points, and those labelled by N, C and S are given in Table \ref{tab:cellnodes}. The masses allow for the problem formulation of the equations of motion for each node of the structure using the rotational symmetry of the model. This choice of the nodes in the reference cell uses the inherent five-fold rotational symmetry of the icosahedron-dodecahedron lattice. 

The numbered nodal points of the reference cell, as well as the nodes labelled by N, C and S, shown in Fig. \ref{fig:models}a are given by
\begin{table}[h!]
    \centering
    \begin{tabular}{c|c|c|c}
        Node & $x$ & $y$ & $z$ \\
        \hline
        $1$ & $r_1$ & $0$ & $\sqrt{\rho_D^2-r_1^2}$ \\
        \hline
        $2$ & $r_2$ & $0$ & $\sqrt{\rho_D^2-r_2^2}$ \\
        \hline
        $3$ & $2\rho_I/\sqrt{5}$ & $0$ & $-\rho_I/\sqrt{5}$ \\
        \hline
        $4$ & $(2\rho_I/\sqrt{5})\cos{(\frac{\pi}{5})}$ & $(2\rho_I/\sqrt{5})\sin{(\frac{\pi}{5})}$ & $\rho_I/\sqrt{5}$ \\
        \hline
        $5$ & $r_2\cos{(\frac{\pi}{5})}$ & $r_2\sin{(\frac{\pi}{5})}$ & $-\sqrt{\rho_D^2-r_2^2}$ \\
        \hline
        $6$ & $r_1\cos{(\frac{\pi}{5})}$ & $r_1\sin{(\frac{\pi}{5})}$ & $-\sqrt{\rho_D^2-r_1^2}$ \\
        \hline
        $N$ & $0$ & $0$ & $\rho_I$ \\
        \hline
        $C$ & $0$ & $0$ & $0$ \\
        \hline
        $S$ & $0$ & $0$ & $-\rho_I$ \\
    \end{tabular}
    \caption{The node locations of the generating reference cell associated with the symmetry cell shown in Fig. \ref{fig:models}a.}
    \label{tab:cellnodes}
\end{table}
\FloatBarrier
\noindent where $\rho_I$ and $\rho_D$ are the radii of the circumscribed spheres for the icosahedron and dodecahedron in the icosahedron-dodecahedron lattice, respectively, and 
\begin{equation}
    r_1 = \frac{\rho_D}{\varphi\sqrt{3}\sin(\frac{\pi}{5})}, \quad
    r_2 =  \sqrt{
    \frac{4\rho_D^2-\left( \frac{2\rho_D}{\varphi\sqrt{3}} \right)^2}{2\Big(1+\cos(\frac{\pi}{5})\Big)}
    }.
\end{equation}
where $\varphi=(1+\sqrt{5})/2$ is the golden ratio. {
The notation ${\bf x}_k$ is used for the position vector of the $k$-th nodal mass in the icosahedron-dodecahedron lattice; these position vectors  are used to determine the unit vectors characterising the directions of the elastic links which connect the neighbouring masses. }

A general representation for the unit vector from the $i$-th mass in the $p$-th reference cell to the $j$-th mass in the $q$-th reference cell is given by
\begin{equation}
\mathbf{a}_{i,j}^{(p,q)} = \frac{{\bf L}(\frac{2\pi(q-p)}{5})\mathbf{x}_j-\mathbf{x}_i}{|{\bf L}(\frac{2\pi(q-p)}{5})\mathbf{x}_j-\mathbf{x}_i|},
\end{equation}
where ${\bf L}$ is the three-dimensional rotation matrix, representing a counterclockwise rotation by an angle of $2\pi/5$ about the vertical $z$-axis, and is given by
\begin{equation}
{\bf L}=\begin{pmatrix}
    \cos{(2\pi/5)} & -\sin{(2\pi/5)} & 0 \\
    \sin{(2\pi/5)} & \cos{(2\pi/5)} & 0 \\
    0 & 0 & 1
\end{pmatrix}. \label{rotasymmMat}
\end{equation}
The matrices describing the forces introduced by each elastic link for the generating reference cell take the form
\begin{equation}
\mathcal{A}_{i,j}^{(p,q)} = \mathbf{a}_{i,j}^{(p,q)}\left(\mathbf{a}^{\left(p,q\right)}_{i,j}\right)^T.
\end{equation}
By taking into account the rotational symmetry of the icosahedron-dodecahedron lattice, the matrix  characterising the spring connection between the $i$-th node in the $(p+n)$-th cell and the $j$-th node in the $(q+n)$-th cell can be written as
\begin{equation}
\mathcal{A}_{i,j}^{(p+n,q+n)} = (\mathbf{L}^n)\left[ \mathbf{a}_{i,j}^{(p,q)}\left(\mathbf{a}^{\left(p,q\right)}_{i,j}\right)^T \right](\mathbf{L}^n)^{T}.
\end{equation}
The presence of gyroscopic spinners at each nodal point introduces a coupling of the displacement components, leading to vortex-type oscillations of the icosahedron-dodecahedron lattice. The lattice model considered here represents an active gyroscopic system, with the magnitude of the gyricity ${\Omega} = \alpha \omega,$ where $\alpha$ is the chirality parameter and $\omega$ is the radian frequency. 


Taking into account the above problem formulation and notations, the equations of motion in the time-harmonic regime for the masses in the $n$-th reference cell are
{\small
\begin{equation}
\begin{split}
\omega^2 \mathbf{M}_D \mathbf{U}^{(1,n)} &=
\kappa_D \left[ \A{n}{n}{1}{2} + \A{n}{n-1}{1}{1} + \A{n}{n+1}{1}{1} \right] \\
&+ \kappa_T \left[ \A{n}{n}{1}{4} + \A{n}{n-1}{1}{4} + 
\mathcal{A}^{(n,n)}_{1,N}\left(\mathbf{U}^{(1,n)} - \mathbf{U}^{(N)}\right) \right] \\
&+ i\omega^2 \alpha \mathbf{R} \mathbf{U}^{(1,n)},
\end{split} \label{eeqq1}
\end{equation}

\begin{equation}
\begin{split}
\omega^2 \mathbf{M}_D \mathbf{U}^{(2,n)} &=
\kappa_D \left[ \A{n}{n}{2}{1} + \A{n}{n}{2}{5} + \A{n}{n-1}{2}{5} \right] \\
&+ \kappa_T \left[ \A{n}{n}{2}{3} + \A{n}{n}{2}{4} + \A{n}{n-1}{2}{4} \right] \\
&+ i\omega^2 \alpha \mathbf{R} \mathbf{U}^{(2,n)},
\end{split}
\end{equation}

\begin{equation}
\begin{split}
\omega^2 \mathbf{M}_I \mathbf{U}^{(3,n)} =
\kappa_I &\left[ \A{n}{n-1}{3}{3} + \A{n}{n+1}{3}{3} \right. \\ 
&+ \left. \A{n}{n}{3}{4} + \A{n}{n-1}{3}{4} + \mathcal{A}^{(n,n)}_{3,S}\left(\mathbf{U}^{(3,n)} - \mathbf{U}^{(S)}\right) \right] \\
&+ \kappa_T \left[ \A{n}{n}{3}{2} + \A{n}{n}{3}{5} + \A{n}{n-1}{3}{5} \right. \\ 
&+ \left. \A{n}{n}{3}{6} + \A{n}{n-1}{3}{6} \right] \\
&+ \mathcal{A}^{(n,n)}_{3,C}\left(\mathbf{U}^{(3,n)} - \mathbf{U}^{(C)}\right)
+ i\omega^2 \alpha \mathbf{R} \mathbf{U}^{(3,n)},
\end{split}
\end{equation}

\begin{equation}
\begin{split}
\omega^2 \mathbf{M}_I \mathbf{U}^{(4,n)} =
\kappa_I &\left[ \A{n}{n-1}{4}{4} + \A{n}{n+1}{4}{4} \right. \\
&+ \left. \A{n}{n}{4}{3} + \A{n}{n+1}{4}{3} + \mathcal{A}^{(n,n)}_{4,N}\left(\mathbf{U}^{(4,n)} - \mathbf{U}^{(N)}\right) \right] \\
&+ \kappa_T \left[ \A{n}{n}{4}{1} + \A{n}{n+1}{4}{1} + \A{n}{n}{4}{2} \right. \\ &+ \left. \A{n}{n+1}{4}{2} + \A{n}{n}{4}{5} \right] \\
&+ \kappa_C \mathcal{A}^{(n,n)}_{4,C}\left(\mathbf{U}^{(4,n)} - \mathbf{U}^{(C)}\right)
+ i\omega^2 \alpha \mathbf{R} \mathbf{U}^{(4,n)},
\end{split}
\end{equation}

\begin{equation}
\begin{split}
\omega^2 \mathbf{M}_D \mathbf{U}^{(5,n)} &=
\kappa_D \left[ \A{n}{n}{5}{2} + \A{n}{n+1}{5}{2} + \A{n}{n}{5}{6} \right] \\
&+ \kappa_T \left[ \A{n}{n}{5}{4} + \A{n}{n}{5}{3} + \A{n}{n+1}{5}{3} \right] \\
&+ i\omega^2 \alpha \mathbf{R} \mathbf{U}^{(5,n)},
\end{split}
\end{equation}

\begin{equation}
\begin{split}
\omega^2 \mathbf{M}_D \mathbf{U}^{(6,n)} &=
\kappa_D \left[ \A{n}{n+1}{6}{6} + \A{n}{n-1}{6}{6} + \A{n}{n+1}{6}{3} \right] \\
&+ \kappa_T \left[ \A{n}{n}{6}{3} + \A{n}{n}{6}{5} + \mathcal{A}^{(n,n)}_{6,S}\left(\mathbf{U}^{(6,n)} - \mathbf{U}^{(S)}\right) \right] \\
&+ i\omega^2 \alpha \mathbf{R} \mathbf{U}^{(6,n)},
\end{split}
\end{equation}
}\noindent and the governing equations describing the time harmonic vibrations for the masses along the vertical axis of the reference cell are:
{\small
\begin{equation}
	\begin{split}
		\omega^2 \mathbf{M}_I \mathbf{U}^{(S)} &=
		\kappa_T \left[ \sum_{m={n}}^{n+4}(\mathcal{A}^{(n,m)}_{S,3}\left(\mathbf{U}^{(S)} - \mathbf{U}^{(3,m)}\right) +  \mathcal{A}^{(n,m)}_{S,6}\left(\mathbf{U}^{(S)} - \mathbf{U}^{(6,m)}\right)) \right] \\
		&+ \kappa_I \mathcal{A}^{(n,n)}_{S,C}\left(\mathbf{U}^{(S)} - \mathbf{U}^{(C)}\right)
		+ i\omega^2 \alpha \mathbf{R} \mathbf{U}^{(S)},
	\end{split}
\end{equation}

\begin{equation}
	\begin{split}
		\omega^2 \mathbf{M}_I \mathbf{U}^{(N)} &=
		\kappa_T \left[ \sum_{m={n}}^{n+4}(\mathcal{A}^{(n,m)}_{N,4}\left(\mathbf{U}^{(N)} - \mathbf{U}^{(4,m)}\right) +  \mathcal{A}^{(n,m)}_{N,1}\left(\mathbf{U}^{(N)} - \mathbf{U}^{(1,m)}\right)) \right] \\
		&+ \kappa_I \mathcal{A}^{(n,n)}_{N,C}\left(\mathbf{U}^{(N)} - \mathbf{U}^{(C)}\right)
		+ i\omega^2 \alpha \mathbf{R} \mathbf{U}^{(N)},
	\end{split}
\end{equation}

\begin{equation}
	\begin{split}
		\omega^2 \mathbf{M}_C \mathbf{U}^{(C)} &=
		\kappa_I \left[ \sum_{m={n}}^{n+4}(\mathcal{A}^{(n,m)}_{C,3}\left(\mathbf{U}^{(C)} - \mathbf{U}^{(3,m)}\right) + \mathcal{A}^{(n,m)}_{C,4}\left(\mathbf{U}^{(C)} - \mathbf{U}^{(4,m)}\right)) \right] \\
		&+ \kappa_I \left[ \mathcal{A}^{(n,n)}_{C,S}\left(\mathbf{U}^{(C)} - \mathbf{U}^{(S)}\right) + \mathcal{A}^{(n,n)}_{C,N}\left(\mathbf{U}^{(C)} - \mathbf{U}^{(N)}\right)  \right]
		+ i\omega^2 \alpha \mathbf{R} \mathbf{U}^{(C)},
	\end{split} \label{eeqq9}
\end{equation}
}\noindent where 
$\mathbf{M}_D=m_d\mathbf{I},\mathbf{M}_I=m_I\mathbf{I},\mathbf{M}_C=m_C\mathbf{I},$ where $\mathbf{I}$ is the $3\times 3$ identity matrix, represent the mass matrices of the nodal elements within the dodecahedron, icosahedron and core, respectively, the quantities ${\bf U}^{(i)}(\mathbf{x})$ are the three-dimensional displacement vectors, with $\mathbf{x}$ denoting the position vector, and  $\kappa_D,\kappa_I,\kappa_B$ and $\kappa_C$ are the stiffnesses of the elastic links for the dodecahedron, icosahedron, between the layers of the icosahedron and dodecahedron and from the central mass to the icosahedron, respectively. In the above system of equations, the gyroscopic forces are characterised by the chirality parameter $\alpha$ and the matrix ${\bf R},$ which is defined by  
\begin{equation}
\mathbf{R} = \begin{pmatrix}
    0 & 1 & 0 \\
    -1 & 0 & 0 \\
    0 & 0 & 0
\end{pmatrix},
\end{equation}
that corresponds to the rotational axis of the gyroscopic spinners being aligned with the $z$-axis of the icosahedron-dodecahedron lattice, as shown in Fig. \ref{fig:models}a. The parameter values used for the chiral centred icosahedron-dodecahedron lattice to approximate the physical and geometrical properties of the Earth are given in Table \ref{tab:earth_params}. The choice of parameters for the gyroscopic lattice provide a very good agreement with the frequency ranges of the ground vibrations and tectonic plate motions observed during large earthquakes as discussed in Section \ref{sec:math_formulation}.  
In particular, the central nodal point of the lattice approximates the Earth's core, with the elastic links between the central node and the icosahedron vertices resembling the Earth's region between the core and mantle, which includes the outer core. Furthermore, the icosahedron frame models the Earth's mantle, the elastic links between the icosahedron and dodecahedron consist of the mantle region, which is a combination of semi-liquid and solid material \cite{planetHandbook}, and the dodecahedron frame approximates the crust of the Earth. The approximation of the centred icosahedron-dodecahedron lattice with the Earth layers is depicted in Fig. \ref{physicalEarthicosahedron}a. The arrangement of the masses for the nodal elements in the centre, on the vertices of the icosahedron and on the vertices of the dodecahedron is chosen to resemble the mass distribution within the core, mantle and crust layers of the Earth, respectively. The gyroscopic forces acting on each node of the icosahedron-dodecahedron lattice are chosen to be equal in magnitude and are associated with the Coriolis force  (see also derivation in Section \ref{icos:lattzxsa}). 

\section{Optimisation of the node locations for the icosahedron-dodecahedron lattice}\label{opt:algozasxasx}
The approximation of the centred icosahedron-dodecahedron lattice with the positions of earthquake epicentres, in Section \ref{sec:math_formulation}, was performed using an optimisation in Matlab. This also resulted in a very good alignment with the locations of the nodal points and the tectonic plate boundaries on the Earth, which further validated the orientation of the discrete lattice structure relative to the Earth and the comparison between the motions of the nodal elements with the observed major earthquakes. 

We define a procedure to optimise the node positions of the icosahedron-dodecahedron lattice with the locations of earthquakes. In particular, we look to minimise the distance between the nodal elements in the lattice and locations of earthquakes on the Earth. 
The earthquake dataset is obtained from the U.S. Geological Survey (USGS) earthquake catalogue \url{https://earthquake.usgs.gov/earthquakes/search/} (downloaded on 19/05/2025). The tables of the raw earthquake data is also provided in Table \ref{tab:earthquake-data}. 
In this study, only earthquakes of magnitude $7$ and higher, that have occurred in the past $50$ years, are considered. This selection ensures a more accurate match with the higher-magnitude seismic events, and includes the earthquakes discussed in Sections \ref{Mexico:earthquake} - \ref{turk1}. The raw earthquake dataset is in the longitude-latitude format where the depth data of the epicentre is also included, therefore we convert to Cartesian coordinates and position these earthquakes on concentric spheres with radii linked to the earthquakes' depths. The real world geometrical parameters are also used to construct the different layers of the discrete centred icosahedron-dodecahedron lattice as detailed in Section \ref{newsearthasa12a}, which further strengthens the comparison between the discrete lattice structures and the geophysical phenomena. 

In the optimisation, we preserve the structure of the centred icosahedron-dodecahedron lattice by allowing only rotational adjustments. 
The rotational adjustments consist of an initial rotation around the $z$-axis to characterise the spin, followed by a rotation around the $y$-axis to determine the latitude, and an additional rotation around the $z$-axis to specify the longitude.
 For the optimisation procedure, we define the cost function parametrised by the spin ($\alpha$), longitude ($\gamma$) and colatitude ($\beta$) angles as follows 

\begin{equation} \label{eq:cost-functional}
    J(\alpha,\beta,\gamma) = \sum_{i=1}^{N=32}\min_j||\mathcal{R}(\alpha,\beta,\gamma)\mathbf{v}^{(i)} - \mathbf{x}^{(j)}||_2,
\end{equation}
where $N$ is the number of nodes included from the icosahedron-dodecahedron lattice (note the central node is {\em not} included), $\mathbf{x}^{(j)}$ are the Cartesian coordinates of the earthquake locations (including depth), $\mathbf{v}^{(i)}$ are the Cartesian coordinates of the icosahedron-dodecahedron lattice nodes and 
{\small
\begin{equation}
\mathcal{R}(\alpha,\beta,\gamma) = R_z(\alpha)R_y(\beta)R_z(\gamma)=\begin{pmatrix}
    \cos\alpha\cos\beta\cos\gamma - \sin\alpha\sin\gamma & -\cos\alpha\cos\beta\sin\gamma-\sin\alpha\cos\gamma & \cos\alpha\sin\beta \\
    \sin\alpha\cos\beta\cos\gamma+\cos\alpha\sin\gamma & -\sin\alpha\cos\beta\sin\gamma+\cos\alpha\cos\gamma & \sin\alpha\sin\beta \\
    -\sin\beta\cos\gamma & \sin\beta\sin\gamma & \cos\beta
\end{pmatrix},
\end{equation}
}

\noindent where $R_y(\cdot)$ and $R_z(\cdot)$ are rotation matrices about the $y$- and $z$- axes, respectively.
Figure \ref{fig:models}a shows the reference cell of the centred icosahedron-dodecahedron lattice. We use this parametrisation as the vertical axis of the centred icosahedron-dodecahedron lattice is aligned with the $z$-axis. 
Additionally, the half-plane $Oxz,$ $x>0,$ of the icosahedron-dodecahedron lattice (see, for example Figure \ref{fig:models}a) includes the prime meridian in Greenwich.   
The function (\ref{eq:cost-functional}) provides a quantitative measure of the alignment between the discrete lattice structure and the spatial distribution of earthquake epicentres, presenting a robust framework for analysing large-scale geophysical tectonic activity.


\begin{figure}[h!]
	\centering
	\begin{subfigure}[h!]{0.9\textwidth}
		\centering
		\includegraphics[width=\linewidth]{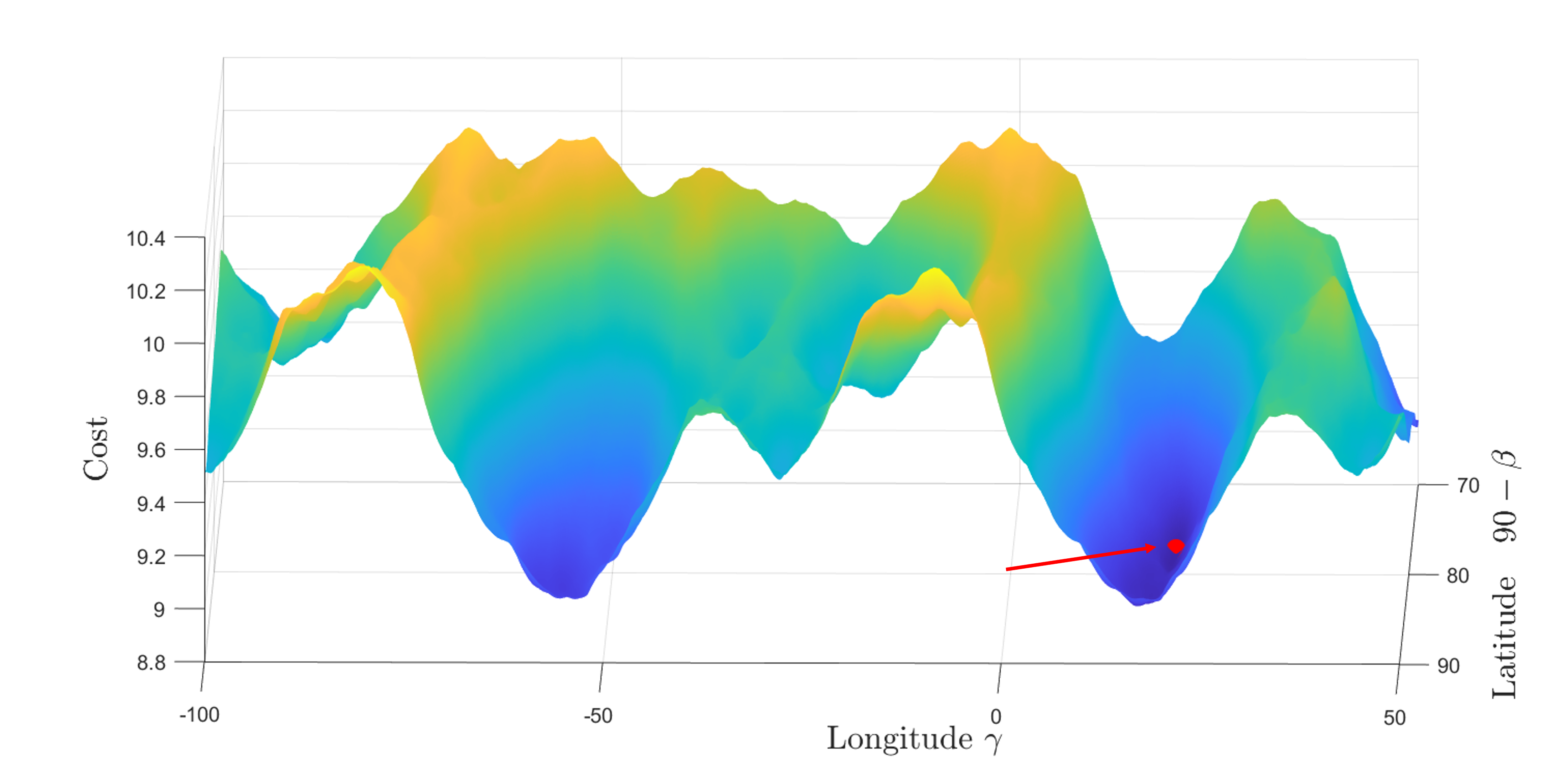}
		\caption{} \label{optmisation_surfaces}
	\end{subfigure} 
	\begin{subfigure}[h!]{0.8\textwidth}
		\centering
		\includegraphics[width=\linewidth]{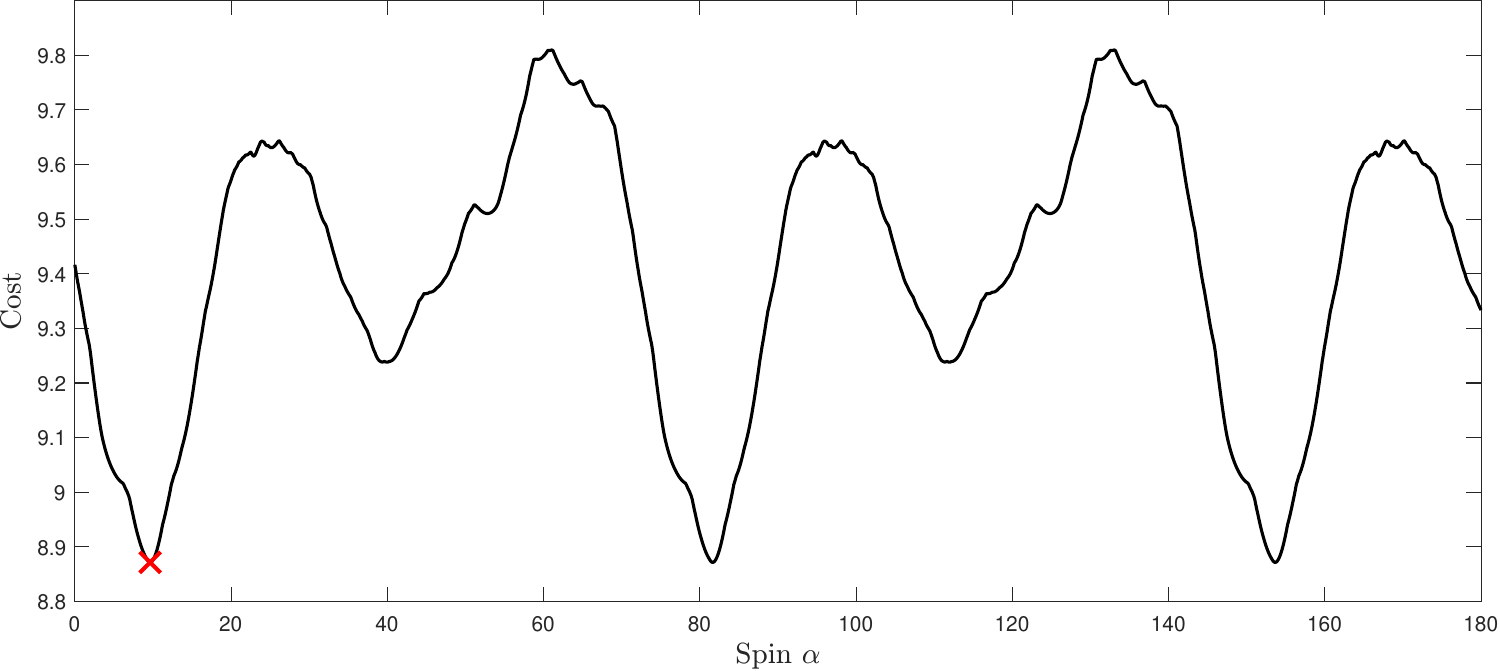}
		\caption{} \label{optimisation_curves}
	\end{subfigure} 
	\caption{
    {
    (a) Cross-sectional surface plot of the cost function (\ref{eq:cost-functional}) with two minimum cost regions for $\alpha=9.64^{\circ}.$ The {\em Tectonic North} point, shown in Figure \ref{fig:structuremap}a with the geographical coordinates $79.60^\circ$ N, $20.72^\circ$ E, is located in the rightmost minimum cost region and marked by a red dot and an arrow, and the centre of rotation of the global low pressure region, observed in Greenland (see Fig \ref{stormsocean}) is positioned in the leftmost minimum region. (b)  Diagram of the optimised choice of the spin angle $\alpha$ versus the cost function with colatitude $\beta = 10.40^\circ$ and longitude $\gamma = 20.72^\circ,$ associated with the Tectonic North. The value $\alpha=9.64^{\circ}$ is marked by a red cross.}}
	\label{}
\end{figure}

We minimise the function (\ref{eq:cost-functional}) using a gradient descent algorithm. In particular, the fmincon function in Matlab has been used with the Sequential Quadratic Programming (SQP) algorithm, due to its efficiency in handling constraints and fast convergence. In addition, the SQP algorithm  uses gradient-based optimisation and quasi-Newton methods, which provides effective methods to obtain an optimal solution with computational efficiency. 
It is important to note that variations in the initial position of the icoshaedron-dodecahedron lattice can lead to numerous optimal solutions. Thus, we employed a multi-start function to explore a range of initial values of the lattice structure, and determined the optimal alignment of the icosahedron-dodecahedron lattice with the earthquake epicentres. 
Although the optimisation method provides an approximation for the orientation of the icosahedron-dodecahedron lattice relative to the earthquakes in the continuum Earth model, it shows excellent agreement between the $33$ nodal elements of the lattice nodes with the tectonic plate boundaries and earthquake distributions, as detailed in Section \ref{sec:math_formulation}.

{
Fig. \ref{optmisation_surfaces} shows the surface plot of the cost function (\ref{eq:cost-functional}) as a function of the longitude, $\gamma,$ and latitude, $90^{\circ}-\beta$, where $\beta$ is the colatitude, for the spin angle $\alpha=9.64^{\circ}.$ 
The longitude and latitude at  the 
point of minimum, 
shown by the arrow in Fig. \ref{optmisation_surfaces}, correspond to the position of the northern node N of the icosahedron-dodecahedron lattice, illustrated in Figure \ref{fig:structuremap}a.
For the longitude and latitude of the Tectonic North in the highlighted minimum point in Fig. \ref{optmisation_surfaces} (at $\gamma = 20.72^\circ$, $\beta = 10.4^\circ$), the corresponding optimal values of the spin angle $\alpha$ are presented in Fig. \ref{optimisation_curves}, where the value $\alpha= 9.64^{\circ}$ has been marked with a cross. The  optimisation procedure results in an  excellent match between the icosahedron-dodecahedron lattice and the boundaries of the tectonic plates, as shown in Fig. \ref{fig:structuremap}b. 
}

 \section{Earthquake dataset tables}\label{data:earthquaket}
 
This section provides the table of the earthquake dataset associated with the optimisation detailed in the previous section. The tables consist of large earthquakes with a magnitude of $7$ and above, and the visualisation of the earthquakes' epicentres is provided in Figure \ref{fig:structuremap}b. The raw earthquake data were sourced from a public dataset, and can be found at  \url{https://earthquake.usgs.gov/earthquakes/search/}.


\begin{landscape}

{\small
    
\begin{longtable}[c]{|l|l|l|l|l|l|l|l|l|l|l|l|l|l|l|l|l|l|l|l|l|l|}

        \caption{{
        The earthquake dataset. 
        }\label{tab:earthquake-data}} \\
         \hline
        date, time & latitude (deg) & longitude (deg) & depth (km) & mag & magType & depthError & magError & locationSource & magSource \\
        \hline
        \endfirsthead
        \\
        \hline
       date, time & latitude (deg) & longitude (deg) & depth (km) & mag & magType & depthError & magError & locationSource & magSource \\
        \hline
        \endhead
        
        2004-12-26T00:58:53.450Z & 3.295 & 95.982 & 30 & 9.1 & mw & ~ & ~ & us & official \\ \hline
        2011-03-11T05:46:24.120Z & 38.297 & 142.373 & 29 & 9.1 & mww & ~ & ~ & us & official \\ \hline
        2010-02-27T06:34:11.530Z & -36.122 & -72.898 & 22.9 & 8.8 & mww & 9.2 & ~ & us & official \\ \hline
        2012-04-11T08:38:36.720Z & 2.327 & 93.063 & 20 & 8.6 & mw & ~ & ~ & us & official \\ \hline
        2005-03-28T16:09:36.530Z & 2.085 & 97.108 & 30 & 8.6 & mww & ~ & ~ & us & official \\ \hline
        2001-06-23T20:33:14.130Z & -16.265 & -73.641 & 33 & 8.4 & mww & ~ & ~ & us & official \\ \hline
        2007-09-12T11:10:26.830Z & -4.438 & 101.367 & 34 & 8.4 & mww & ~ & ~ & us & official \\ \hline
        1977-08-19T06:08:55.510Z & -11.164 & 118.378 & 25 & 8.31 & mw & 3.9 & 0.1 & iscgem & iscgem \\
\hline
    
\end{longtable}}
\end{landscape}



\begin{thebibliography}{70}
{\small 
\bibitem{rundle2003statistical}
Rundle, J., Turcotte, D., Shcherbakov, R., Klein, W. \& Sammis, C. Statistical physics approach to understanding the multiscale dynamics of earthquake fault systems. {\em Reviews Of Geophysics}. \textbf{41} (2003)

\bibitem{moczo2014finite}
Moczo, P., Kristek, J. \& Gális, M. The finite-difference modelling of earthquake motions: Waves and ruptures. (Cambridge University Press,2014)

\bibitem{lu2011large}
Lu, J., Elgamal, A., Yan, L., Law, K. \& Conte, J. Large-scale numerical modeling in geotechnical earthquake engineering. {\em International Journal Of Geomechanics}. \textbf{11}, 490-503 (2011)

\bibitem{minster1974numerical}
Minster, J., Jordan, T., Molnar, P. \& Haines, E. Numerical modelling of instantaneous plate tectonics. {\em Geophysical Journal International}. \textbf{36}, 541-576 (1974)

\bibitem{shearer2019introduction}
Shearer, P. Introduction to seismology. (Cambridge university press, 2019)

\bibitem{ghosh2019role}
Ghosh, A., Holt, W. \& Bahadori, A. Role of large-scale tectonic forces in intraplate earthquakes of central and eastern North America. {\em Geochemistry, Geophysics, Geosystems}. \textbf{20}, 2134-2156 (2019)

\bibitem{bahadori2019geodynamic}
Bahadori, A. \& Holt, W. Geodynamic evolution of southwestern North America since the Late Eocene. {\em Nature Communications}. \textbf{10}, 5213 (2019)

\bibitem{kandiah2024controlling}
Kandiah, A., Jones, I., Movchan, N. \& Movchan, A. Controlling the motion of gravitational spinners and waves in chiral waveguides. {\em Scientific Reports}. \textbf{14}, 1203 (2024)

\bibitem{kandiah2025dispersion}
Kandiah, A., Jones, I., Movchan, N. \& Movchan, A. Dispersion and asymmetry of chiral gravitational waves in gyroscopic mechanical systems. Part 1: Discrete lattice strips. {\em Quarterly Journal Of Mechanics And Applied Mathematics}. \textbf{78}, hbaf004 (2025)

\bibitem{kandiah2025dispersion2}
Kandiah, A., Jones, I., Movchan, N. \& Movchan, A. Dispersion and asymmetry of chiral gravitational waves in gyroscopic mechanical systems. Part 2: Continuum asymptotic models in equatorial and polar regions. {\em Quarterly Journal Of Mechanics And Applied Mathematics}. \textbf{78}, hbaf005 (2025)

\bibitem{deen2017first}
Deen, M., Wielandt, E., Stutzmann, E., Crawford, W., Barruol, G. \& Sigloch, K. First observation of the Earth's permanent free oscillations on ocean bottom seismometers. {\em Geophysical Research Letters}. \textbf{44}, 10-988 (2017)

\bibitem{aki1980attenuation}
Aki, K. Attenuation of shear-waves in the lithosphere for frequencies from 0.05 to 25 Hz. {\em Physics Of The Earth And Planetary Interiors}. \textbf{21}, 50-60 (1980)

\bibitem{tanimoto2005oceanic}
Tanimoto, T. The oceanic excitation hypothesis for the continuous oscillations of the Earth. {\em Geophysical Journal International}. \textbf{160}, 276-288 (2005)

\bibitem{kobayashi1998continuous}
Kobayashi, N. \& Nishida, K. Continuous excitation of planetary free oscillations by atmospheric disturbances. {\em Nature}. \textbf{395}, 357-360 (1998)

\bibitem{suda1998earth}
Suda, N., Nawa, K. \& Fukao, Y. Earth's background free oscillations. {\em Science}. \textbf{279}, 2089-2091 (1998)

\bibitem{webb2007earth}
Webb, S. The Earth’s ‘hum’ is driven by ocean waves over the continental shelves. {\em Nature}. \textbf{445}, 754-756 (2007)




\bibitem{carbone2021mathematical}
Carbone, V., Piersanti, M., Materassi, M., Battiston, R., Lepreti, F. \& Ubertini, P. A mathematical model of lithosphere–atmosphere coupling for seismic events. {\em Scientific Reports}. \textbf{11}, 8682 (2021)


\bibitem{de2016statistical}
Arcangelis, L., Godano, C., Grasso, J. \& Lippiello, E. Statistical physics approach to earthquake occurrence and forecasting. {\em Physics Reports}. \textbf{628} pp. 1-91 (2016)

\bibitem{arrowsmith2010seismoacoustic}
Arrowsmith, S., Johnson, J., Drob, D. \& Hedlin, M. The seismoacoustic wavefield: A new paradigm in studying geophysical phenomena. {\em Reviews Of Geophysics}. \textbf{48} (2010)


\bibitem{anagnos1988review}
Anagnos, T. \& Kiremidjian, A. A review of earthquake occurrence models for seismic hazard analysis. {\em Probabilistic Engineering Mechanics}. \textbf{3}, 3-11 (1988)

\bibitem{kagan1994observational}
Kagan, Y. Observational evidence for earthquakes as a nonlinear dynamic process. {\em Physica D: Nonlinear Phenomena}. \textbf{77}, 160-192 (1994)

\bibitem{historic1930}
Historic Natural Events. Nature 125, 149 (1930). https://doi.org/10.1038/125149a0

\bibitem{the1884}
The Low Barometer of January 26, 1884. Nature 30, 58–59 (1884). https://doi.org/10.1038/030058a0

\bibitem{carta2019wave}
Carta, G., Jones, I., Movchan, N. \& Movchan, A. Wave polarization and dynamic degeneracy in a chiral elastic lattice. {\em Proceedings Of The Royal Society A}. \textbf{475}, 20190313 (2019)

\bibitem{carta2018elastic}
Carta, G., Nieves, M., Jones, I., Movchan, N. \& Movchan, A. Elastic chiral waveguides with gyro-hinges. {\em The Quarterly Journal Of Mechanics And Applied Mathematics}. \textbf{71}, 157-185 (2018)

\bibitem{allison2024mechanical}
Allison, F., Selsil, Ö., Haslinger, S. \& Movchan, A. A mechanical analogue of electromagnetic induction for waves in a chiral elastic structure. {\em Proceedings A}. \textbf{480}, 20240372 (2024)

\bibitem{garau2018interfacial}
Garau, M., Carta, G., Nieves, M., Jones, I., Movchan, N. \& Movchan, A. Interfacial waveforms in chiral lattices with gyroscopic spinners. {\em Proceedings Of The Royal Society A: Mathematical, Physical And Engineering Sciences}. \textbf{474}, 20180132 (2018)

\bibitem{carta2014dispersion}
Carta, G., Brun, M., Movchan, A., Movchan, N. \& Jones, I. Dispersion properties of vortex-type monatomic lattices. {\em International Journal Of Solids And Structures}. \textbf{51}, 2213-2225 (2014)

\bibitem{brun2012vortex}
Brun, M., Jones, I. \& Movchan, A. Vortex-type elastic structured media and dynamic shielding. {\em Proceedings Of The Royal Society A: Mathematical, Physical And Engineering Sciences}. \textbf{468}, 3027-3046 (2012)

\bibitem{carta2020one}
Carta, G., Colquitt, D., Movchan, A., Movchan, N. \& Jones, I. One-way interfacial waves in a flexural plate with chiral double resonators. {\em Philosophical Transactions Of The Royal Society A}. \textbf{378}, 20190350 (2020)

\bibitem{kandiah2023effect}
Kandiah, A., Jones, I., Movchan, N. \& Movchan, A. Effect of gravity on the dispersion and wave localisation in gyroscopic elastic systems. {\em Mechanics Of Heterogeneous Materials}. pp. 219-274 (2023)

\bibitem{carta2017deflecting}
Carta, G., Jones, I., Movchan, N., Movchan, A. \& Nieves, M. “Deflecting elastic prism” and unidirectional localisation for waves in chiral elastic systems. {\em Scientific Reports}. \textbf{7}, 26 (2017)

\bibitem{tallarico2017tilted}
Tallarico, D., Movchan, N., Movchan, A. \& Colquitt, D. Tilted resonators in a triangular elastic lattice: chirality, Bloch waves and negative refraction. {\em Journal Of The Mechanics And Physics Of Solids}. \textbf{103} pp. 236-256 (2017)

\bibitem{kandiah2023gravity}
Kandiah, A., Movchan, N. \& Movchan, A. Gravity-induced waveforms in chiral non-periodic waveguides. {\em International Journal Of Solids And Structures}. \textbf{285} pp. 112528 (2023)

\bibitem{sato2012seismic}
Sato, H., Fehler, M. \& Maeda, T. Seismic wave propagation and scattering in the heterogeneous earth. (Springer, 2012)

\bibitem{benioff1961excitation}
Benioff, H., Press, F. \& Smith, S. Excitation of the free oscillations of the Earth by earthquakes. {\em Journal Of Geophysical Research}. \textbf{66}, 605-619 (1961)

\bibitem{jeans1923propagation}
Jeans, J. The propagation of earthquake waves. {\em Proceedings Of The Royal Society Of London. Series A, Containing Papers Of A Mathematical And Physical Character}. \textbf{102}, 554-574 (1923)

\bibitem{alterman1959oscillations}
Alterman, Z., Jarosch, H. \& Pekeris, C. Oscillations of the Earth. {\em Proceedings Of The Royal Society Of London. Series A. Mathematical And Physical Sciences}. \textbf{252}, 80-95 (1959)

\bibitem{artru2004acoustic}
Artru, J., Farges, T. \& Lognonné, P. Acoustic waves generated from seismic surface waves: Propagation properties determined from Doppler sounding observations and normal-mode modelling. {\em Geophysical Journal International}. \textbf{158}, 1067-1077 (2004)

\bibitem{lachugin2005earth}
Lachugin, K. Is the Earth a big crystal? Based on the work by N.F. Goncharov, V.A. Makarov, V.S. Morozov. (Zakharov, Moscow, 224 pages, 2005)

\bibitem{hughes2025}
Hughes, A. Earth's core may be changing shape and it has scientists puzzled. {\em BBC Science Focus}. (2025)

\bibitem{gubbins1981rotation}
Gubbins, D. Rotation of the inner core. {\em Journal Of Geophysical Research: Solid Earth}. \textbf{86}, 11695-11699 (1981)

\bibitem{montagner2008normal}
Montagner, J. \& Roult, G. Normal modes of the earth. {\em Journal Of Physics: Conference Series}. \textbf{118}, 012004 (2008)


















\bibitem{roult2010observation}
Roult, G., Roch, J. \& Clévédé, E. Observation of split modes from the 26th December 2004 Sumatra-Andaman mega-event. {\em Physics Of The Earth And Planetary Interiors}. \textbf{179}, 45-59 (2010)

\bibitem{vidale2025annual}
Vidale, J., Wang, W., Wang, R., Pang, G. \& Koper, K. Annual-scale variability in both the rotation rate and near surface of Earth’s inner core. {\em Nature Geoscience}. pp. 1-6 (2025)

\bibitem{yang2023multidecadal}
Yang, Y. \& Song, X. Multidecadal variation of the Earth’s inner-core rotation. {\em Nature Geoscience}. \textbf{16} pp. 182-187 (2023), https://doi.org/10.1038/s41561-022-01112-z

\bibitem{ChiralMultiStructures}
Nieves, M., Carta, G., Jones, I., Movchan, A. \& Movchan, N. Vibrations and elastic waves in chiral multi-structures. {\em Journal Of The Mechanics And Physics Of Solids}. \textbf{121} pp. 387-408 (2018), https://www.sciencedirect.com/science/article/pii/S0022509618303570

\bibitem{chiralElasticChain}
Jones, I., Movchan, N. \& Movchan, A. Two-Dimensional Waves in A Chiral Elastic Chain: Dynamic Green's Matrices and Localised Defect Modes. {\em The Quarterly Journal Of Mechanics And Applied Mathematics}. \textbf{73}, 305-328 (2021,1), https://doi.org/10.1093/qjmam/hbaa014

\bibitem{subich2018higher}
Subich, C. Higher-order finite volume differential operators with selective upwinding on the icosahedral spherical grid. {\em Journal Of Computational Physics}. \textbf{368} pp. 21-46 (2018)

\bibitem{williamson1968integration}
Williamson, D. Integration of the barotropic vorticity equation on a spherical geodesic grid. {\em Tellus}. \textbf{20}, 642-653 (1968)

\bibitem{williamson2007evolution}
Williamson, D. The evolution of dynamical cores for global atmospheric models. {\em Journal of the Meteorological Society of Japan. Ser. II}. \textbf{85} pp. 241-269 (2007)

\bibitem{satoh2014non}
Satoh, M., Tomita, H., Yashiro, H., Miura, H., Kodama, C., Seiki, T., Noda, A., Yamada, Y., Goto, D., Sawada, M. \& Others The non-hydrostatic icosahedral atmospheric model: Description and development. {\em Progress In Earth And Planetary Science}. \textbf{1} pp. 1-32 (2014)

\bibitem{fisher1943world}
Fisher, I. A world map on a regular icosahedron by gnomonic projection. {\em Geographical Review}. \textbf{33}, 605-619 (1943)

\bibitem{chao2000coseismic}
Chao, B. \& Gross, R. Coseismic excitation of the Earth’s polar motion. {\em International Astronomical Union Colloquium}. \textbf{178} pp. 355-367 (2000)

\bibitem{wang2024inner}
Wang, W., Vidale, J., Pang, G., Koper, K. \& Wang, R. Inner core backtracking by seismic waveform change reversals. {\em Nature}. \textbf{631}, 340-343 (2024)

\bibitem{cambiotti2016residual}
Cambiotti, G., Wang, X., Sabadini, R. \& Yuen, D. Residual polar motion caused by coseismic and interseismic deformations from 1900 to present. {\em Geophysical Journal International}. \textbf{205}, 1165-1179 (2016)

\bibitem{mccarthy1996path}
McCarthy, D. \& Luzum, B. Path of the mean rotational pole from 1899 to 1994. {\em Geophysical Journal International}. \textbf{125}, 623-629 (1996)

\bibitem{song1996seismological}
Song, X. \& Richards, P. Seismological evidence for differential rotation of the Earth's inner core. {\em Nature}. \textbf{382}, 221-224 (1996)

\bibitem{wang2022seismological}
Wang, W. \& Vidale, J. Seismological observation of Earth’s oscillating inner core. {\em Science Advances}. \textbf{8}, eabm9916 (2022)

\bibitem{yao2019temporal}
Yao, J., Tian, D., Sun, L. \& Wen, L. Temporal change of seismic Earth's inner core phases: Inner core differential rotation or temporal change of inner core surface?. {\em Journal Of Geophysical Research: Solid Earth}. \textbf{124}, 6720-6736 (2019)

\bibitem{greiner2000influence}
Greiner-Mai, H., Jochmann, H. \& Barthelmes, F. Influence of possible inner-core motions on the polar motion and the gravity field. {\em Physics Of The Earth And Planetary Interiors}. \textbf{117}, 81-93 (2000)

\bibitem{zhang2005inner}
Zhang, J., Song, X., Li, Y., Richards, P., Sun, X. \& Waldhauser, F. Inner core differential motion confirmed by earthquake waveform doublets. {\em Science}. \textbf{309}, 1357-1360 (2005)

\bibitem{mathews1991forced}
Mathews, P., Buffett, B., Herring, T. \& Shapiro, I. Forced nutations of the Earth: Influence of inner core dynamics: 1. Theory. {\em Journal Of Geophysical Research: Solid Earth}. \textbf{96}, 8219-8242 (1991)

\bibitem{song1997anisotropy}
Song, X. Anisotropy of the Earth's inner core. {\em Reviews Of Geophysics}. \textbf{35}, 297-313 (1997)

\bibitem{planetHandbook}
Ladders, K. \& Fegley, J. The Earth and the Moon. {\em The Planetary Scientist’s Companion}. (1998) ,12, https://doi.org/10.1093/oso/9780195116946.003.0006

\bibitem{mcdonough1995composition}
McDonough, W. \& Sun, S. The composition of the Earth. {\em Chemical Geology}. \textbf{120}, 223-253 (1995)

\bibitem{o2005mantle}
O'Neill, H. Mantle Composition. {\em The Mantle And Core: Treatise On Geochemistry, Volume 2}. \textbf{2} pp. 1 (2005)

\bibitem{BirdTectonicPlates}
Bird, P. An updated digital model of plate boundaries. {\em Geochemistry, Geophysics, Geosystems}. \textbf{4} (2003), https://agupubs.onlinelibrary.wiley.com/doi/abs/10.1029/2001GC000252

\bibitem{mccann1979seismic}
McCann, W., Nishenko, S., Sykes, L. \& Krause, J. Seismic gaps and plate tectonics: Seismic potential for major boundaries. {\em Earthquake Prediction And Seismicity Patterns}. pp. 1082-1147 (1979)

\bibitem{frohlich1992earthquake}
Frohlich, C. \& Apperson, K. Earthquake focal mechanisms, moment tensors, and the consistency of seismic activity near plate boundaries. {\em Tectonics}. \textbf{11}, 279-296 (1992)

\bibitem{khattri1987great}
Khattri, K. Great earthquakes, seismicity gaps and potential for earthquake disaster along the Himalaya plate boundary. {\em Tectonophysics}. \textbf{138}, 79-92 (1987)

\bibitem{berryman2012major}
Berryman, K., Cochran, U., Clark, K., Biasi, G., Langridge, R. \& Villamor, P. Major earthquakes occur regularly on an isolated plate boundary fault. {\em Science}. \textbf{336}, 1690-1693 (2012)

\bibitem{sykes1981repeat}
Sykes, L. \& Quittmeyer, R. Repeat times of great earthquakes along simple plate boundaries. {\em Earthquake Prediction: An International Review}. \textbf{4} pp. 217-247 (1981)

\bibitem{chao1996seismic}
Chao, B., Gross, R. \& Han, Y. Seismic excitation of the polar motion, 1977–1993. {\em Pure And Applied Geophysics}. \textbf{146} pp. 407-419 (1996)

\bibitem{rochester1973earth}
Rochester, M. The Earth's rotation. {\em EOS, Transactions American Geophysical Union}. \textbf{54}, 769-780 (1973)

\bibitem{eubanks1993variations}
Eubanks, T. Variations in the orientation of the Earth. {\em Contributions Of Space Geodesy To Geodynamics: Earth Dynamics}. \textbf{24} pp. 1-54 (1993)

\bibitem{gross2000excitation}
Gross, R. The excitation of the Chandler wobble. {\em Geophysical Research Letters}. \textbf{27}, 2329-2332 (2000)

\bibitem{brzezinski2002oceanic}
Brzeziński, A. \& Nastula, J. Oceanic excitation of the Chandler wobble. {\em Advances In Space Research}. \textbf{30}, 195-200 (2002)

\bibitem{dahlen1975influence}
Dahlen, F. \& Smith, M. The influence of rotation on the free oscillations of the Earth. {\em Philosophical Transactions Of The Royal Society Of London. Series A, Mathematical And Physical Sciences}. \textbf{279}, 583-624 (1975)

\bibitem{castro2016review}
Castro, R., P\'erez-Campos, X., Z\'u\~niga, R., Ram\'irez-Guzm\'an, L., Aguirre, J., Husker, A., Cu\'ellar, A. \& S\'anchez, T. A review on advances in seismology in Mexico after 30 years from the 1985 earthquake. {\em Journal Of South American Earth Sciences}. \textbf{70} pp. 49-54 (2016)

\bibitem{flores2007seismic}
Flores-Estrella, H., Yussim, S. \& Lomnitz, C. Seismic response of the Mexico City Basin: A review of twenty years of research. {\em Natural Hazards}. \textbf{40} pp. 357-372 (2007)

\bibitem{beck1986factors}
Beck, J. \& Hall, J. Factors contributing to the catastrophe in Mexico City during the earthquake of September 19, 1985. {\em Geophysical Research Letters}. \textbf{13}, 593-596 (1986)

\bibitem{katayama2006earthquake}
Katayama, T. Earthquake disaster mitigation and earthquake engineering in Japan–a review with a special emphasis on the Kobe earthquake and its impact. {\em Journal Of Disaster Research}. \textbf{1}, 11-24 (2006)

\bibitem{morozov2021mechanism}
Morozov, V. \& Manevich, A. Mechanism of Rupture Formation of the Hanshin–Awaji Earthquake (Kobe, Japan) January 17, 1995, M 6.9. {\em Doklady Earth Sciences}. \textbf{499} pp. 654-660 (2021)

\bibitem{wald1996slip}
Wald, D. Slip history of the 1995 Kobe, Japan, earthquake determined from strong motion, teleseismic, and geodetic data. {\em Journal Of Physics Of The Earth}. \textbf{44}, 489-503 (1996)

\bibitem{li1998delineation}
Li, Y., Aki, K., Vidale, J. \& Alvarez, M. A delineation of the Nojima fault ruptured in the M7. 2 Kobe, Japan, earthquake of 1995 using fault zone trapped waves. {\em Journal Of Geophysical Research: Solid Earth}. \textbf{103}, 7247-7263 (1998)

\bibitem{utkucu2003slip}
Utkucu, M., Nalbant, S., McCloskey, J., Steacy, S. \& Alptekin, Ö. Slip distribution and stress changes associated with the 1999 November 12, Düzce (Turkey) earthquake (Mw= 7.1). {\em Geophysical Journal International}. \textbf{153}, 229-241 (2003)

\bibitem{motosaka2002ground}
Motosaka, M. \& Somer, A. Ground motion directionality inferred from a survey of minaret damage during the 1999 Kocaeli and Düzce, Turkey earthquakes. {\em Journal Of Seismology}. \textbf{6} pp. 419-430 (2002)

\bibitem{akyuz2002surface}
Akyuz, H., Hartleb, R., Barka, A., Altunel, E., Sunal, G., Meyer, B. \& Armijo, V. Surface rupture and slip distribution of the 12 November 1999 Duzce earthquake (M 7.1), North Anatolian fault, Bolu, Turkey. {\em Bulletin Of The Seismological Society Of America}. \textbf{92}, 61-66 (2002)

\bibitem{duman2005step}
Duman, T., Emre, O., Dogan, A. \& Ozalp, S. Step-over and bend structures along the 1999 Duzce earthquake surface rupture, North Anatolian fault, Turkey. {\em Bulletin Of The Seismological Society Of America}. \textbf{95}, 1250-1262 (2005)

\bibitem{lognonne1998computation}
Lognonné, P., Clévédé, E. \& Kanamori, H. Computation of seismograms and atmospheric oscillations by normal-mode summation for a spherical earth model with realistic atmosphere. {\em Geophysical Journal International}. \textbf{135}, 388-406 (1998)

\bibitem{kanamori1978quantification}
Kanamori, H. Quantification of earthquakes. {\em Nature}. \textbf{271}, 411-414 (1978)


\bibitem{schneider2006general}
Schneider, T. The general circulation of the atmosphere. {\em Annu. Rev. Earth Planet. Sci.}. \textbf{34}, 655-688 (2006)



\bibitem{loboda2010theoretical}
Loboda, O. \& Goncharuk, V. Theoretical study on icosahedral water clusters. {\em Chemical Physics Letters}. \textbf{484}, 144-147 (2010)

}
\end{thebibliography}

\end{document}